\documentclass[useAMS,usenatbib,twoside,a4paper,iop,numberedappendix,appendixfloats]{emulateapj} 
\usepackage{graphicx,mathrsfs}
\slugcomment{{\sc Accepted for publication in ApJ on October 20, 2013 issue}}

\usepackage{epstopdf}
\usepackage{appendix}
\usepackage{color}

\newcommand{\Msol}{\mbox{$M_{\odot}$}}

\newcommand{\Halpha}{\mbox{H$\alpha$} }
\newcommand{\Hbeta}{\mbox{H$\beta$} }

\def\deg      {{\ifmmode^\circ\else$^\circ$\fi}}

\shorttitle{ZENS III.  Galaxy Photometric Measurements and Resolved Color Properties of Satellites}
\shortauthors{Cibinel et al.} 

\begin{document}

\title{The Zurich ENvironmental Study of Galaxies in Groups along the Cosmic Web. III. \\ 
Galaxy Photometric Measurements and the Spatially-Resolved Color Properties \\ of Early- and Late-Type Satellites in Diverse Environments \footnotemark[\dag]}
  
\author{A. Cibinel,\altaffilmark{1,$\star$}, 
C. M. Carollo\altaffilmark{1,$\star$}, 
S. J. Lilly\altaffilmark{1},
S. Bonoli\altaffilmark{2},
F. Miniati\altaffilmark{1}, 
A. Pipino\altaffilmark{1},
J. D. Silverman\altaffilmark{3}, 
J. H. van Gorkom\altaffilmark{4},
E. Cameron\altaffilmark{1}, 
A. Finoguenov\altaffilmark{5}, 
P. Norberg\altaffilmark{6},
Y. Peng\altaffilmark{1},
C. S. Rudick\altaffilmark{1}
}
 
\altaffiltext{$\star$}{E-mail: \texttt{cibinel@phys.ethz.ch, marcella@phys.ethz.ch}}
\altaffiltext{1}{Institute of Astronomy, ETH Zurich, CH-8093 Zurich, Switzerland}
\altaffiltext{2}{Institute for Theoretical Physics, University of Z\"urich, Winterthurerstrasse 190, CH-8057 Z\"urich, Switzerland}
\altaffiltext{3}{Kavli Institute for the Physics and Mathematics of the Universe (WPI), Todai Institutes for Advanced Study, The University of Tokyo, 5-1-5 Kashiwanoha, Kashiwa, 277-8583, Japan}
\altaffiltext{4}{Department of Astronomy, Columbia University, New York, NY 10027, USA}
\altaffiltext{5}{Max-Planck-Institut f\"ur extraterrestrische Physik, D-84571 Garching, Germany}
\altaffiltext{6}{Institute for Computational Cosmology, Department of Physics, Durham University, South Road, Durham DH1 3LE, UK}
\altaffiltext{$\dagger$}{Based on observations collected at the European Southern Observatory, La Silla Chile. Program ID 177.A-0680}

\begin{abstract}

We present  photometric measurements for the galaxies -- and when possible their bulges
 and disks -- in the $0.05<z<0.0585$  groups of the Zurich Environmental Study (ZENS); these measurements 
include $(B-I)$ colors, color gradients and maps, color dispersions,  as well as stellar masses and star-formation rates.  
 The  ZENS galaxies are classified  into quenched, moderately star-forming, and strongly star-forming using 
 a combination of spectral features and  FUV-to-optical colors; this approach optimally distinguishes
quenched systems from dust-reddened star-forming galaxies.  The latter contribute up to $50\%$ to the $(B-I)$ ``red sequence" at $\sim10^{10}\Msol$.   
At fixed morphological or spectral type,  we find that  galaxy stellar masses are largely independent of
environment, and especially of halo mass. 
As a first utilization of our photometric database, we study, at fixed stellar mass and Hubble type,  how $(B-I)$ colors, color gradients and color dispersion of 
{\it disk satellites} depend on group mass $M_\mathrm{GROUP}$, group-centric distance $R/R_{200}$  and large-scale structure overdensity $\delta_\mathrm{LSS}$. 
The strongest environmental trend is found for \emph{disk-dominated satellites} with $M_\mathrm{GROUP}$ and $R/R_{200}$. 
At $M \lesssim10^{10} M_\odot$, disk-dominated satellites  are  redder in the  inner regions of the groups than in the outer parts. 
At $M \gtrsim10^{10} M_\odot$, these satellites  have shallower  color gradients in higher mass groups and in the cores of groups compared with 
lower mass groups and the outskirts of groups.  Stellar population analyses and semi-analytic models suggest that disk-dominated satellites undergo 
 quenching of star formation in their outer disks, on timescales $\tau_\mathrm{quench}\sim2$ Gyr, as they progressively move inside the group potential.

\end{abstract}



\section{Introduction}

A key open question in galaxy evolution is   which of the many characteristics of a given galaxy's ``environment" most strongly affects its ability to form new stars.  Different environments which galaxies experience vary from conditions within their host group/cluster halo (related to the total mass of the halos), to their  precise location within these halos (related to the distance from the group/cluster center), to their  rank within halos  as   central  or satellite galaxies,  to the underlying density field of  the cosmic large-scale filamentary structure (LSS).   All these different aspects of  environment  may be in principle important.  

To make headway in understanding  which environment has an impact on galactic evolution,   we have undertaken the  \emph{Zurich Environmental Study} (ZENS, \citealt{Carollo_et_al_2013a}, hereafter Paper I), a comprehensive study of detailed galaxy properties as a function of group halo mass,  projected group-centric distance,  LSS density,  and  central/satellite ranking within the group potentials. ZENS uses the same local galaxy sample to perform the environment-vs-environment comparisons, which eliminates any bias when interpreting the results.  Detailed structural properties for the ZENS galaxies and their bar, bulge and disk components have been presented in Cibinel et al.\ (2013, hereafter Paper II).  In this third paper in the ZENS sequence we focus on photometric galactic diagnostics, such as e.g., stellar masses and star formation rates (SFRs).

The history of the growth of stellar mass in the Universe is imprinted in the  properties of star forming and passively-evolving (or, better, ``quenched") galaxy populations at different epochs, and, at each epoch, in different environments.  
Many  works have suggested that high SFRs are rarer in dense environments, and that galaxy groups and clusters may host a  higher fraction of quenched  early-type galaxies  relative to the field \citep[e.g.][]{Oemler_1974, Dressler_1980,Balogh_et_al_2004,Croton_et_al_2005}. Relatively recently,  however,   with the advent of   large spectroscopic surveys such as the 2dFGRS  \citep{Colless_et_al_2001} and the  SDSS \citep{York_et_al_2000} at low redshifts,  and others at high redshifts (e.g., zCOSMOS, \citealt{Lilly_et_al_2007,Lilly_et_al_2009};  DEEP2, \citealt{Davis_et_al_2005,Newman_et_al_2013}), it has become clear 
that galaxy properties, including SFRs, strongly depend on galaxy stellar mass \citep[e.g.][]{Brinchmann_et_al_2004,Tremonti_et_al_2004,Baldry_et_al_2006,van_der_Wel_2008,Peng_et_al_2010}. This requires that  environmental  effects   are carefully disentangled  from those of galaxy mass both in the implementation and in the interpretation of observational studies.
 
Analyses of the SDSS have suggested that, over about two dex of galaxy mass, the fraction of red galaxies rises with increasing environmental  density  (Baldry et al.\ 2006; Bamford et al.\ 2009; measured however through a proxy  that is  degenerate with group-centric distance, see Paper I; 
\citealt{Peng_et_al_2012, Woo_et_al_2013}).   \citet{Woo_et_al_2013} have furthermore shown that the fraction of quenched galaxies appears more strongly correlated with halo mass at fixed galaxy stellar mass  than viceversa. The interpretation of those results is however made fuzzy by the fact that they can be explained, at least qualitatively, in terms of a hierarchical increase in halo mass, without an increase in stellar mass for quenched objects.  There is furthermore evidence that the galaxy mass function may vary itself with local environmental density \citep{Baldry_et_al_2006,Bundy_et_al_2006,Bolzonella_et_al_2010,Kovac_et_al_2010}.  

Based on a phenomenological approach to galaxy evolution that is constructed around the basic continuity equation for galaxies,   \citet{Peng_et_al_2010} have  shown that the quenching transition of galaxies from the blue cloud  to the red sequence is due to a fully-separable combination of two processes, one linked to the stellar mass of a galaxy (or a proxy), independent of the environment, and the other linked to the environment, independent of the galaxy stellar mass.  The former  ``mass-quenching" process dominates at high stellar masses above $\sim10^{10.5} M_\odot$ and has a rate proportional to the SFR; it is this process that seems to control the mass functions of both active and quenched galaxies.  The second ``environment-quenching" process dominates at lower stellar masses \citep{Peng_et_al_2012}.
 Which physics is responsible for mass-quenching and environment-quenching is still debated. 
 
 A number of studies have postulated the existence of a typical halo mass scale, of about $10^{12}\Msol$, above which the interplay between shock heating and feedback from, e.g., AGNs, makes the cooling of the intergalactic medium and subsequent star-formation progressively inefficient   \citep[e.g.][]{Dekel_Birnboim_2006,Cattaneo_et_al_2006,Khochfar_Ostriker_2008,Cen_2011}. 
Others have highlighted  a possible environmental dependence of the halo assembly histories as the origin of the  
relation between galaxy  star formation activity   and environment   \citep[e.g.,][]{Sheth_Tormen_2004,Gao_et_al_2005,Maulbetsch_et_al_2007,Hahn_et_al_2007,Hahn_et_al_2009,Crain_et_al_2009}.  

\citet{Peng_et_al_2012}  find that their environment-quenching in the general galaxy population is confined to processes acting on the satellite population (i.e., their environment-quenching is a ``satellite-quenching" process); the  fraction of satellite galaxies that experience such quenching in their analysis ranges from $30\%$ to $75\%$, depending on the local density, but not at all on halo mass (above $10^{12}$M$_{\odot}$). This indicates that environmental effects may thus be relevant as galaxies enter group-sized halo potentials and there become satellites of the central galaxy. Analyses of SDSS groups have shown that, at fixed stellar mass, satellites in groups are indeed on average redder than central galaxies \citep{van_den_Bosch_2008,Skibba_2009,Weinmann_et_al_2009}.
An analysis of groups out to $z\simeq0.8$ in the zCOSMOS field demonstrated that such difference between central and satellite galaxies is still observed also at these higher redshift \citep{Knobel_et_al_2012}.
 Within a cold dark matter (CDM) framework, suppression of star-formation in galaxies once these are accreted into bigger haloes may happen e.g., due to ram pressure stripping of their cold gas \citep{Gunn_Gott_1972} or  removal of their hot gas  \citep{Larson_et_al_1980, Balogh_et_al_2000}. 

The precise mass scale at which halo physics may start influencing the star formation properties of satellite galaxies is however not well known.  Numerical simulations   show that  galaxies are depleted of their cold and hot gas as they enter the virialized regions of relatively small groups (e.g., \citealt{Kawata_Mulchaey_2008,Feldmann_et_al_2011}).   \citet{Peng_et_al_2012} have argued that many of the key evolutionary processes on satellites, namely the specific star-formation rates (i.e., SFR$/M_\mathrm{galaxy}$; sSFR)  of star-forming galaxies, the mass-quenching process and the environment-quenching (at fixed density) are all strikingly independent of halo mass  above $10^{12}$M$_{\odot}$.
Galaxies  may thus be quenched as they become satellites in small groups,  well before these assemble into larger cluster-sized structures.  

Observational diagnostics that have been heavily studied to understand  how star forming galaxies are transformed into red-and-dead stellar systems have been typically limited to  integrated galaxy properties (e.g., total galaxy colors and SFRs).  Also, morphological analyses of the satellite and central galaxy populations have been typically based on a separation into early- and late-type galaxies by means of a threshold in either galaxy concentration or S\'ersic index $n$ \citep{Sersic_1968}, which leads to  mixing of galaxies with very different bulge-to-disk ratios, arguably the most physical diagnostics to classify galaxy structure and morphology (see Paper II).  

Open questions thus remain whether and how the strength of environmental effects varies between  galaxy populations with different bulge-to-disk ratios, and whether and to what extent the environmental forcing affects differentially the bulge or disk components of galaxies.  To test the different physical scenarios, it is furthermore imperative that, in addressing these questions,  the  different aspects of the environment of a galaxy are carefully disentangled. We will directly tackle these issues in a number of forthcoming ZENS papers.  

In the present third paper in the ZENS sequence we introduce  the methodology  adopted to measure the photometric properties of ZENS galaxies, and the tests that we have performed to assess the impact of systematic  errors on these properties. The photometric measurements  include galaxy SFRs, stellar masses,  colors, color gradients, two-dimensional color maps, as well as e.g., bulge/disk colors and masses, etc.  With the purpose of presenting a simple first utilization of these photometric measurements (coupled with the structural galaxy measurements of  Paper II   and with the  environmental diagnostics  of Paper I),  we also discuss  the $(B-I)$ colors and color gradients of {\it satellite} galaxies as a function of  galaxy mass at constant Hubble type, and of group halo mass,  group-centric distance  and  LSS  density.   

After a brief summary of  ZENS in Section \ref{sec:surveyData}, this paper is structured as follows. 
The bulk of the paper  describes the photometric analyses performed on the ZENS galaxy sample: Stellar masses, both for the entire galaxies and their bulge and disk components, and mass completeness limits, are derived using synthetic templates and broad-band photometry in Sections \ref{sec:masses}-\ref{sec:bulgeDiskmasses}. 
Section \ref{sec:ActiveQuiescent} presents the criteria  that we have used to classify strongly and moderately  star-forming galaxies, and quenched galaxies, 
 based on the combination of  spectral features, far- and near-UV (FUV/NUV) versus optical colors, and  fits to the galaxy Spectral Energy Distributions (SEDs). We also highlight in this Section the importance of the comparison of multiple star formation diagnostics, to avoid misclassifications in the spectral types of galaxies.
We discuss the derivation and associated systematic biases of resolved color maps and radial color profiles in Section  \ref{sec:ColorMaps}.
Section \ref{vars} analyses, above our stellar mass completeness limits,  variations in median  galaxy stellar mass of satellites and centrals  as a function of group mass, group-centric distance and LSS density.
 Section \ref{sec:trendswithmass} shows  the dependence of colors, color gradients and color scatter  at fixed stellar mass for the different Hubble types determined in Paper II.
  The final Section \ref{sec:results} 
presents our first environmental  study  of the presented measurements, i.e.,  the dependence of  colors, color  gradients and color scatter in {\it disk satellites} on the different  environmental parameters. We summarize  the paper in Section \ref{sec:summary}. In Appendix \ref{sec:MassAccuracy} we provide  supplementary information on the tests we performed to assess the reliability of  galaxy properties derived from our SED fits. Throughout the paper we will assume the following values for the relevant cosmological parameters: $\Omega_m=0.3$, $\Omega_{\Lambda}=0.7$ and $h=0.7$. Magnitudes are in the AB-system \citep{Oke_1974}.

We remind the reader that all measurements for the 1484 galaxies in the 141 ZENS groups are made public in the ZENS catalog\footnote{The ZENS catalog is also downloadable from: http://www.astro.ethz.ch/research/Projects/ZENS.} published with Paper I.


\section{A summary of ZENS} \label{sec:surveyData}

ZENS uses 141 groups of galaxies selected from the 2dFGRS Percolation-Inferred Galaxy 
 Group (\emph{2PIGG}) catalogue \citep{Eke_et_al_2004}. 
The groups were randomly selected from all groups in the 2PIGG catalog 
which are located at 0.05$<z<$0.0585 and have at least five spectroscopically-confirmed galaxy members.
New $B-$ and $I-$band  observations were taken with the Wide Field Imager (WFI) camera 
mounted at the MPG/ESO 2.2m Telescope, reaching 
 a background-limited depth of $\mu(B)=27.2$ mag arcsec$^{-2}$ and $\mu(I)=25.5$ mag arcsec$^{-2}$ over an  area of 1 arcsec$^2$. 
  
In \textsc{ZENS} we focus our attention on the following measures of galactic environment: 
\begin{enumerate}
\item the mass of the group halo in which galaxies reside, $M_\mathrm{GROUP}$;
\item  the distance of a galaxy from the  center of the host group, normalized to the characteristic group radius
 $R_{200}$;
\item the large scale (over)density field $\delta_\mathrm{LSS}$, quantified through the overdensity  of 2PIGG \emph{groups} 
 at the ZENS field positions, within a circular projected area 
 of size equal to the comoving distance to the 5th nearest 2PIGG group, i.e., $\delta_\mathrm{LSS}=\frac{\rho_\mathrm{LSS}(z)-\rho_m}{\rho_m}$ (with $\rho_\mathrm{LSS}(z)$  the density of 2PIGG groups out to the 5th 
nearest neighboring group,  and $\rho_m$ the mean projected density in the 2dFGRS volume);
\item the rank of the galaxy within the host group, i.e., whether the galaxy is the central galaxy or a satellite galaxy in orbit around the central.
\end{enumerate}  

Attention was paid to ascertain whether the ranking of galaxies as centrals and satellites  led to a self-consistent  solution for the central galaxies (required to be the most massive, within the errors on the estimated stellar masses, as well as compatible with being the geometrical and velocity centroids of the groups). Groups with a well-identified central galaxy were defined as ``relaxed", while groups in which  no galaxy member satisfied all requirements to be considered a central galaxy were defined to be ``unrelaxed".

We refer to Paper I for further details on the ZENS sample and our environmental definitions, and to Paper II for details on the structural measurements and morphological classification of the ZENS galaxies, as well as on the data reduction and calibration of the WFI data.


\section{Stellar masses for the ZENS galaxies} \label{sec:masses}

To estimate the stellar masses (hereafter $M_\mathrm{galaxy}$ or, when unambiguous, simply $M$ for simplicity) for the ZENS galaxies we  fitted
 synthetic spectral templates to their broad-band photometric data.
To this end we employed our  code  \emph{ZEBRA+} 
(\emph{The Zurich Extragalactic Bayesian Redshift Analyzer Plus}, \citealt{Oesch_et_al_2010}),
an upgrade of  the  public version of the ZEBRA package presented in \citet{Feldmann_et_al_2006}.  Relative to ZEBRA, the upgraded ZEBRA+ code  enables the use of synthetic templates for the estimation of stellar masses, ages and metallicities. 
In the following we describe in detail the photometric data and the model templates used to derive the galaxy stellar masses.

\subsection{Photometry} \label{sec:Phot}

In order to constrain at best the template fits, we combined all available photometry for the sample galaxies,  including our new WFI $B$ and $I$ imaging data  as well as data from the literature.  Specifically, for each galaxy  we used, when available,  SDSS $u$, $g$, $r$, $i$, $z$ \citep{Abazajian_et_al_2009},  2MASS $J$, $H$, $K$  \citep{Skrutskie_et_al_2006} and  $GALEX$ \citep{Martin_et_al_2005} NUV and FUV magnitudes.
The fraction of ZENS galaxies  with observations at any given wavelength is summarized in Table \ref{tab:Photometry}.

\begin{deluxetable}{cc}
\tablecaption{Fraction of ZENS galaxies with available multi-band photometry} 
\tablehead{ \colhead{Band}  & \colhead{Coverage}    }
\startdata
FUV, NUV ($GALEX$) &   89$\%$ \\
$B$, $I$  (ESO/WFI) &  100$\%$\\
$u$, $g$, $r$, $i$, $z$  (SDSS) &   29$\%$\\
$J$, $H$,$K$ (2MASS) &   98$\%$\\
\enddata\label{tab:Photometry}
\tablecomments{The fraction of ZENS galaxies with available photometric data from  the  listed surveys.}
\end{deluxetable}  

We  calibrated the data products from these surveys following the instructions provided by the respective teams. 
For the SDSS we used the legacy imaging data with the ``best"  reduction from the Data Release 7 (DR7) and for 2MASS we use the Atlas data product.
In the case of $GALEX$ we used the GR6 All-sky Survey (AIS) tiles  or, when available, the deeper Medium Survey (MIS) intensity maps.
In the case of more than one AIS observation, we used the data having the longer exposure time. 

We measured the magnitudes in all pass-bands using an elliptical aperture equal to $2\times R_{p_{MAX}}$, with $R_{p_{MAX}}$ 
the larger of the two Petrosian semi-major axes $R_{p_{I}}$ and $R_{p_{B}}$ measured with SExtractor
 \citep{Bertin_et_Arnouts_1996} on the WFI $I$- and $B$-band images. 
When available we used the sky-background estimates or maps provided by the survey pipelines, otherwise the background was subtracted using the mode of the intensity in the pixels located outside the Petrosian aperture. A comparison between the fluxes we measured and those found in the survey's public catalogues is shown in Figure \ref{fig:PhotometryComp} of Appendix \ref{sec:MassAccuracy}, together with a discussion on possible biases arising from differences in the average point-spread function (PSF) in the different
passbands. 

All the magnitudes were corrected for galactic extinction using $E(B-V)$ values from the survey catalogues, if provided, or obtained from the dust maps of \citet{Schlegel_et_al_1998}, assuming a \citet{Cardelli_et_al_1989} extinction law.
For correcting  $GALEX$ magnitudes,  we applied the relations $A_\mathrm{FUV}/E(B-V)=8.376$ and $A_\mathrm{NUV}/E(B-V)=8.741$ as derived by \citet{Wyder_et_al_2005}.
2MASS magnitudes were converted into the AB system by applying the following transformations which we calculated using the IRAF package {\ttfamily synphot} and the
composite Vega spectrum released by STSCI\footnote{http://www.stsci.edu/hst/observatory/cdbs/calspec.html}: 
$J_{AB}=J_{VEGA}+0.894$, $H_{AB}=H_{VEGA}+1.367$ and $K_{AB}=K_{VEGA}+1.837$.

Before calculating the masses we ran ZEBRA+  in the \emph{photometry-check mode} in order to correct for possible systematics in the calibration of the input catalogue.  With this option it is  possible to detect residuals between the best-fit template and the galaxy fluxes that are independent of template type itself  (see \citealt{Feldmann_et_al_2006} for more details). This check revealed that there was a small offset in the SDSS $i-$ and $z$-bands equal to -0.03 and 0.04 magnitudes, respectively. 
We  applied these small shifts to the SDSS $i$- and $z$-band before proceeding to the derivation of masses for our galaxies.

\subsection{Synthetic Template Library} 

To model the spectral energy distribution of the ZENS galaxies we used a set of  Bruzual \& Charlot (2003; BC03) synthetic models with a \citet{Chabrier_2003} initial mass function (IMF), truncated at $0.1\Msol$ and  $100\Msol$ and based on the Padova 1994 evolutionary tracks. 
We used star formation histories (SFHs) described by  exponentially declining SFRs $\psi(t)=\psi_0\tau^{-1}e^{-\frac{t}{\tau}}$, with different values for the characteristic time $\tau$, as well as by a constant SFR (CSF, $\tau\rightarrow \infty$).
The normalization factor  $\psi_0$ was set to 1 by imposing a total mass of $1\Msol$ at time $t \rightarrow \infty$.
 Each SFH was sampled with 900 templates of metallicities $Z$ ranging between 0.004 (1/5$Z_\odot$) and 0.04 (2$Z_\odot$) and ages between 10 Myr and 12 Gyr.
 
Reddening produced by dust was also taken into account:   a dust correction was allowed in the ZEBRA+ fits, and often required for galaxies morphologically classified as spirals or irregular.   The dust extinction was assumed to be described by a  \citet{Calzetti_et_al_2000} relation and the corresponding reddening $E(B-V)$ was allowed to vary between 0 and 0.5, in 27 fine steps.  A substantial dust attenuation, $E (B-V)>0.25$,  was permitted only for star-forming templates, i.e., templates with $\mathrm{age}/\tau<4$  (see also \citealt{Pozzetti_et_al_2007,Ilbert_et_al_2010}) and for galaxies not spectrally-classified as ``quenched" (see Section \ref{sec:ActiveQuiescent}). Galaxies with a best-fit template with $\mathrm{age}/\tau>4$ and $E (B-V)>0.25$ were fitted using a dust grid with a maximum value of  $E (B-V)=0.25$. Galaxies morphologically classified as elliptical and S0s required no dust in their SED fits.
A  complete summary of the parameter space covered by the adopted templates is presented in Table \ref{tab:ZEBRA}. 
Knowing the spectroscopic redshift of our galaxies,  we ran ZEBRA+ in the maximum likelihood mode keeping the redshift fixed to the right values.  The best fit templates from ZEBRA+ were also used to compute the $k$-correction for each galaxy.
At the redshift of the ZENS groups, the $k$-corrections are small but not negligible: in $B-$band these are typically 0.15mag, and they increase up to  0.3mag in the NUV, depending on the galaxy intrinsic SED (see also \citealt{Blanton_Roweis_2007}.)

\begin{deluxetable}{cc}
\tablecaption{Template library used to derive stellar masses and (s)SFRs with ZEBRA+.} 
\tablehead{ \colhead{Parameter}  & \colhead{Range of Values}    }
\startdata
$Z$ & 0.004, 0.007, 0.01, 0.02 ($Z_\odot$), 0.03, 0.04 \\
$\tau$ (Gyr) & 0.05, 0.1, 0.3, 0.6, 1, 1.5, 2, 3, ..., 10,  $\infty$ \\
Age (Gyr) & 0.01--12  \\
$E(B-V)$ & 0--0.5 \\   
\enddata
\tablecomments{\label{tab:ZEBRA}The properties of the synthetic spectra used to derive  galaxy stellar masses and SFRs through fits to the  galaxy SEDs. From  top to bottom: metallicity, exponentially-declining SFH time-scale, stellar age (sampled in 150 logarithmic bins) and dust reddening 
(sampled in 27 steps, spaced by 0.01 mag in the range $E(B-V)=0 \rightarrow 0.1$ and by 0.025 mag at higher reddening).
Value of dust extinction $E(B-V)>0.25$ were allowed only in  models
with $\mathrm{age}/\tau<4$ for galaxies classified as non-quenched (see Section \ref{sec:ActiveQuiescent}).}
\end{deluxetable}   

SFRs were computed as   $\mathrm{SFR}=M_\mathrm{galaxy}/t_A$ for  constant SFR models, and 
$\mathrm{SFR}=\frac{M_\mathrm{galaxy}}{\left(1-e^{-t_A/\tau}\right)}\frac{e^{-t_A/\tau}}{\tau}$
for exponentially declining SFHs (with $t_A$   the age of the best fit model.) 

In our ZENS  analyses we will adopt as or fiducial estimates for the stellar   masses those obtained from the integration of
the galaxies' SFHs. 
We estimate that the ``actual" stellar masses, that are obtained by removing the 
fraction of mass that is returned to the interstellar-medium by the dying stars, are on average smaller  by about -0.2 dex  than the adopted  ``integrated'' stellar masses.
An assessment of the uncertainty and reliability of the masses derived with ZEBRA+ is given in 
Appendix \ref{sec:MassAccuracy}, where we compare our results with previously published data
and present further considerations on  other possible systematics. 

\subsection{Mass Completeness of the ZENS Galaxy Sample} \label{sec:MassCompleteness}

\begin{figure}[htbp]
\begin{center}
\includegraphics[width=86mm]{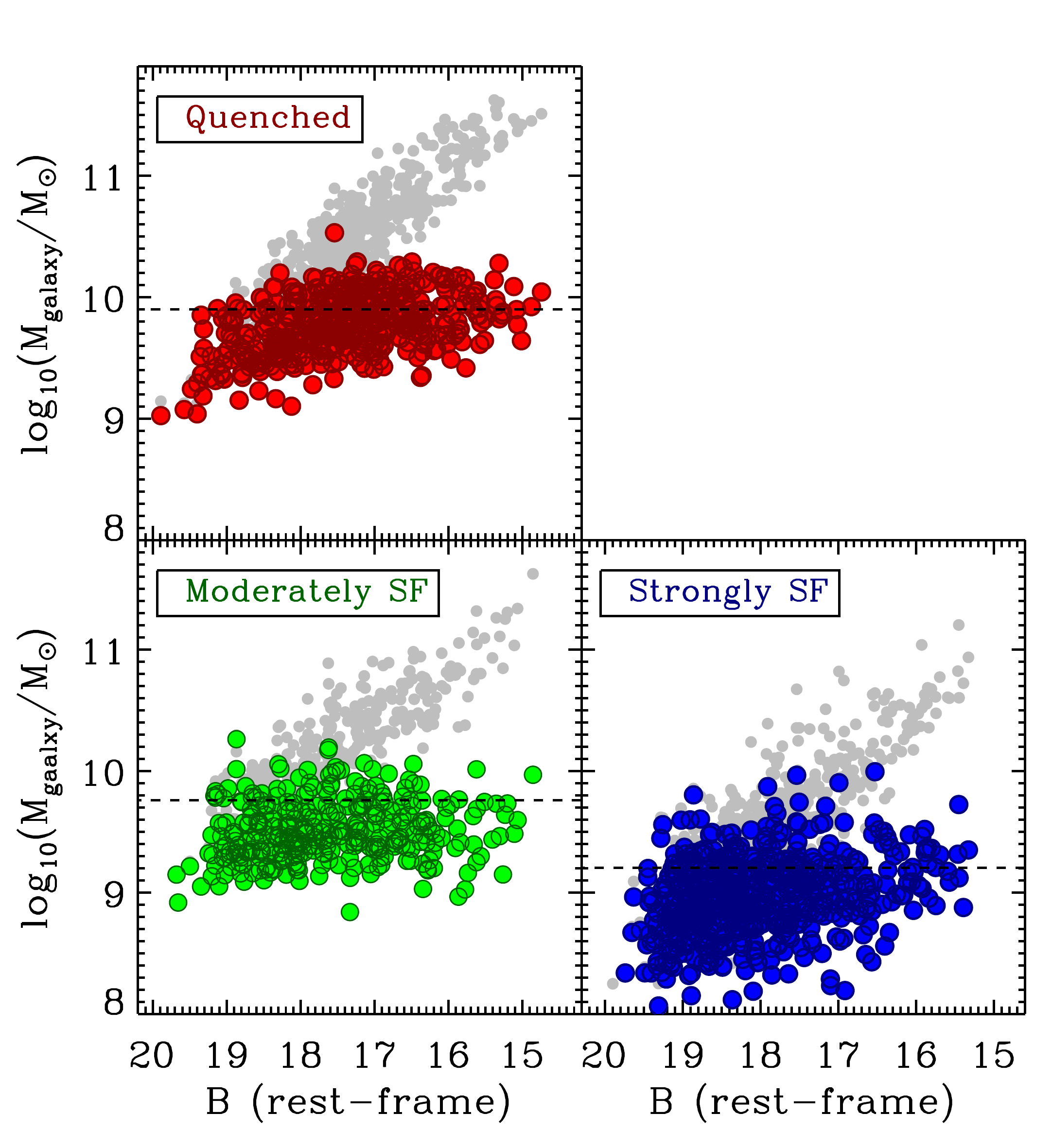}
\end{center}
\caption{\label{fig:MassCompl}Illustration of the mass completeness derivation 
for galaxies of the three different spectral types defined in Section \ref{sec:ActiveQuiescent}.
The panels show galaxy stellar mass as a function of   apparent $B$-band magnitude, separately for quenched, moderately star-forming and strongly star-forming galaxies.
Small gray dots are galaxy masses inferred from the ZEBRA+ best fits to the photometric data of ZENS galaxies;
 large black symbols (colored in the online version)  show the stellar masses $\log(M_{lim})= \log(M)+0.4(b_j-b_{j,lim})$ obtained by fading  galaxies to the 2dFGRS limiting magnitude (red, green and blue are respectively used for quenched, moderately star forming 
 and strongly star forming galaxies in the online version). The dashed horizontal lines mark the mass completeness limits,  
 defined, separately  for each spectral type, as the mass below which lie 85$\%$ of the $M_{lim}$ for the 30$\%$ faintest galaxies of that given type (see text).\\
(A color version of this figure is available in the online journal.) }
\end{figure}

Given the apparent magnitude selection of the 2dFGRS and the 2PIGG group catalog,
the  raw ZENS galaxy sample is not a mass-complete sample: 
the $b_j\lesssim 19.45$ magnitude
limit of the survey translates into a 
minimum observed mass that depends on the intrinsic SED of the galaxies and  is thus 
different for different galaxy types.  Galaxies with the highest mass-to-light ratio ($M/L$) thus set the mass completeness limit of our survey.
We infer this limit empirically from the observed $M/L$ of the ZENS galaxies, following an approach similar to that described in  \citet{Pozzetti_et_al_2010}.
Specifically, for each galaxy  we compute the mass that it would have, given its $M/L$, once reduced in brightness to the survey limiting magnitude, 
i.e. $\log(M_{lim})= \log(M)+0.4(b_j-b_{j,lim})$.
As discussed in \citet{Pozzetti_et_al_2010}, deriving the mass limits using the entire galaxy sample can lead to  values that are too conservative and not representative of the masses of the typical galaxies with luminosity close the survey limit. For each morphological or spectral galaxy type we therefore used the 30$\%$ faintest galaxies to calculate the
completeness mass. This is defined as the mass below which lie 85$\%$ of the $M_{lim}$ for the faint galaxies, i.e., the mass \emph{above}
which 85$\%$ of these galaxies  still fulfill the criterion  $b_j<19.45$.
This is illustrated in Figure \ref{fig:MassCompl} for the three  galaxy spectral types in which 
we have classified our ZENS sample in Section \ref{sec:ActiveQuiescent}.  

The resulting completeness limits  are, precisely,  $\log(M/\Msol)>$9.90, $\log(M\Msol)>$9.78 and $\log(M\Msol)>$9.24 for quenched, moderately star-forming   and  strongly star-forming galaxies. Folding in the statistical mix of spectroscopic type in our morphological bins (defined in Paper II),  the mass completeness limits for galaxies of different morphological types are $\log(M/\Msol)>$10.04 for E/S0 galaxies, $\log(M/\Msol)>$9.93 for bulge dominated spiral galaxies, $\log(M/\Msol)>$9.78 for intermediate spirals and $\log(M/\Msol)>$9.55 for late spirals and irregulars. 
In the rest of this paper we will homogenize these limits and use $M=10^{10}\Msol$ as the mass limit for the spectral and morphological early-type galaxies, and $M=10^{9.5}\Msol$ for star forming and morphological late-type galaxies.

\begin{figure}[htbp]
\begin{center}
\includegraphics[width=67mm,angle=90]{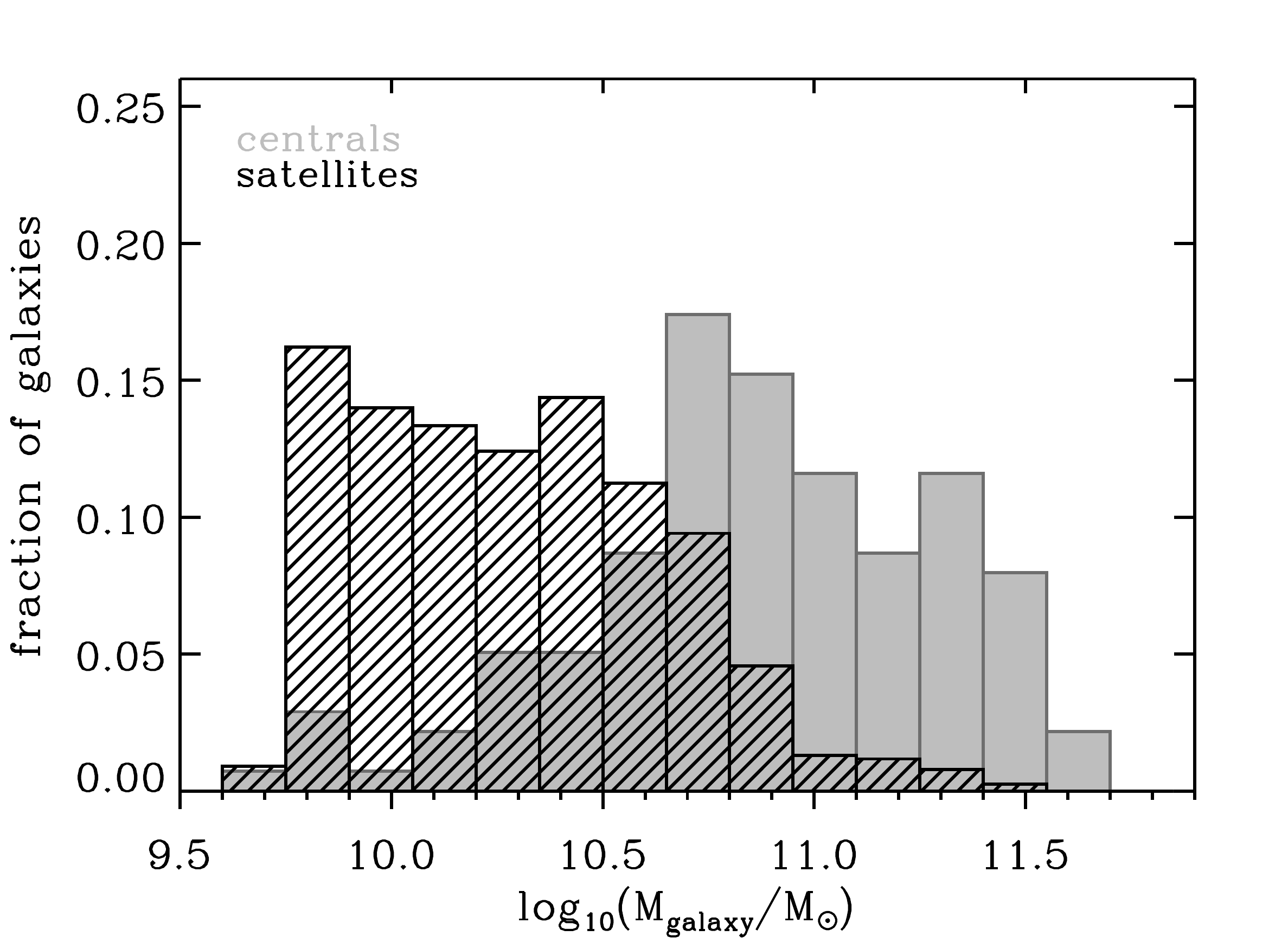}
\end{center}
\caption{\label{fig:MassCensSats} Distributions of galaxy stellar masses for central (filled, gray histogram) 
and satellite galaxies (black, hatched histogram), over the mass range spanned by the central population. Each histogram is normalized to its sum.}
\end{figure}

We finally stress  that, as a consequence of their definition and of our ZENS sample selection, the distributions of galaxy masses for central and satellite galaxies  over the common mass range are 
substantially different, as shown explicitly in  Figure \ref{fig:MassCensSats};  furthermore, these two populations overlap basically only at 
mass scales $\sim10^{10}-10^{10.7}\Msol$. We will thus take this bias into account when comparing the properties of centrals and satellites in our ZENS studies.

 \section{Derivation of stellar masses for the bulges and disks of two-component galaxies}\label{sec:bulgeDiskmasses}

In addition to deriving the stellar masses for the entire galaxies as described above, we also estimated stellar masses separately for the  bulge and disk components of disk galaxies with bulge/disk decompositions available from Paper II through \textsc{GIM2D} \citep{Simard_et_al_2002} decompositions.
To this purpose, we converted the $(B-I)$ colors of bulges and disks  into an $I$-band mass-to-light  ratio (M/L) by using a set of stellar population models 
from the BC03 library, constructed with a Chabrier IMF. We adopted as our fiducial M/L for the galaxy components the median M/L obtained from fitting all the synthetic spectra.

We employed  simple stellar population models (SSP) as well as exponentially decreasing
SFH, with characteristic time scales of $\tau=1,3,6$ Gyr.
For each SFH we used the range of metallicities  $Z$=0.004, 0.005, 0.008, 0.01, 0.02, 0.03, 0.05, in units such that 0.02 is the solar metallicity $(Z_\odot)$.
 Another set of models, based on the same grid of parameters,  was 
  reddened with  a Calzetti dust law and  $E(B-V)$=0.05, 0.1, 0.15, 0.2, 0.4, 0.5. Dust-reddened models with $E(B-V)>0.2$ were used only for galaxies classified as dusty-star forming  (see Section \ref{sec:ActiveQuiescent}).

A $k$-correction is needed to convert the bulge and disk apparent magnitudes into   rest-frame luminosities. Whereas for the entire galaxies we can calculate these corrections from the UV to near-infrared (NIR) SED fits of ZEBRA+, such information is not available for the bulges and disks  for which we have only the $B$- and $I$-band photometry.
For this reason, we performed a linear fit to the relation between the apparent $(B-I)$ color and the ZEBRA+ $k$-corrections as obtained for the entire galaxies,  and we used this average relation to infer the $k$-corrections for the galaxy components. 
Consistently with the definition adopted for the galaxies, we consider in the following   total stellar masses for bulges and disks, 
which  include also the fraction of baryons which is returned as gas to  the ISM as stars evolve.

\begin{figure*}
\begin{center}
\includegraphics[width=90mm,angle=90]{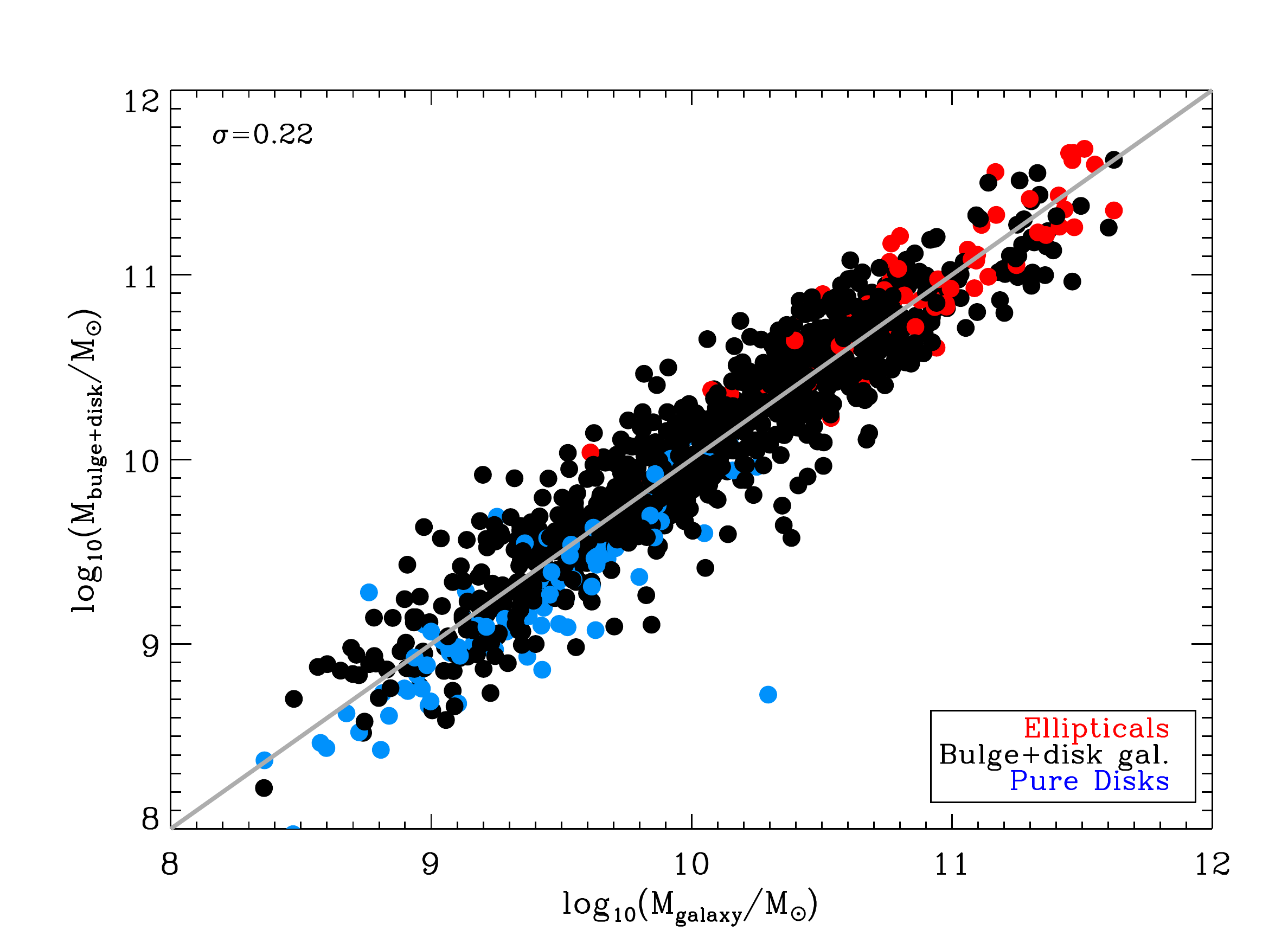}
\end{center}
\caption{\label{fig:MassComp} Comparison between  galaxy stellar masses derived from the SED fits to the galaxies photometric data,
and galaxy stellar masses  obtained by summing up the separate contributions of their  bulges and disks.  Bulge and disk stellar masses are inferred from the   GIM2D bulge+disk decompositions of Paper II, using the
 $(B-I)$ colors of these  components as proxies for their $M/L$ values ($(B-I)$ colors from the single S\'ersic fits are used for elliptical galaxies).
Elliptical galaxies are shown in black (red in the online version),  and single-exponential (i.e., ``pure") disk galaxies in dark gray (blue in the online version); 
galaxies with  both a bulge and a disk component are shown in gray (black in the online version). The two mass estimates show  a very good agreement within a scatter  of $\sigma=0.22$ dex.\\
(A color version of this figure is available in the online journal.) }
\end{figure*}

To assess the reliability of the derived masses of bulges and disks, in Figure \ref{fig:MassComp} we compare the total galaxy 
stellar masses derived in Section \ref{sec:masses} with the sum of the stellar masses of  their bulge and  disk components.
We highlight with red and blue symbols respectively elliptical galaxies and pure disk galaxies, i.e., galaxies for which a two-component GIM2D fit returns as the best fit  a model with B/T=0.
The figure shows a very good agreement  with a scatter of  $\sim$0.2 dex, comparable with the
average error on the total galaxy mass estimate (see the Appendix).

\section{Definition of galaxy spectral types: Quenched, moderately star-forming and strongly star-forming galaxies}\label{sec:ActiveQuiescent}

The split between quenched  and star-forming galaxies is a rather arbitrary one, for many reasons. First, it is plausible to expect that, in nature, star formation activity is a continuum property of galaxies, and the distinction between the two populations is not a dichotomy. Second, observational biases such as threshold of detectability of SFRs in any given survey can affect the classification.   
 In ZENS we try to clean, as much as possible, our quenched/star-forming classification of galaxies in the sample, by adopting an approach that combines three different probes of star formation activity, namely $(1)$ the original 2dF spectra, $(2)$ FUV-NUV-optical colors, and  $(3)$ the SFR estimates from the SED fits to the galaxy photometry described above.  The main philosophy that we adopt is to automatically classify a galaxy as star-forming or quenched when all three diagnostics agree on the classification, and to explore all diagnostics to understand what causes discrepancies when these arises. We end up with a three-bin classification in quenched, moderately star forming and strongly star forming galaxies, the last two classes being broadly diversified by the level of the sSFR measured by the SED  fits. We detail our classification procedure below.
 
\subsection{First constraint to the spectral classification: Line features in the 2dFGRS spectra}

The 2dFGRS team provides for each galaxy the spectral indicator $\eta$  \citep{Madgwick_et_al_2002} which
 characterizes the activity of the galaxy. 
The parameter $\eta$ correlates by definition
with the intensity of the emission lines (most strongly with the equivalent width of $\Halpha$); however,  it does not contain detailed information on which species are in emission nor their relative strength, aspects which are important to identify mixed types or galaxies in which the emission is powered by an AGN rather than by  star formation.  Furthermore,  we worried that the fibers of the 2dFGRS spectrograph, having a fixed diameter of  $2^{\prime\prime}-2.16^{\prime\prime}$, probe a different  fraction of the galaxy light depending on the galaxy's apparent radius.
 Whereas for a good fraction of the ZENS galaxies the $2^{\prime\prime}$  aperture is large enough to cover a significant part of the galaxy, for
 others the spectrum is only probing the inner most region, leaving scope for misclassifications.
 We thus individually inspected  the 2dFGRS spectra of the ZENS galaxies  to study the  emission and absorption line features in the galaxy spectra. 

A relative  flux calibration for the  spectra was obtained by using the average response function and telluric corrections
 available on the 2dFGRS web-site\footnote{These are the \emph{eff-pop-syn-mid.dat} and \emph{eff-telluric.dat} files, respectively. We did not attempt an absolute flux calibration of the spectra, which is beyond our purposes.}. 
A boxcar-averaged continuum (over 250 pixels)  was subtracted from the spectra before performing the visual  inspection.
The wide wavelength coverage of the 2dFGRS spectra and the low redshifts of our galaxies allow us to look for a number of star-formation tracers, i.e.,  hydrogen $\Hbeta$  and $\Halpha$ Balmer lines,  and  forbidden transitions of [OII] $\lambda 3727$,
 [OIII] $\lambda 5007$, and [NII] $\lambda 6584$.

Based on the spectral information, we classified a galaxy as \emph{quenched} if its spectrum satisfies one of the following conditions: 
a) none of the aforementioned lines is
 observed in emission, or  b) some [OII] /[NII] emission is present but no $\Hbeta$ or $\Halpha$ is detected.
It is not unusual to observe [OII]  or [NII] emission in ellipticals or in galaxies on the red sequence \citep[e.g.][]{Caldwell_1984,Phillips_et_al_1986, Macchetto_et_al_1996}, but
the presence of this ionized gas is often a consequence of active galactic nuclei (AGNs) or excitation in low-ionization nuclear emission line regions  (LINER)  rather than a signature of  star-formation spread across the entire galaxy \citep{Yan_et_al_2006, Annibali_et_al_2010, Lemaux_et_al_2010}.

Galaxies were classified as  star-forming if they showed  emission in the above species, particularly  $\Hbeta$ and/or $\Halpha$.
Emission-line spectra  can  have however large contributions from  AGNs  \citep{Heckman_1980,Baldwin_et_al_1981,Veron_Cetty_1997,Kauffmann_et_al_2003}. 
A widely used method to identify AGNs,  first proposed by \citet{Baldwin_et_al_1981} and 
revised in several later works \citep[e.g.][]{Veilleux_Osterbrock_1987,Tresse_et_al_1997}, relies on the relative strength of the [NII]/$\Halpha$ and [OIII]/$\Hbeta$ line ratios. 
We used this diagnostic to identify AGN candidates in our sample; 
in these galaxies, the classification in quenched or star-forming systems relied entirely on the color-color criteria.

In the next Section we describe how we further split star forming galaxies into two classes of, respectively, {\it strongly star forming} and {\it moderately star forming} systems. The latter identify  a transition galaxy type with modest star formation activity in between that of  strongly star-forming   and   quenched galaxies. This transition class is defined on the basis of  an NUV-optical color cut; an a posteriori check of the 2dF spectra of these galaxies shows indeed that they have typically intermediate  spectra between those
 of strongly star-forming and quenched galaxies. 

\begin{figure}[htbp]
\begin{center}
\includegraphics[width=95mm]{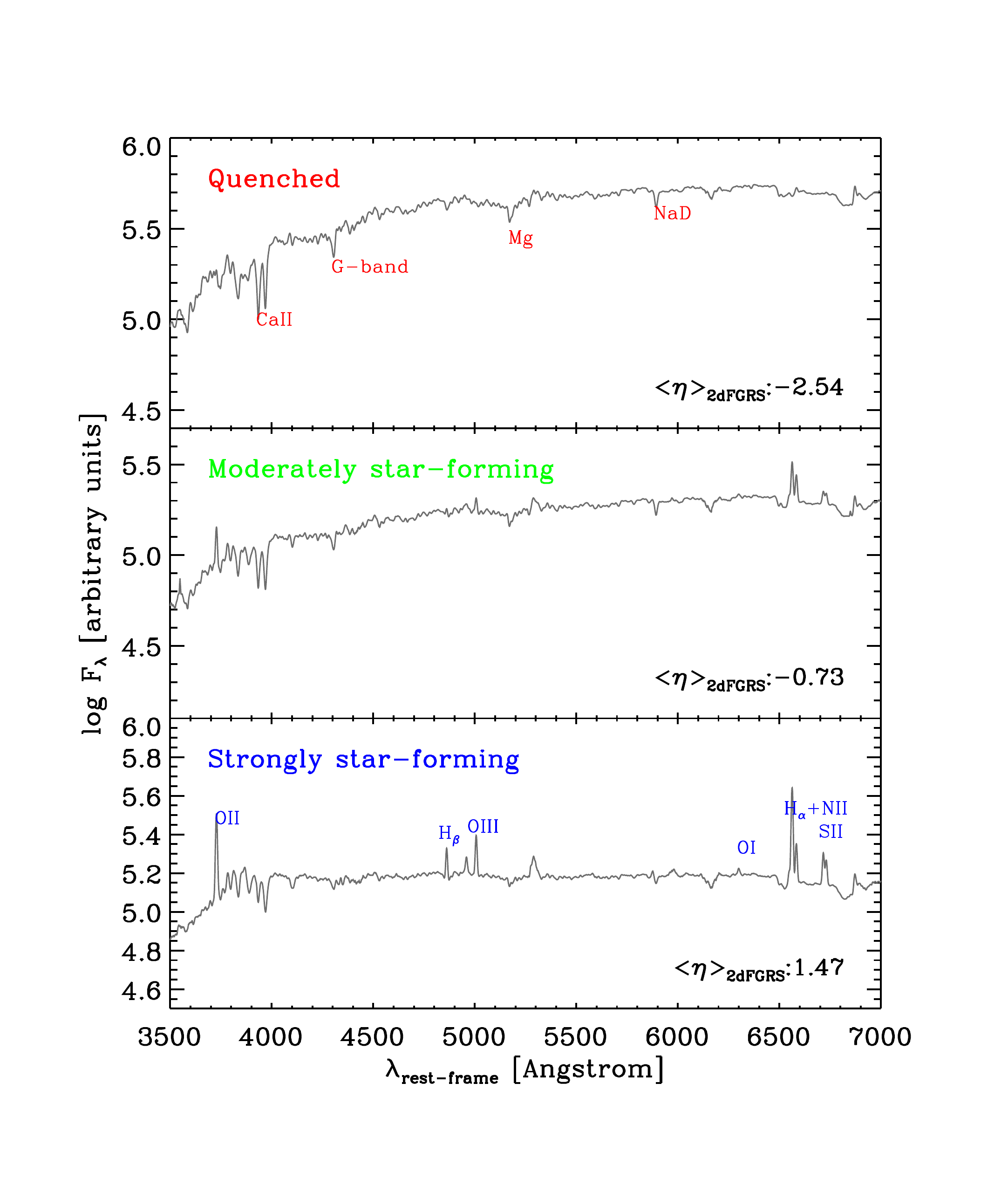}
\end{center}
\caption{\label{fig:SpecTypes} Stacked spectra of the ZENS galaxies classified as quenched, moderately star-forming 
and strongly star-forming. The main emission/absorption features are indicated on top of the 
upper and lower panels.
Quenched galaxies are characterized by strong stellar absorption lines and an underlying red continuum;  star-forming galaxies have a blue continuum and intense nebular emission.
The moderately star-forming type have properties which are intermediate between these previous
two classes.  Each composite spectra is obtained from $>300$ individual spectra. 
The flux is given in logarithm of raw counts, hence the absolute normalization is arbitrary.
The median value of the spectral parameter $\eta$ provided with the 2dFGRS database for the three spectral types is indicated on the bottom right of the panels. Note that in the original 2dFGRS classification scheme, quenched galaxies have $\eta<-1.4$, star-forming galaxies $\eta>-1.4$ and starbursts $\eta>3.5$. \\
(A color version of this figure is available in the online journal.) }
\end{figure}

The stacked spectra of the three spectral types are shown in Figure \ref{fig:SpecTypes}.  Clearly visible is the transition from a spectrum dominated by stellar absorption
 for the quenched galaxies, to a spectrum dominated by nebular lines for  the strongly star-forming galaxies.  
The moderately star-forming galaxies present weaker emission lines   than those observed in the strongly star-forming galaxies; as also highlighted below, they  also have a redder continuum in comparison to the latter class.  In Figure \ref{fig:SpecTypes} we also indicate the median value of the 2dFGRS spectral parameter $\eta$ for the three classes: in about $90\%$ of the galaxies we find a good agreement between our classification and the one provided by the 2dFGRS team, with the remaining discrepancies due to spectral misidentifications, as discussed above.

\subsection{Second constraint to the spectral classification: FUV-NUV-optical broad-band colors}\label{sec:color_color_cut}

\begin{figure*}[htbp]
\begin{center}
\includegraphics[width=90mm,angle=90]{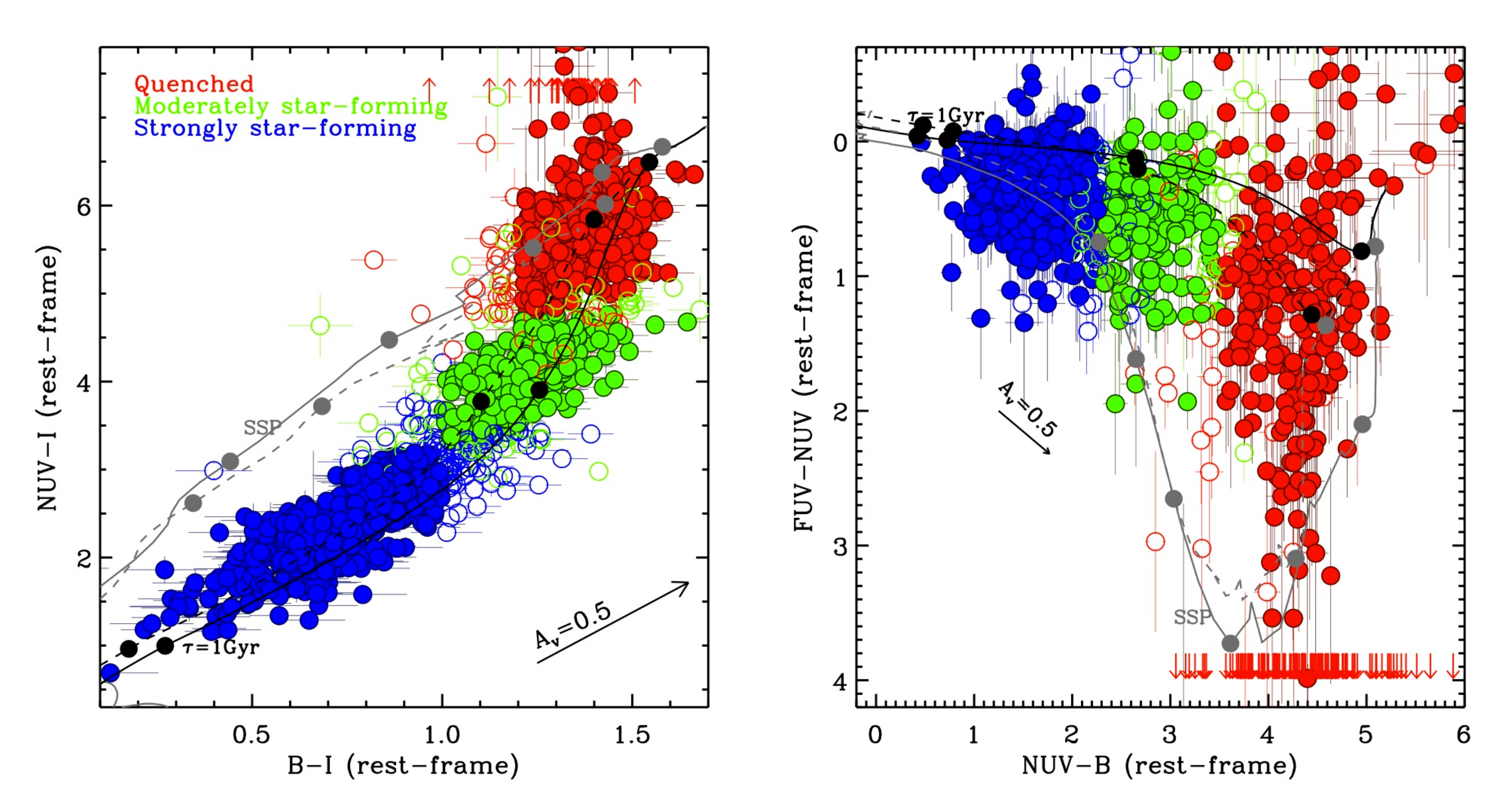}
\end{center}
\caption{\label{fig:color-color} 
Color-color diagrams used, together with the original 2dF spectra,  to classify  ZENS galaxies in the three spectral types of strongly star-forming, moderately star-forming and quenched galaxies.
\emph{Left panel}: Rest-frame $(NUV-I)$-$(B-I)$ color-color diagram. 
Quenched galaxies are shown in red circles, moderately star-forming galaxies in  green and strongly star-forming
galaxies in blue. Filled points show galaxies which satisfy all  color and spectral criteria used to define these classes;  empty symbols are galaxies that do not meet some of the constraints but are assigned a given class after visual inspection of the spectral and imaging data. 
Red arrows show quenched galaxies that are undetected in the $NUV$; these are positioned at an arbitrary value along the vertical axis for visualization purposes.
The lines are  predictions for the color evolution of  two synthetic models from the BC03 library, i.e.,  an SSP (gray) and  an exponentially declining star-formation history with an e-folding time of 1 Gyr (black lines), with a metallicity of $Z=Z_{\odot}$ (solid lines) and  $Z=0.4Z_{\odot}$ (dashed lines). From  left to   right, the large black circles indicate the colors of these stellar populations at the ages of 0.5, 1, 5 and 10 Gyr.  
The effect of  a $A_V=0.5$  dust reddening vector  is shown with an  arrow. 
Conversion to rest-frame colors is obtained using the $k-$correction values given by the ZEBRA+ best-fit templates.
\emph{Right panel:} $(FUV-NUV)$-$(NUV-B)$ color-color diagram that we use to  further check our spectral classification  of ZENS galaxies. Symbols and lines are  as in the left panel. Quenched galaxies that are undetected in the FUV are represented as arrows and are positioned at an arbitrary value along the vertical axis for visualization purposes.} 
\end{figure*}

The combination of FUV/NUV and optical/NIR colors has been shown to be a powerful tool
to disentangle dust-reddened star-forming galaxies from  red-and-dead quenched galaxies \citep{GildePaz_et_al_2007,Haines_et_al_2008, Williams_et_al_2009, Bundy_et_al_2010}.
As an independent validation of the spectral classification of the ZENS galaxies presented above,
 we analyzed the two color-color diagrams illustrated in Figure \ref{fig:color-color}, i.e., the ($NUV-I$)-($B-I$) and the ($FUV-NUV$)-($NUV-B$) diagrams. To this purpose we used the aperture photometry described in Section \ref{sec:Phot}; 
rest-frame magnitudes were obtained using the $k$-corrections derived from the ZEBRA+ best-fit SEDs.  

We used BC03  models to determine suitable color cuts in these diagrams for a three-bin  spectral  classification as discussed above. Specifically, the condition for galaxies to be classified as quenched was set to:  $NUV-I>4.8$, $NUV-B > 3.5$ and $B-I > 1.2$; these color cuts identify  $>$5 Gyr passively evolving SSP models (gray lines in the figure).
Strongly star-forming galaxies were defined to have  $NUV-I<$3.2, 
$NUV-B < 2.3$ and $B-I<1.0$, i.e., cuts that are consistent with the colors of  1-5 Gyr-old stellar populations with a $\tau=1$ Gyr exponentially declining star-formation history  (black lines).
Moderately star-forming galaxies are shown with green points in Figure \ref{fig:color-color} and were selected to have NUV-optical colors that fall between the two bracketing classes above. 

The reddening arrow  in the ($NUV-I$)-($B-I$) diagram moves galaxies parallel to the sequence defined in this diagram by the star-forming galaxies, but away from the region occupied by  the quenched galaxies. 
In contrast with the  degeneracy observed for optical colors, the $NUV$-optical color-color diagram is thus particularly  effective  in discriminating dusty star-forming galaxies from quenched  galaxies. Note that  we do not use the FUV passband as a classification criterion, but its availability enables us  to define a FUV-NUV color in the vertical axis of the right panel of Figure  \ref{fig:color-color}, which gives a further handle to verify  the robustness of the different spectral types.

The filled symbols in Figure \ref{fig:color-color}  show galaxies that satisfy both the spectral {\it and}  color conditions that are required for qualifying as quenched and moderately/strongly star forming galaxies.  
In the light of non-zero photometric errors,   however, we expect that the sharp color cuts that we  assumed  introduce  some degree of  contamination from one to another of the different spectral classes. Most misclassifications are likely to  show up as galaxies that do not satisfy simultaneously both spectral {\it and} color criteria qualifying a specific spectral class. Galaxies that did not satisfy simultaneously both classification criteria  are shown as empty symbols in Figure  \ref{fig:color-color};  their  spectra and images were visually  inspected before they were finally assigned to a given spectral class.  The categories of such objects are: $(i)$  galaxies with a  quenched   spectrum but ``blue" colors relative to normal quenched galaxies  (there are 61 such galaxies in the entire ZENS sample);   $(ii)$ galaxies with a strongly  star-forming spectrum but with  redder colors than normal star forming galaxies;  $(iii)$ a few galaxies, classified as moderately star-forming on the basis of their  spectral features, that however do not satisfy simultaneously all $NUV-B$, $NUV-I$ and
$B-I$  color selection criteria; and, finally, $(iv)$ a few (21) galaxies that nominally would  fall in the ``quenched'' region of the color-color planes, but have clear  emission in the $H_{\alpha}$ and are assigned to the moderately star-forming class. 

Several factors contribute to category ``$(i)$", i.e., to the detection of 61 absorption-spectrum galaxies with blue colors. About 40\% of these systems are simply scattered in the NUV-optical color diagrams by photometric errors, and we assess they are, at a visual inspection, rather normal quenched galaxies. In seven of these galaxies, instead,  the analysis of  the WFI $B$ and $I$ images  shows that the fiber-spectra catch mostly the light of a quenched bulge, and leave  low level star formation in an outer disk undetected. 
For two other galaxies the spectra present line ratios  that are indicative of AGN activity.
In the remaining cases, however,  the blue colors in these  absorption-spectrum galaxies   may interestingly indicate either a minor episode of star-formation, which is causing the integrated color to become slightly bluer, or, passively-evolving populations that are on average younger than those of their redder counterpart. 
We thus further inspected the 2dFGRS spectra  of these galaxies  to search for  possible ``post-starbursts" features, specifically the presence of strong Balmer absorption  lines (especially H$_{\delta}$) but no significant [OII] emission \citep[e.g.][]{Dressler_Gunn_1983,Couch_Sharples_1987,Zabludoff_et_al_1996,Poggianti_et_al_1999,Poggianti_et_al_2009}.
 We adopted as the threshold for ``strong" Balmer  absorption an H$_{\delta}$ value equal to 1.5$\times$ the median  value of 1.67$\AA$  measured for the sample of normal quenched galaxies. 
Twelve galaxies  were found with  such post-starburst  spectra;  these galaxies are accordingly  flagged in the ZENS catalog published with Paper I. 
The reminder of the blue-but-quenched galaxies did not formally satisfy this quantitative criterion. They are also however flagged in the catalog, not as ``post-starburst", but as blue-but-quenched systems, to keep memory of their peculiarity. We postpone a detailed study of these galaxy populations to a future ZENS analysis.
 
\begin{figure}[htbp]
\begin{center}
\includegraphics[width=60mm,angle=90]{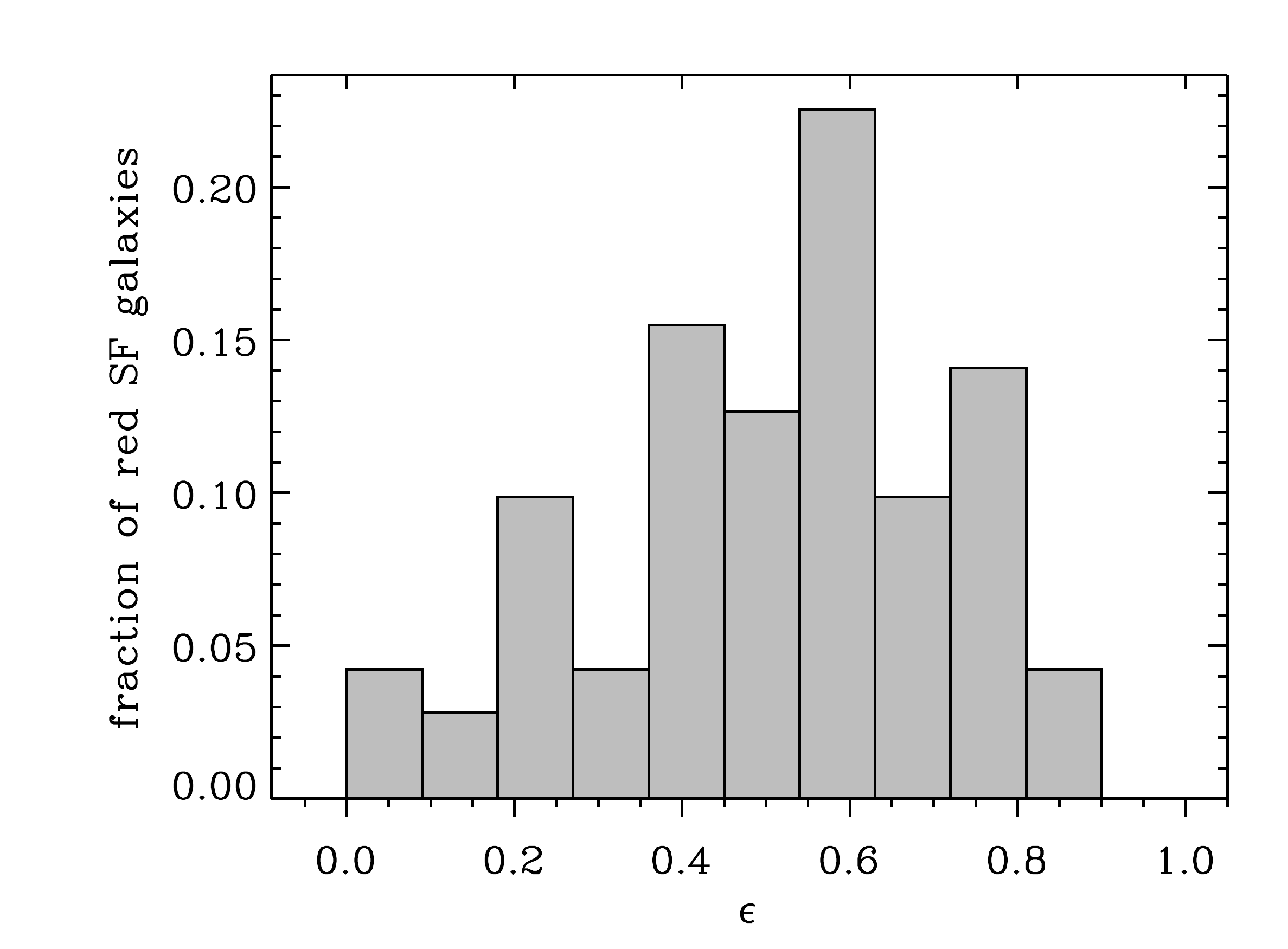} 
\end{center}
\caption{\label{fig:RedSF_ellip} Distribution of ellipticities for 
galaxies that are classified as strongly star-forming based on the 2dF spectra, but which have red optical or NUV-optical colors.
The  majority of these galaxies have large ellipticities; also at a visual inspection, they are disk galaxies with large inclination angles. Thus we  attribute the reddening of their colors to  dust obscuration.} 
\end{figure}

We finally  note that ``red"  galaxies with a star-forming spectrum  (i.e., galaxies in category $(ii)$ above),  are nevertheless generally bluer than the genuinely ``quenched'' galaxy population. 
Most of these red star-forming galaxies  are disks with a relatively high inclination, i.e. ellipticity $\epsilon > 0.4$ (see Figure \ref{fig:RedSF_ellip}). The red colors of these object are  a consequence of dust obscuration.

Summarizing, the combination of the spectroscopic information discussed in the previous Section, with the (FUV and) NUV-optical photometric diagnostics discussed in this Section, leads to a  robust spectral classification for the ZENS galaxies in the three classes of quenched, moderately star forming and strongly star forming galaxies. Note however that this classification does not provide a quantitative estimate for the SFRs of the galaxies, for which we  rely on the ZEBRA+ SED fits. We therefore performed a further test of consistency between the galaxy spectroscopic classes (based on the analysis of spectral features and colors) and the SFRs (based on the ZEBRA+ SED fits), to ensure self-consistency between  SFR estimates and spectral classification of the ZENS galaxy sample.

\subsection{Quantitative SFRs from ZEBRA+ SED fits} \label{sec:SpectraMorph}
 
\begin{figure*}[htbp]
\begin{center}
\includegraphics[width=105mm]{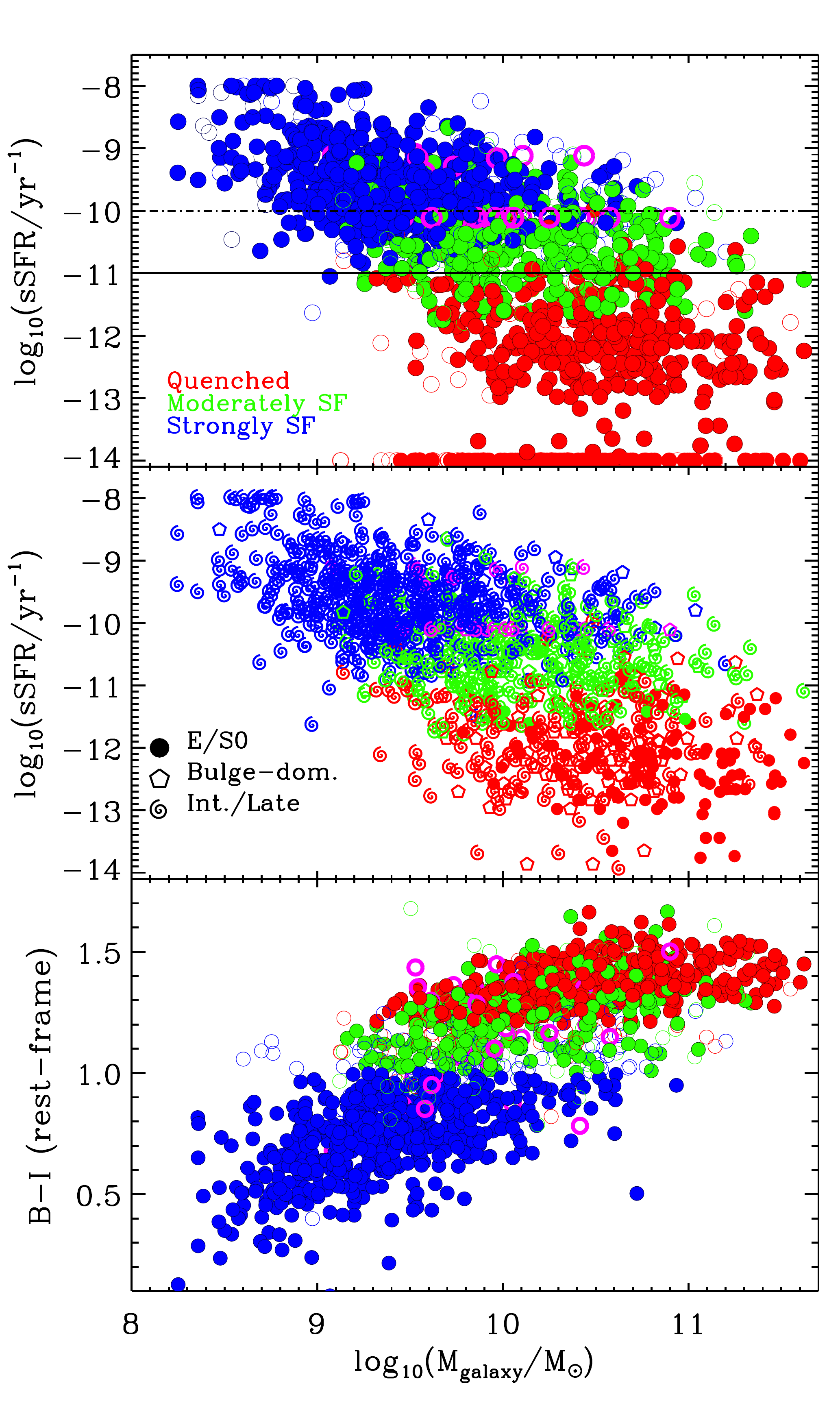} 
\end{center}
\caption{\label{fig:sSFRtypes}  \emph{Top}: Location of the three spectral types  on the sSFR vs. galaxy stellar mass relation.  Symbols and colors are as in Figure \ref{fig:color-color}. The horizontal dashed and solid lines mark the inverse of the current age of the Universe and ten times the current age of the Universe, respectively. The 31 galaxies with ``enforced" star-forming SED templates discussed in the text are identified with magenta (empty) symbols.  Galaxies plotted at a constant $\log_{10}($sSFR/yr$)=-14$ are quenched systems for which the best fit SED model resulted in very low star-formation rates (SFR$<10^{-4}\Msol$yr$^{-1}$). 
\emph{Middle}:  Symbols in this panel correspond to different morphological types: ellipticals and S0s are shown with circles, pentagons are for bulge-dominated spiral galaxies and intermediate-/late-type disks are shown with spiral symbols; color are as before. \emph{Bottom}: $(B-I)$ color vs.\, galaxy stellar mass. Symbols and colors are as in the top panel of the figure.} 
\end{figure*}

 The  top panel of Figure \ref{fig:sSFRtypes} shows the relation between
sSFR and galaxy stellar mass $M_\mathrm{galaxy}$, with highlighted in red, green and blue the three spectral classes discussed above (color coding as in Figure \ref{fig:color-color}). 
Not surprisingly and reassuringly, there is a very good agreement between the quantitative SED-based SFR  estimates and the assigned spectral types. Galaxies classified as quenched based on their absorption spectral features and/or red NUV-optical colors  have generally  $\log$(sSFR/yr) $\lesssim-11$, i.e., at their current SFRs, it would take more than 10 times the present age of the Universe ($t_U$)  to double their mass. Strongly star-forming galaxies have $\log$(sSFR/yr)$ \gtrsim-10$, i.e., mass-doubling times much shorter than $t_U$;  moderately star-forming galaxies fall in between the  two bracketing classes. It should be noted that the sample in Figure \ref{fig:sSFRtypes} includes quenched galaxies, and is not complete in stellar mass at the low end, and so the steep slope of the sSFR-mass relation indicated in the Figure should not be taken to indicate a highly negative value of $\beta$ in the relation $sSFR \propto M^\beta$ (c.f. \citealt{Peng_et_al_2010}).

We highlight that in the ZENS sample, there are 31 galaxies that are classified as moderately or strongly star-forming systems based on their spectral features and/or colors, but whose initial, unconstrained ZEBRA+ best fits resulted in incorrectly  low specific star-formation rates ($\log$(sSFR)<$-11.8$).
We explored the origin of this inconsistency. For 17 of these ``problematic" galaxies no $GALEX$ photometry is available, and thus  
for them a reduced  number of passbands were used to perform the SED fits, which are unconstrained at UV wavelengths. 
All but two of  the remaining galaxies  were found to have
 high inclinations thus the too low sSFR are likely the result of an underestimation of the  amount of dust reddening in such galaxies 
 (the typical reddening is $E(B-V)=0.08$ in the unconstrained fits).
For all these galaxies, we thus re-run our ZEBRA+ fits constraining the latter to a star-forming SED model, specifically by only using constantly star-forming templates (csf) or templates 
with an exponentially decreasing star-formation history with $\tau=2, 4, 6, 8 $ Gyr and ages between 0.4 and 4 Gyr.
The same metallicity and reddening  as in Table \ref{tab:ZEBRA} were used for these `constrained'  ZEBRA+ runs.
This led to revised SFR, stellar mass and sSFR estimates  (magenta symbols in Figure \ref{fig:sSFRtypes}) for these 31 galaxies,
 which are flagged  in the ZENS catalog that we have published in Paper I.  
As expected, the revised stellar masses are slightly larger (on average by 0.2 dex) than the mass estimates providing the clearly incorrect (i.e., too low) sSFRs. The column in the ZENS catalog that is used as a flag to identify these galaxies lists the ``incorrect" galaxy stellar mass values for these 31 galaxies (and a dummy entrance for the remaining ones); the column listing SFRs and sSFRs incorporate the ``correct" values, obtained with the star-forming SED templates of the constrained ZEBRA+ fits.
When appropriate, we will test our analysis  excluding these galaxies from our sample.  We also however warn against a blind use of SED-fit-based estimates of SFRs and stellar masses, without suitable checks that these agree with the global colors and/or spectra of individual galaxies.

\subsection{Spectral types versus morphological types}

We used the morphological classes determined in Paper II to examine the morphological mix within each of the spectral type defined above. These morphological mixes are listed in  Table \ref{tab:MorphoSpec} and shown in the middle panel of Figure \ref{fig:sSFRtypes}, which again plots the sSFR-mass relation as before, with highlighted in different symbols galaxies of the various morphological classes (with colors still representing the spectral type as before).   As expected, there is  a broad correlation between morphological and spectral types, i.e., E/S0 galaxies are largely quenched, and morphologically-classified late-type  disks are mostly strongly star-forming galaxies.
Intermediate-type disks  distribute themselves almost 
equally in the three spectral classes,  and a third of the bulge-dominated spiral
galaxies shows some star-formation activity.  
We discuss in   detail in Paper IV (Carollo et al.\ 2013b in preparation) the morphological properties of quenched   galaxies and the dependence of the quenched fraction on the different environments that we study in ZENS.

\begin{figure*}[htbp]
\begin{center}
\includegraphics[width=85mm,angle=90]{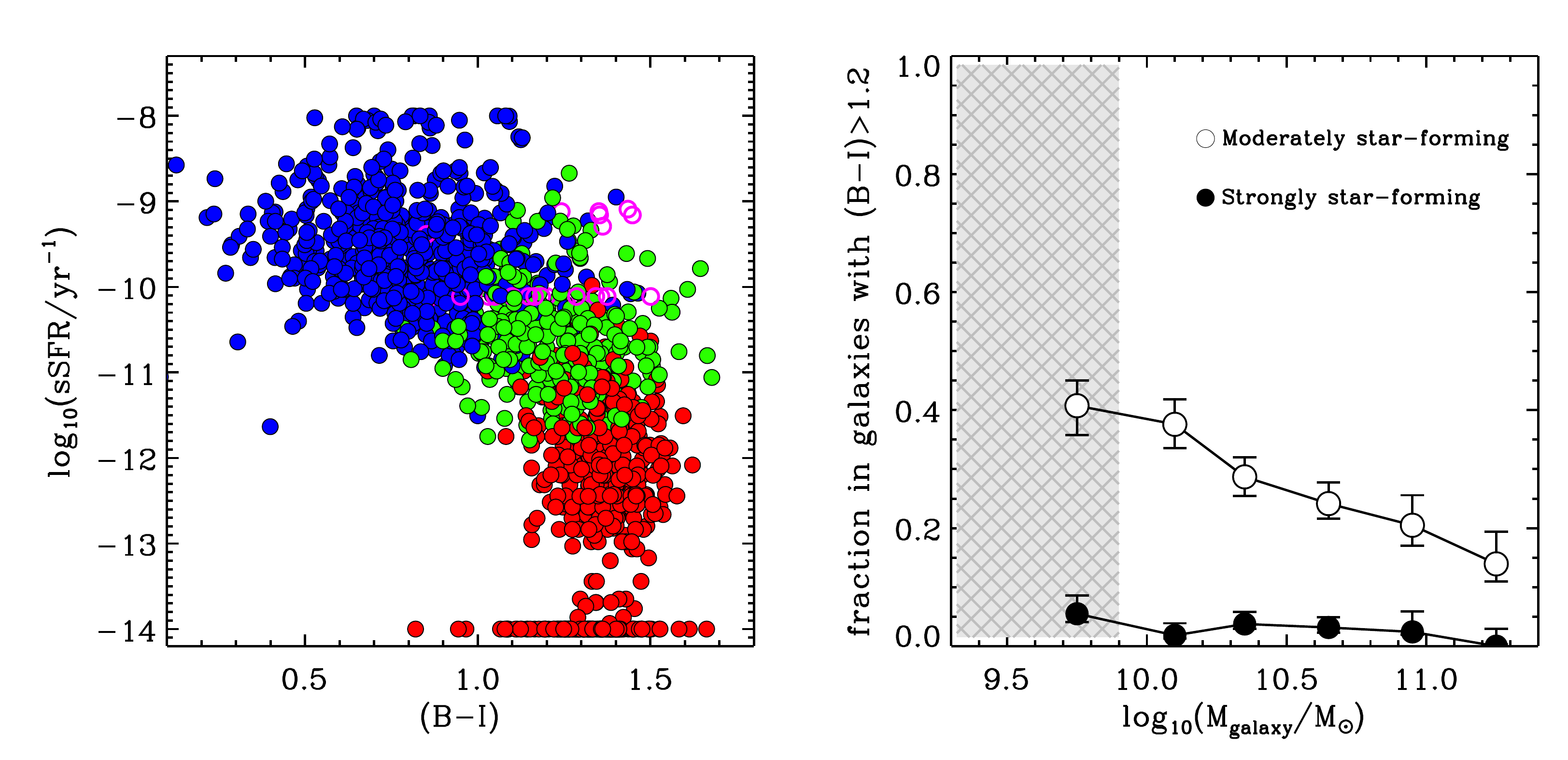} 
\end{center}
\caption{\label{fig:RScontamination}\emph{Left}: sSFR as a function of galaxy integrated $(B-I)$ color. Symbols are color-coded according to the spectral type as in Figure \ref{fig:sSFRtypes} (galaxies plotted at a constant $\log_{10}($sSFR/yr$)=-14$ are systems for which the best fit SED model resulted in very low star-formation rates, SFR$<10^{-4}\Msol$yr$^{-1}$).  \emph{Right}: Fraction of galaxies with  $(B-I)>1.2$ (i.e., contaminants to the ``optical red-sequence" of quenched galaxies) that are classified as strongly star-forming (filled symbols) or moderately star-forming (empty symbols).
The gray shaded area show the region of mass incompleteness  for quenched galaxies in the ZENS sample.} 
\end{figure*}
   
 \subsection{An Important Truism: Red Galaxies Are Not All Quenched Galaxies}
 
  Finally, the bottom panel of  Figure \ref{fig:sSFRtypes} explicitly shows that, although  quenched galaxies segregate
into a tight optical ``red sequence", defined by our  color cut $(B-I)>1.2$,  and moderately- 
and strongly star-forming galaxies are respectively mostly
 located in   ``green-valley" and  ``blue-cloud" regions, 
both    star-forming classes have $(B-I)$ colors that overlap with the red sequence of quenched galaxies.

This is further illustrated in the left panel of Figure  \ref{fig:RScontamination}, which shows the sSFR versus integrated $(B-I)$ color, and, explicitly,  in the right panel of the same figure, which quantifies
the contamination of star-forming galaxies to the ``red sequence", as defined through a simple optical color, in our case the $(B-I)$ color.
Specifically, the right panel  of Figure  \ref{fig:RScontamination} shows, as a function of stellar mass, the fraction of galaxies that are classified as either strongly star-forming or moderately star-forming and have colors $(B-I)>1.2$ (normalized to the total number of galaxies with such $(B-I)$ colors).
We note that the relative importance of dust-reddened galaxies increases  with decreasing stellar mass; at our completeness mass limit for quenched galaxies of 
 $10^{10}\Msol$ star forming galaxies contribute up to  $\sim50\%$  of the nominal $(B-I)$ red sequence (assuming that the reddening is in all  cases attributable to dust effects only;  a population with relatively low sSFR with respect to those expected for a given stellar mass may also  be present within the \emph{star-forming} class). This result is supported by independent analyses  at similar  mass scales
 (e.g.  \citealt{Bell_et_al_2004,Davoodi_et_al_2006,Haines_et_al_2008,Maller_et_al_2009}).    

\begin{deluxetable*}{cccc}
\tablewidth{0pt}
\tabletypesize{\small}
\tablewidth{0pt}
\tablecaption{Morphological mix for  the three spectral  classes in the \textsc{ZENS} sample} 
\tablehead{ \colhead{Morphological Type}  & \colhead{Quenched (\%)}  &  \colhead{Moderately-SF  (\%)} & \colhead{Strongly-SF  (\%)}}
\startdata
Ellipticals &   100  &  ...   &   ... \\
S0s &   91  &   7  &   1 \\
Bulge-dominated spirals &   67  &   22  &   11  \\
Intermediate-type disks &   44  &   29  &   27  \\
Late-type disks &   3  &   25  &   72  \\
Irregulars &  3 &  3 &   94  \\
\enddata\label{tab:MorphoSpec}
\tablecomments{For each of the morphological types of Paper II, the table lists  
 the corresponding fractions for the different spectral types discussed in Section \ref{sec:ActiveQuiescent}.}
\end{deluxetable*}


\section{$(B-I)$ two-dimensional color maps and radial  color gradients}\label{sec:ColorMaps}

To study the variation of stellar populations and distribution of star-forming regions within galaxies, we used the WFI $B$ and $I$ images to construct  a two-dimensional $(B-I)$ color map for each galaxy.
Since the $B$ and $I$ filter observations have different PSFs, we cross-convolved each passband with the PSF of the other passband.
To prevent spurious color gradients arising from  errors on the  registration of the two images, we rebinned the original pixels in 7x7 sub-pixels before performing the registration of the $B$- and $I$-images using the IRAF task {\ttfamily imalign}.
A correction for Galactic dust extinction was  applied to the  color maps.

\subsection{Voronoi tessellation}

It is customary to perform a binning or smoothing of the spatial elements in astronomical images in  order to homogenize the signal-to-noise ratio (S/N) across the sources \citep[e.g.][]{Sanders_Fabian_2001,Cappellari_Copin_2003,Ebeling_et_al_2006,Zibetti_et_al_2009}. 
To increase the S/N of the ZENS WFI color maps in  the outer regions of  galaxies, where the flux from the sky background is dominant, we performed an adaptive local binning of pixels using a \emph{Voronoi tessellation} (VT)  approach. 

The idea behind the method is to group adjacent pixels into bigger units that have a minimum scatter
around a desired S/N  (\emph{uniformity requirement}). The tessellation should further 
satisfy a \emph{topological} and \emph{morphological} requirement, i.e., create a partition without gaps 
or overlapping regions and a maximum roundness of the bins should also be attained.
In the VT technique a density distribution of the S/N 
over the area of the galaxy is defined, $\rho(r)=(S/N)^2(r)$, and binning to a constant S/N is then ensured by determining regions of equal mass according to $\rho$.
The VT method is often used to maximize the reliability of astrophysical information from noisy data (albeit at the expense of angular resolution; see e.g. \citealt{Ferreras_et_al_2005,Daigle_et_al_2006}).  

To perform the VT on the $(B-I)$ color maps of the ZENS galaxies we used the publicly available IDL codes
of \citet{Cappellari_Copin_2003}\footnote{http://www-astro.physics.ox.ac.uk/$\sim$mxc/idl/} and in particular
we adopted the generalization (\emph{weighted voronoi tessellation (WVT)}) proposed by \citet{Diehl_Statler_2006}.
For the code to work, along with the color map it is necessary to provide an estimate of the S/N of each pixel. 
We computed the error on each pixels using the cross-convolved images as $ \sigma=1.0857\frac{\sigma_{flux}}{<I>} $.
Here $<I>$ is the intensity in counts measured from the source.
 The flux error is the sum of poisson fluctuations in the number of photons from the source plus random noise from the sky, i.e., 
$ \sigma_{flux}= \sqrt{<I>/g_{eff}+\sigma_{sky}^2} $ where
$g_{eff}$ is the effective gain of the image and $\sigma_{sky}$ the standard deviation of the background;
the use of the effective gain takes into account that poisson statistics applies 
to the total number of  electrons coming from the sources.  
Error on the pixels of color maps were then obtained by adding in quadrature the errors in the two filters.
 
Pixels with very low S/N, which would affect the robustness of the algorithm, were excluded from the binning by imposing a minimum threshold of S/N$^2=0.05$, found to optimize the results. Pixels belonging to the background, defined as pixels at a distance $>$1.5  Petrosian radii from the center of the galaxy,  were also rejected by using a mask image.
A target S/N of 10 was chosen to construct the binned color maps. 
      
\begin{figure*}
\begin{center}
\includegraphics[width=120mm]{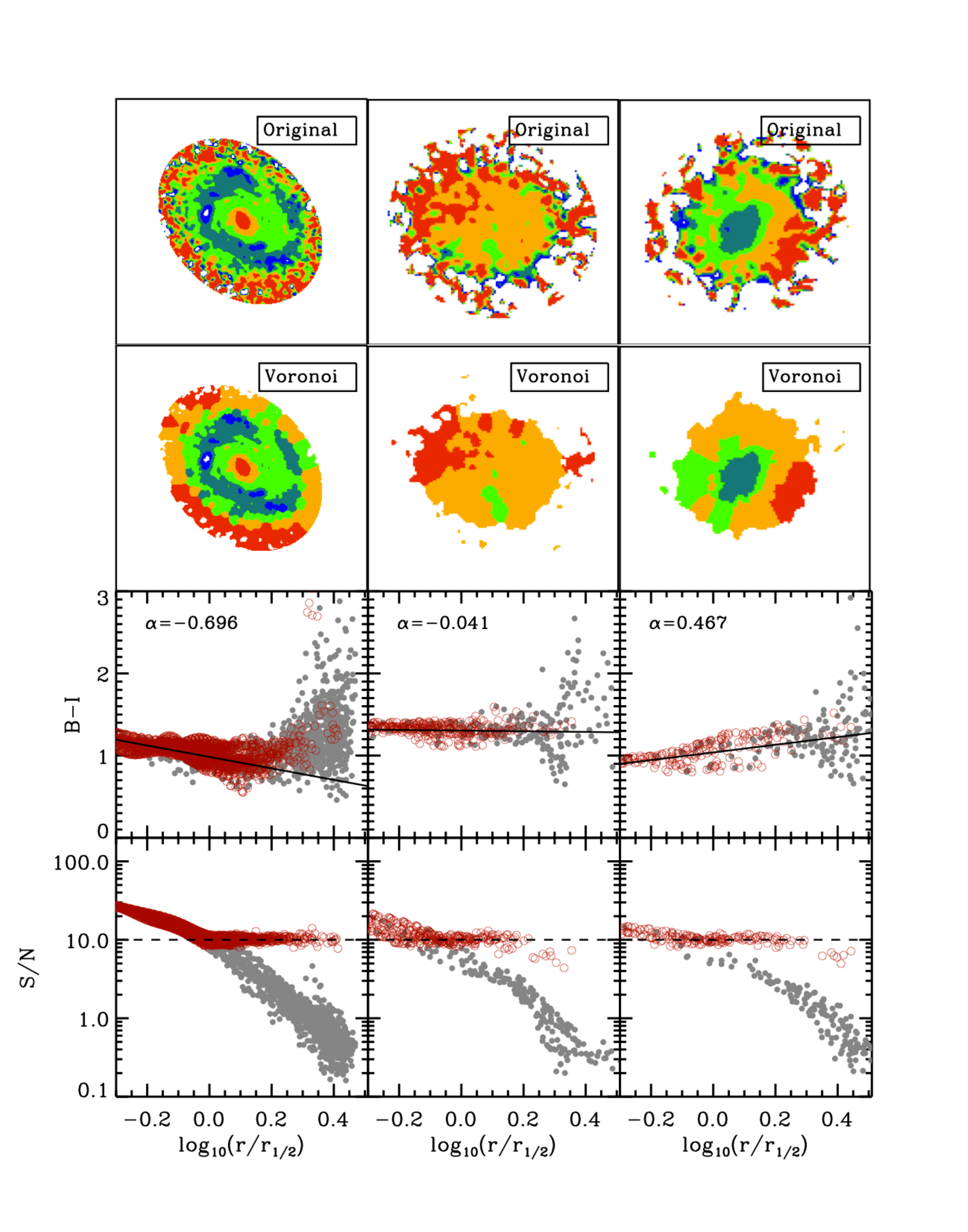}
\end{center}
\caption{\label{fig:VoronoiEx}
Examples of the  Voronoi-tessellated color maps for three of the ZENS galaxies. 
From left to right the figure shows  a galaxy with negative color gradient (i.e., outer regions bluer than the nucleus), one with a flat profile and one with positive gradient (outskirts redder than the nucleus). 
The top panels show the original un-tessellated color maps;  the second row from the top shows the Voronoi-tessellated maps.
The next two rows are, respectively, the azimuthally-averaged color and signal-to-noise radial profiles.
 Gray symbols correspond to the profiles derived from the original color maps (1 in 20  points   plotted for clarity);   black empty points (red in the online version)  are the values obtained from the tessellated color maps.
The lines in the third row panels are the best-fit linear relations to the tessellated profiles;  the dashed lines in the bottom panels show the target S/N used for the tessellation.
Radii are normalized to the galaxy half-light radius.\\
(A color version of this figure is available in the online journal.) }
\end{figure*}

 Figure \ref{fig:VoronoiEx} shows the comparison between the original, un-tessellated color maps and the VT maps,
 together with the azimuthally-averaged S/N and color radial profiles derived from the original and tessellated images.
The three galaxies in the figure are representative of the variety of observed color profiles: 
a galaxy with a negative or  ``normal" color gradient (redder center, bluer outskirts), a galaxy with a flat color profile and a galaxy  with a positive or ``inverted"color gradient (bluer center, redder outskirts).
The VT clearly enables  a more  robust measurement of the color distribution of galaxies at large radii, and  removes  the  high frequency fluctuations associated with the noise in the original color maps, while retaining substantial information on lower frequency, physical  color variations within galaxies.

\subsection{Azimuthally Averaged Radial $(B-I)$ Color Profiles}\label{sec:Color-Grad}

Azimuthally-averaged $(B-I)$ color profiles for the ZENS galaxies were obtained both from the tessellated color maps and from the GIM2D $B$ and $I-$band single-component S\'ersic fit parameters derived and calibrated in Paper II (to which we refer for details of the fits and their calibration against residual PSF, ellipticity and concentration biases).  The advantage of the former approach is that color profiles from the color maps are a straight measurement from the data and are thus not affected by data-model mismatches; the advantage the latter approach is the substantial removal of observational biases, including PSF smearing, which enables in principle to study color profiles also within the PSF radius. 

Logarithmic color gradients were in both cases calculated by fitting the linear relation 
$(B-I)=(B-I)_{r_{1/2}}+\alpha \log (r/r_{1/2})$
to the color radial profiles. The slope $\alpha=\Delta(B-I)/\Delta(\log r) \equiv \nabla(B-I)$ defines 
 what we will refer to as the ``radial color gradient";  $(B-I)_{r_{1/2}}$ defines the color at the galaxy half-light semi-major axis ($r_{1/2}$). These  $(B-I)_{r_{1/2}}$ colors were converted from observed to
rest-frame   using the $k$-correction derived from the ZEBRA+ fits. 
 
Unless obvious, we will distinguish in the following and future papers between gradient parameters derived from the GIM2D analytic fits and those obtained from the VT color maps by adding the subscripts ``GIM2D" and ``Voronoi" to the symbols $\nabla(B-I)$ and $(B-I)_{r_{1/2}}$. Note that the agreement between the parameters of the color profiles obtained with the two approaches was found to be generally good; where systematic differences were identified,  we studied the origin of such discrepancies and derived recipes to tie the two measurements to a common scale, as described in Section \ref{sec:common scale}.     

We adopt as our fiducial measurements in our future ZENS analyses the radial color profiles $\nabla(B-I)_\mathrm{GIM2D}$ based on the analytical S\'ersic fits; these fits were performed within the radial range  $0.1r_{1/2} -  2.5r_{1/2}$.  We will use however the parameters $\nabla(B-I)_\mathrm{Voronoi}$ and $(B-I)_\mathrm{r_{1/2},Voronoi}$  derived from the VT color maps for the  $\sim 5\%$ of ZENS galaxies for which no analytical fit could be derived, and also as a  check of the analytical  color profiles at relatively large radii ($>FWHM_\mathrm{PSF}$).   

The empirical  ``Voronoi" radial color profiles are affected by observational biases, and require therefore some corrections. We detail below the main issues involving these  estimates for the radial color profiles, and the approaches that we have taken to minimize spurious observational effects in their estimates.

Finally, we note that the VT color maps  contain  more information than one-dimensional gradients; in particular, they enable us to study also the color r.m.s.  dispersion (scatter) around the smooth average color profiles within galaxies. 
This was defined and computed as $\sigma (B-I)=\sqrt{\sum _i \xi_i^2/N}$, with $\xi$ residuals with respect to the azimuthally-smoothed radial color profile. 

In the last part of this paper we present a  quick look into the colors, color gradients and color scatter properties of disk satellite galaxies of all bulge-to-disk ratios in different environments.
 
 \subsubsection{Color Profiles and Color Scatter from the VT Color Maps: Corrections for Systematic Effects}\label{sec:common scale}
 
Fits to the color profiles based on the VT color maps were limited between $r_\mathrm{PSF}<r<r_{MAX}$, with $r_{MAX}$ either $2.5\times r_{1/2}$ for sufficiently high S/N maps, or the maximum radius over which the tessellation could be performed.
The inner profile within the radius $r_\mathrm{PSF}$, equal to the largest between the  $B$ and $I$ seeing  FWHM, was masked out to minimize the effects of  PSF-blurring.  We assumed that, on large scales, the radial distribution of color in the galaxies is axisymmetric and joined together all the Voronoi bins that are crossed by the same ellipse of given semi-major axis when deriving the color profiles.
To compute color gradients using the Voronoi maps, we masked out all Voronoi cells with colors $>5\sigma$ away from 
a bi-weighted mean of the colors of all cells,
or $>3.5\sigma$ away from the best-fit  linear regression to all cell colors as a function of radius. We checked that these masking criteria exclude very bright compact star-forming regions/clusters and residuals from not  perfectly cleaned nearby companions, without eliminating broader-scale fluctuations associated with internal variations in star-formation activity.  Radial color gradients $\nabla(B-I)_\mathrm{Voronoi}$  and $(B-I)_\mathrm{r_{1/2},Voronoi}$  colors were then calculated with a Levenberg-Marquardt least-squares fit to the remaining Voronoi cells data.

\begin{figure}[htbp]
\begin{center}
\includegraphics[width=65mm,angle=90]{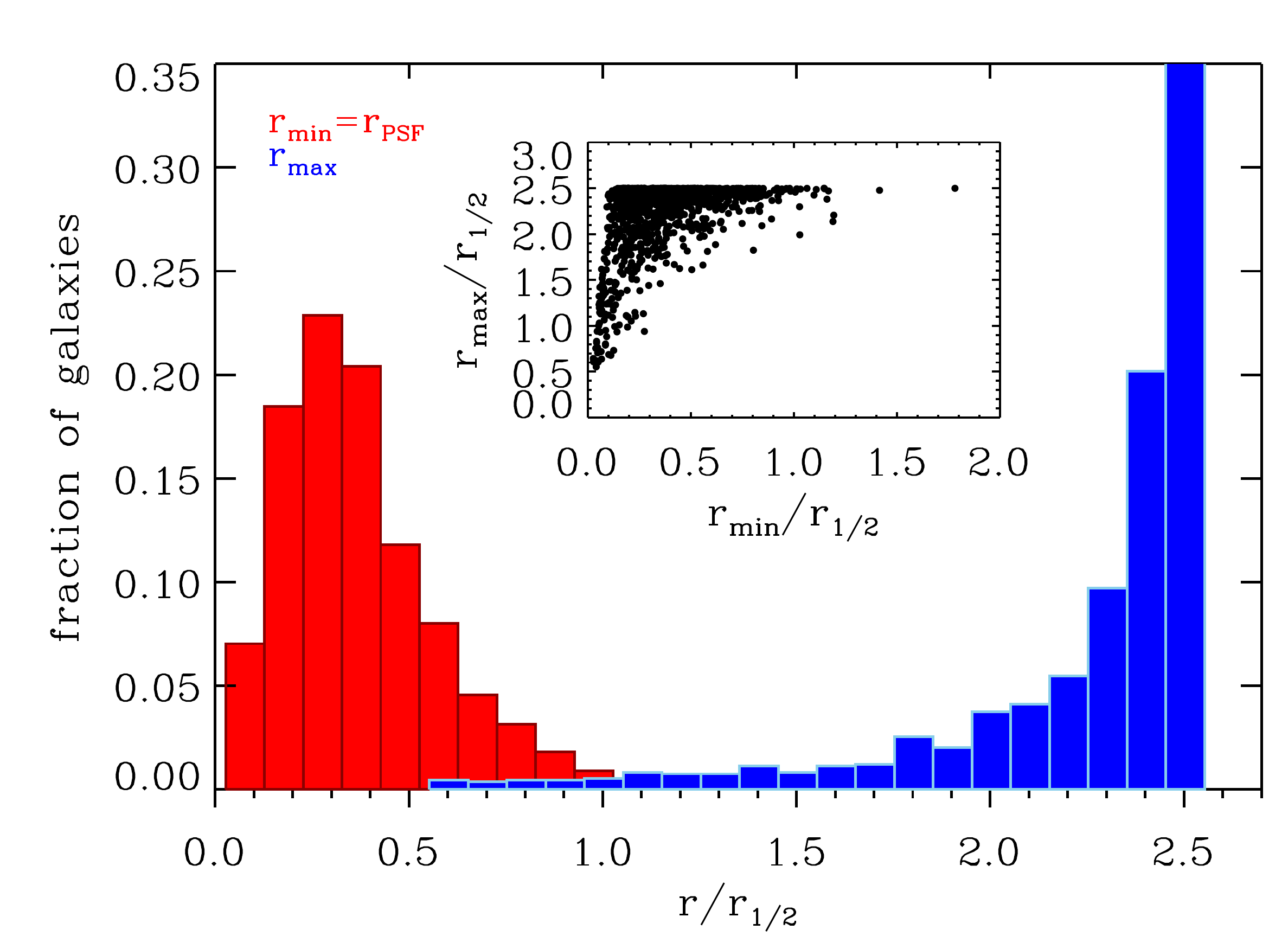}
\end{center}
\caption{\label{fig:GradRadii}
Distribution of minimum and maximum radii, in units of the global galaxy half-light radius, used to fit the azimuthally-averaged color profiles derived from the Voronoi-tessellated color maps. The minimum radius $r_\mathrm{PSF}$ is defined as the maximum between the $B$- and $I$-band PSF size.  The inset shows the comparison between the minimum and maximum radius used for each galaxy (in units of the galaxy half-light radius).\\
(A color version of this figure is available in the online journal.) }
\end{figure}

As anticipated, the  VT-based radial color gradients  require some corrections. Not least, galaxy to galaxy variations prevent  an homogeneous radial sampling of the color gradients among the entire sample:  the minimum radius that is probed depends on the seeing under which a given galaxy was observed,  and the maximum radius is often set by the noise of the image and the galaxy brightness and intrinsic size. Figure \ref{fig:GradRadii} shows the distribution of these minimum and maximum radii employed in the calculation of the color gradients from the Voronoi maps, in units of the galaxies half-light radii. 
Rather comforting is that, for most of the sample, the fitted region spans a range of radii from well below $r_{1/2}$ up to $\sim 2.5\times r_{1/2}$;  in only $8\%$ and $4\%$ of the galaxies the minimum radius is larger than 0.7$r_{1/2}$ and the maximum radius  is smaller than $1.3r_{1/2}$, respectively.
Note that as shown on the inset in Figure \ref{fig:GradRadii} these two troublesome situations, however, never occur at the same time.

\begin{figure}[htbp]
\begin{center}
\includegraphics[width=77mm,angle=90]{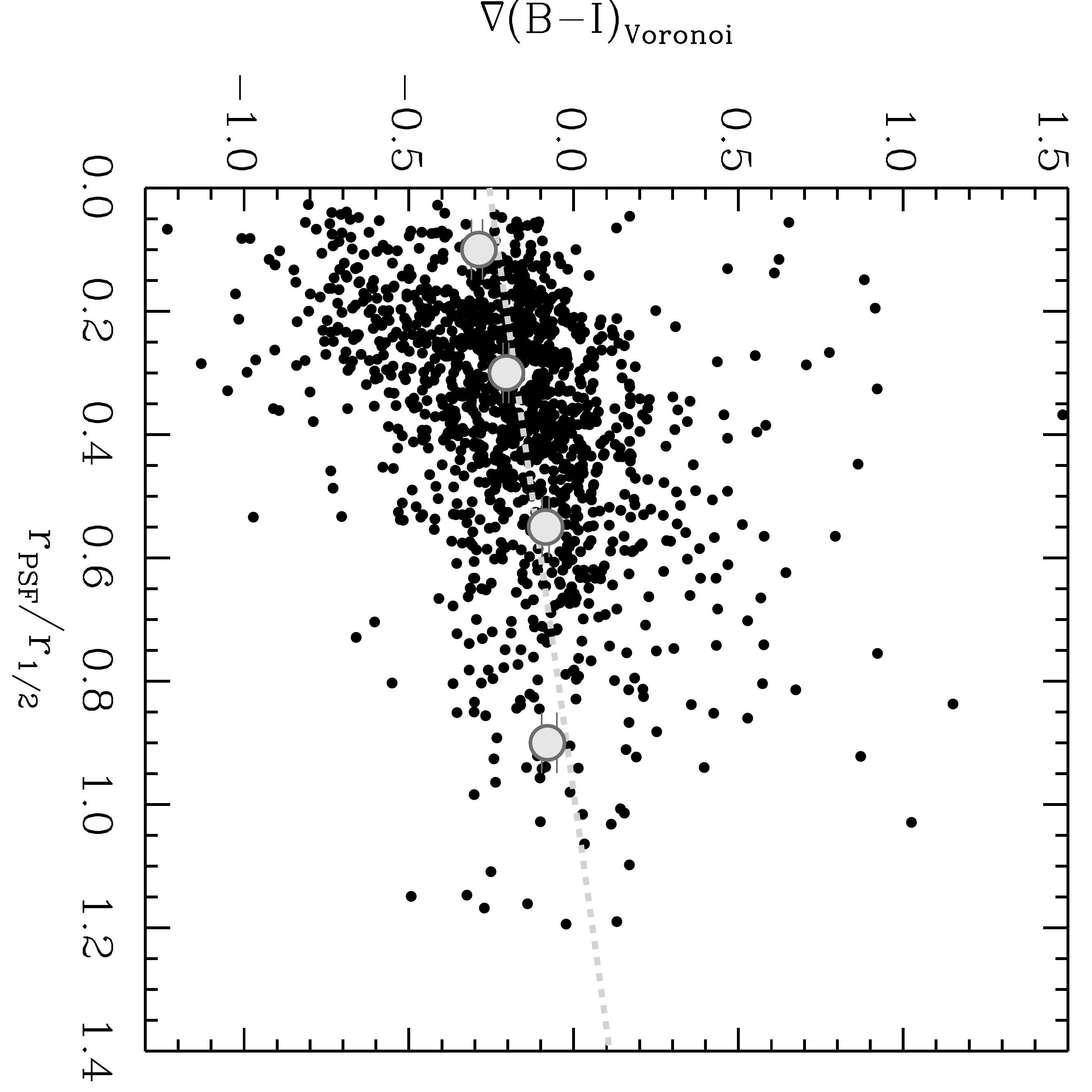}
\end{center}
\caption{\label{Fig:GradPSF}
Color gradients $\nabla(B-I)_\mathrm{Voronoi}$, i.e., derived by fitting  the azimuthally-averaged color profiles obtained from the Voronoi tessellated color maps, as a function of the minimum
radius used in the fits (in units of the half-light radius). Large gray points indicate  the average color gradient in four bins of $r_\mathrm{PSF}/r_{1/2}$; the dashed line is the result of the $3\sigma$-clipped linear least-squares fit  used to statistically correct the  raw  $\nabla(B-I)_\mathrm{Voronoi}$ values.}
\end{figure}

Also,  PSF blurring and  consequent masking of  the central galactic  regions (by one PSF's FWHM) remain a source of concern for the reliability and stability of the derived $\nabla(B-I)_\mathrm{Voronoi}$ values.
In Figure \ref{Fig:GradPSF} we plot the gradients $\nabla(B-I)_\mathrm{Voronoi}$ as a function of $r_\mathrm{PSF}$, in units of  the galaxies half-light radii. 
There is a correlation between the measured gradients $\nabla(B-I)_\mathrm{Voronoi}$ and $r_\mathrm{PSF}/r_{1/2}$:
not surprisingly, the gradients become flatter with increasing  FWHM of the PSF.
To minimize this effect, at least in a statistical manner, we derived an empirical correction function by fitting a linear relation to the $\nabla (B-I)$ vs.\, $r_\mathrm{PSF}/r_{1/2}$ relation, which is shown with a dashed line in Figure \ref{Fig:GradPSF}. This empirical correction was then applied  to all the Voronoi gradients.

\begin{figure*}[htbp]
\begin{center}
\includegraphics[width=90mm,angle=90]{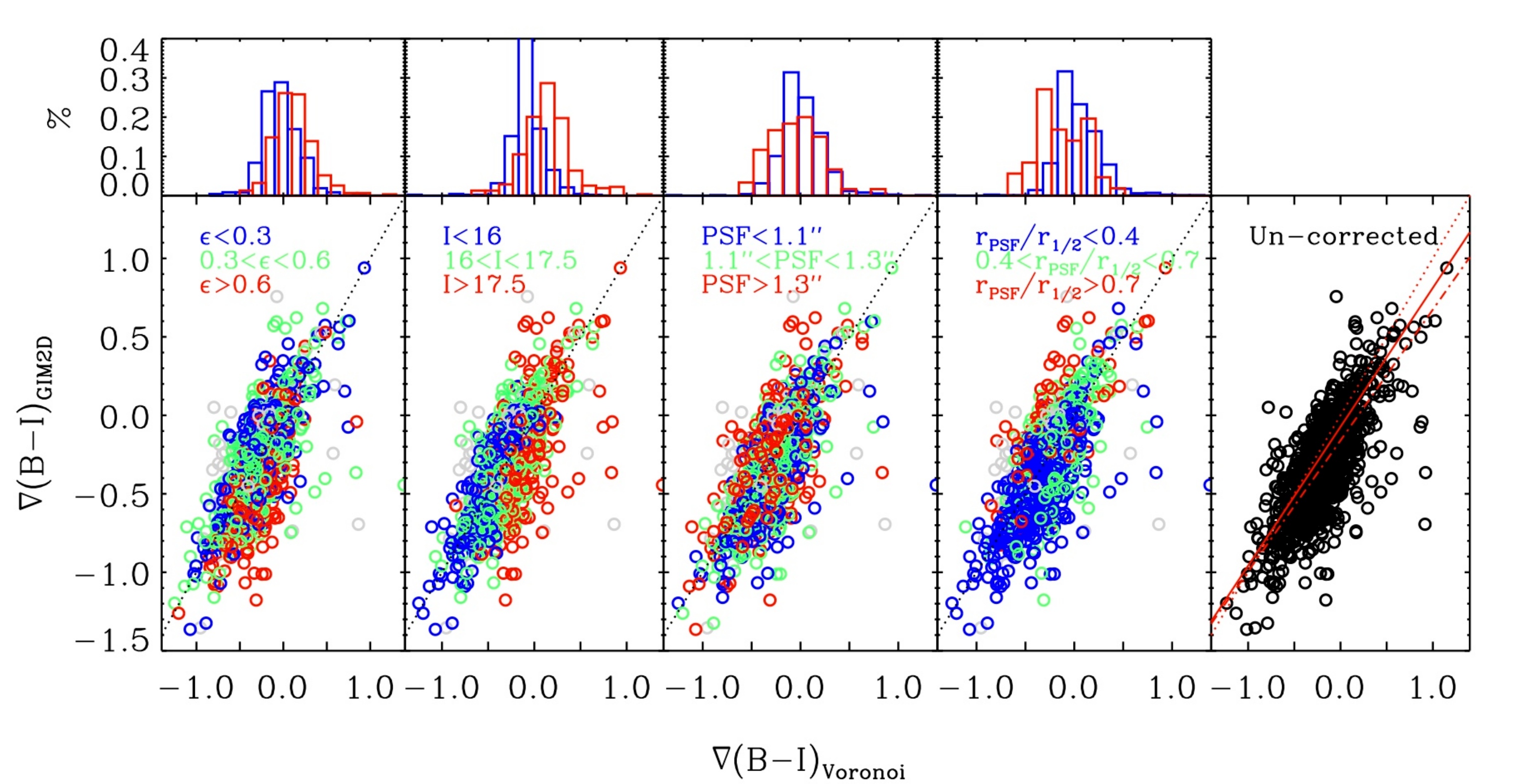}
\end{center}
\caption{\label{fig:GradComp} Comparison between  color gradients $\nabla(B-I)_\mathrm{Voronoi}$ derived from  fits to the average color profiles obtained from the Voronoi tessellated color maps,  and  color gradients obtained from the analytical GIM2D S\'ersic fits to the $B-$ and $I-$band surface brightness profiles. 
ZENS galaxies are color-coded according to a number of observational diagnostics: from left to right, these are galaxy  ellipticity,  $I$-band magnitude,  the maximum size between the $B$- and $I$-band PSF and minimum radius   used in the fits to the color profile. In the last panel the black points show  the values of the Voronoi gradients before applying  the empirical correction that takes into account the effect of masking the central $r_\mathrm{PSF}$ region. The histograms on top of each panel show the distributions of differences between Voronoi and GIM2D gradients in the lowest (blue) and highest (red) bin of the given parameters. Dotted lines indicate the identity line. In the right-most panel,
 solid and dashed-dotted   lines  are the best-fit to the corrected for central masking  and uncorrected $\nabla(B-I)_\mathrm{Voronoi}$ measurements. Grey symbols identify ``troublesome" galaxies with  possible  bright  star  contamination to their photometry or whose fits to the color profiles are restricted within  1.3 half-light radii or less.}
\end{figure*}

\begin{figure*}[htbp]
\begin{center}
\includegraphics[width=90mm,angle=90]{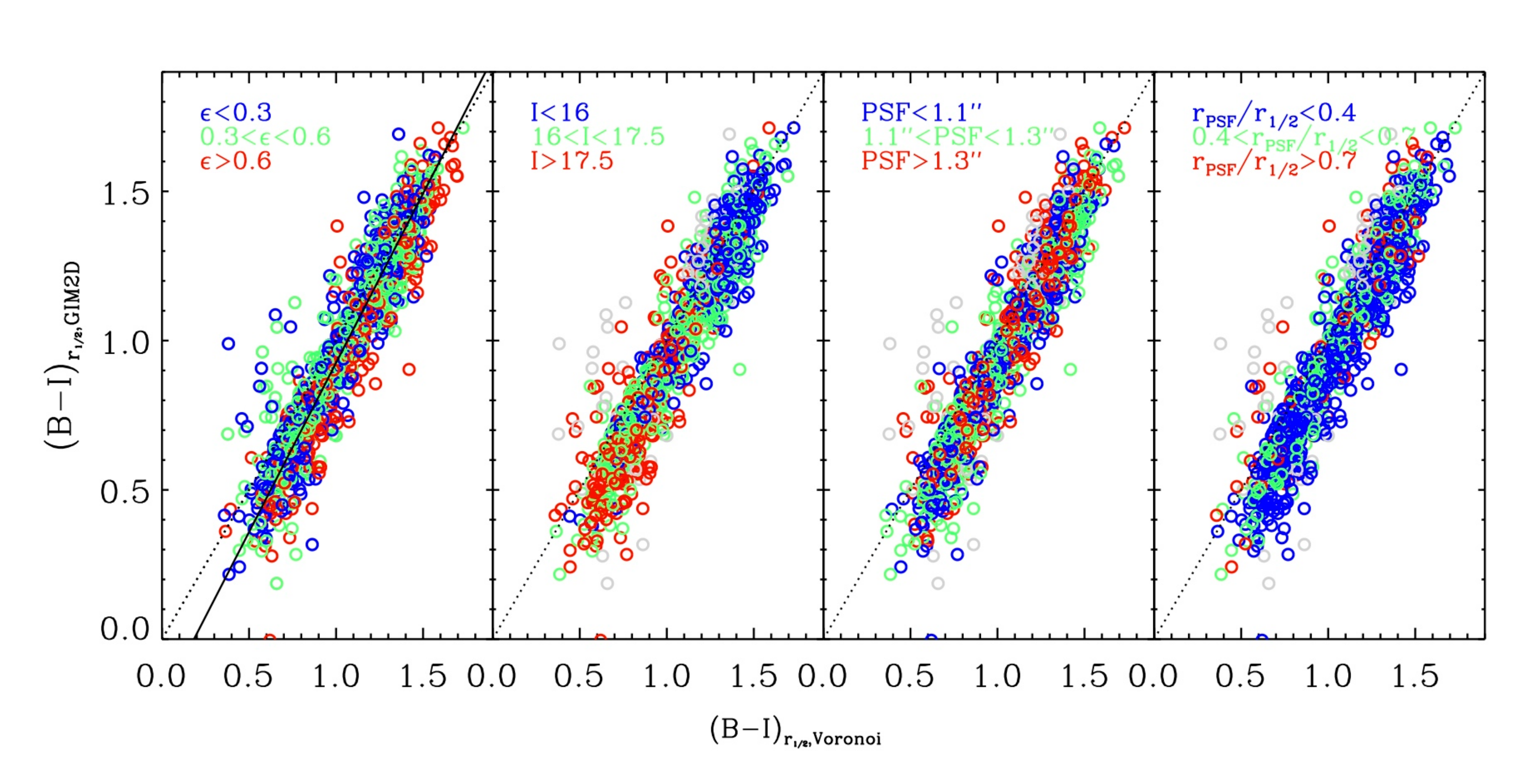}
\end{center}
\caption{\label{fig:GradCompCenCol} As in Figure \ref{fig:GradComp}, but for the $(B-I)_{r_{1/2}}$ measurements.  The dotted and solid lines show the identity line and the fit to the points, respectively.}
\end{figure*}

Figures \ref{fig:GradComp} and \ref{fig:GradCompCenCol}  show respectively the comparison between the
corrected  $\nabla(B-I)_\mathrm{Voronoi}$ color gradients and the $(B-I)_\mathrm{r_{1/2},Voronoi}$ colors  with the corresponding GIM2D measurements. For reference, we also show as black points the uncorrected Voronoi gradients in the last panel of Figure \ref{fig:GradComp}. 
Overall the two methods give qualitatively consistent results, particularly for the $(B-I)_{r_{1/2}}$ values.
Concerning the color gradients, the above correction ameliorates the situation, but  
 the Voronoi-based gradients tend to remain slightly  shallower than the GIM2D-based gradients.
To understand the origin of such residual difference we searched for potential systematic biases with galaxy inclination, magnitude, minimum radius used in the fits, and size of the PSF.

 There is  indeed a bias with galaxy inclination: galaxies with $\epsilon>0.6$ lie systematically below the identity line in the first panel of Figure \ref{fig:GradComp}, 
 mostly as a consequence of projection effects. Along the semi-minor axis, where projection effects are strongest, for example, a given Voronoi cell integrates a larger range of physical radii, causing color information to be smeared out and color profiles to be slightly flattened at these high inclinations.
The difference between the Voronoi and GIM2D profiles depends also on  galaxy magnitude: the difference is larger for fainter  ($I>17.5$) galaxies, whose $\nabla(B-I)_\mathrm{Voronoi}$ and $(B-I)_\mathrm{r_{1/2},Voronoi}$ are respectively flatter and $\sim 0.1$ mag redder  than the corresponding ``GIM2D" measurements.  This is  most likely due to the contribution from the night sky noise, which becomes more significant at the low surface brightness regime and longer wavelengths.
After the application of the correction delineated  above,
no strong biases were however observed with the absolute or relative size of the PSF in either the Voronoi gradients or   the Voronoi $(B-I)_{r_{1/2}}$ colors. 

Given the above, and in order to recover measurements of color gradients and colors at the half-light radius for galaxies without GIM2D fits (and minimize the bias in these measurements), we calculated the median residual differences between Voronoi and GIM2D color gradients and colors at the half-light radii, as a function of galaxy inclination and magnitude,  and applied these
additional correction functions to those galaxies for which no analytic color profile is available. 

We finally highlight that,  since  $(i)$ color gradients of highly-inclined (disk) galaxies as well as  marginally resolved galaxies ($r_{1/2}<1.5\times PSF$) are, even when using GIM2D estimates, subject to larger errors, and $(ii)$
high inclinations and small sizes may exacerbate physical biases introduced by absorption of light by interstellar dust, in our ZENS analyses, including those discussed in the following sections of this paper, 
we  check  that results  hold when such  galaxies are excluded from the sample. Similarly, we also check that results also hold when excluding galaxies with no GIM2D fits, i.e., galaxies for which we  use the Voronoi gradients and colors (calibrated as above) in our analyses. 

 \section{Variations in median satellite/central galaxy stellar mass across environmental bins}\label{vars}

Given $(1)$  the main goal of ZENS, i.e.,  to study differential behaviors in similar galaxy populations as a function of several environmental diagnostics,  and $(2)$, the known sensitivity of galaxy properties to galaxy stellar mass, it is essential to keep under control, in any given galaxy sample,  possible variations of galaxy mass distributions in the different environmental bins, as any such variation may induce spurious  environmental effects. Consequently, before proceeding with our   environmental analyses, we  investigated whether  the stellar mass distributions of ZENS galaxies of 
different morphological or spectral types vary with the environmental diagnostics derived in Paper I.  

\begin{figure*}[htp]
\begin{center}
 \includegraphics[width=130mm]{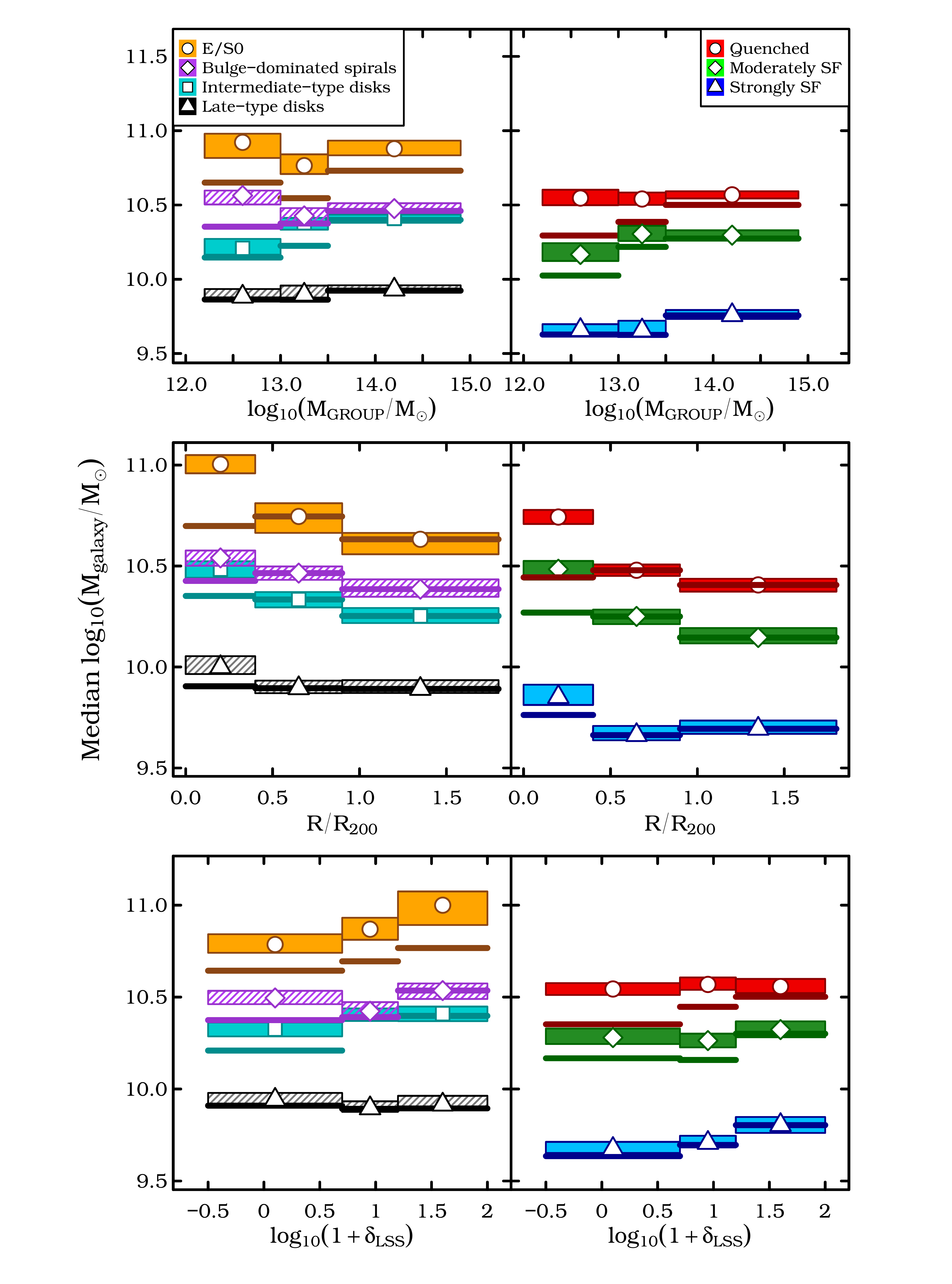} 
\end{center}
\caption{\label{fig:MassGal_MassGrp} Median galaxy stellar mass as a function of environment for galaxies of different morphological (left) and spectral types (right; see legends for details). 
 From top to bottom the considered environments are  group halo mass,   distance from the
 group center and  large scale density.
  Median values are calculated above the mass 
completeness of each spectral/morphological type.
Error bars, shown with hatched areas, represent the standard deviation of the median masses computed from the 16th and 84th percentiles. Colored boxes with white open symbols indicate samples including {\it all} galaxies, independent of their rank as centrals or satellites. The thick solid lines in matching colors are the results for {\it satellites only} samples.\\
(A color version of this figure is available in the online journal.) }
\end{figure*}

Figure \ref{fig:MassGal_MassGrp} plots, from top to bottom, the median galaxy mass of the  various morphological and spectral types as a function of  group mass $M_\mathrm{GROUP}$, distance from the group center (in units of $R_{200}$) and large scale overdensity $\delta_\mathrm{LSS}$, considering only galaxies above the completeness limit of each type (i.e., the mass complete samples that will be used in our ZENS analyses).  
We initially include {\it all} galaxies of a given type in this analysis, independent of whether they are central  or satellite galaxies.
Overplotted in Figure \ref{fig:MassGal_MassGrp},  with thick lines in matching colors for the various galaxy samples, are the median galaxy masses of the  various morphological and spectral types for the corresponding {\it satellites only} samples. The behavior of central galaxies can be inferred by comparing those of total and satellite-only samples.

The median masses of the different morphological and spectral types for all galaxies are sufficiently constant over the three environments to be confident that our analyses, at a \emph{fixed} galaxy type, will  not be biased
 by   strong  trends between galaxy stellar mass and any of the ZENS environmental parameters. 
 Small variations of galaxy stellar mass with the environments are present for some of the galaxy populations: there is  a modest $\sim 0.1$ dex increase in the stellar mass of moderately and strongly star-forming galaxies  with increasing group mass from  $\sim10^{12.5}\Msol$ to $M\gtrsim10^{14}\Msol$;  for these spectral type, a similar mass trends is observed with increasing LSS (over)density.

For morphologically-split galaxy samples including centrals plus satellites, the intermediate-type disks show a dependence with $M_\mathrm{GROUP}$ increasing their mass by 0.2 from low to high mass groups. A variations of galaxy mass is also detected with $\delta_\mathrm{LSS}$ for  E/S0 galaxies, whose median mass  is larger by 0.2 dex at high LSS densities relative to similar galaxies at low $\delta_\mathrm{LSS}$ values.  There is furthermore an increase of the typical stellar mass of all spectral and morphological types when moving closer to the group centers. This latter trend weakens however  substantially
when  central galaxies are excluded, indicating that the effect is mostly due to the addition of the central galaxies and it is largely absent from the  satellite population (see also \citealt{Biviano_et_al_2002,Pracy_et_al_2005,von_der_Linden_et_al_2010}).
For the other two environments, the trends for satellite galaxies match quite reasonably those discussed above for all galaxies, independent of their satellite/central raking.  

These results  support a scenario in which  galaxy stellar mass are, at fixed morphological or spectral type,  largely insensitive to the environment, and especially to the halo mass.
A  constancy of the galaxy stellar mass of the different galaxy types with environmental density has been reported  by a number of authors. The morphologically-split galaxy mass functions for the COSMOS field given by \citet{Pannella_et_al_2009}  show consistency in over- and underdense regions. 
\citet{Bolzonella_et_al_2010} and \citet{Kovac_et_al_2010}, using the zCOSMOS survey data \citep{Lilly_et_al_2007,Lilly_et_al_2009},  find remarkably similar mass functions for  morphological and spectral  early- and late-type galaxies, across the lowest and highest density quartiles up to redshifts of order $z\sim1$.  Analogous results are obtained by \citet{Peng_et_al_2010} on the local SDSS sample of blue galaxies. 
In the \citet{Peng_et_al_2010} formalism, mass-quenching is the only mass-dependent quenching term (environment-quenching is independent of stellar mass) and it is therefore this process that establishes and controls the mass-function of the surviving star-forming galaxies, and therefore also of the resulting passive galaxies. The observed constancy of the characteristic $M_*$ with redshift, back to z$\sim$3 or higher, sets the required form of mass-quenching.  Furthermore, as discussed in \citet{Peng_et_al_2012} the fact that satellites (over a wide range of halo mass) and central galaxies have globally a very similar $M_*$ indicates that mass-quenching operates in a very similar way in centrals and satellites. 

Note  that some  contrasting evidence has been reported. The mass function of star-forming galaxies measured by \citet{Giodini_et_al_2012}, for example, has an identical shape in the field and in a sample of X-ray selected groups, but shows a mild increase of the characteristic $M_*$ mass in low mass  groups with respect to more massive systems. 
\citealt{Calvi_et_al_2012} also report a change in the morphology-mass relation for binary systems, clusters and field galaxies.

This concludes our detailed presentation of the derivation of the photometric diagnostics for the ZENS galaxies. The measurements presented in the sections above will be used in future ZENS papers for detailed analyses of the stellar population and star formation properties of different galaxy populations as a function of different  environmental parameters.
In the reminder of this paper  we present a first utilization of some of these measurements, and briefly investigate $(1)$  global behavior of colors, color gradients and color dispersions of all galaxies (i.e., centrals plus satellites)  at fixed stellar mass and fixed  Hubble type (from Paper II); and $(2)$ colors, color gradients and color dispersions of {\it disk satellites} in different environments.

\section{Colors, color gradients and color dispersions within galaxies: Dependence on galaxy stellar mass and Hubble type}\label{sec:trendswithmass}

\begin{figure*}[htbp]
\begin{center}
\includegraphics[width=100mm]{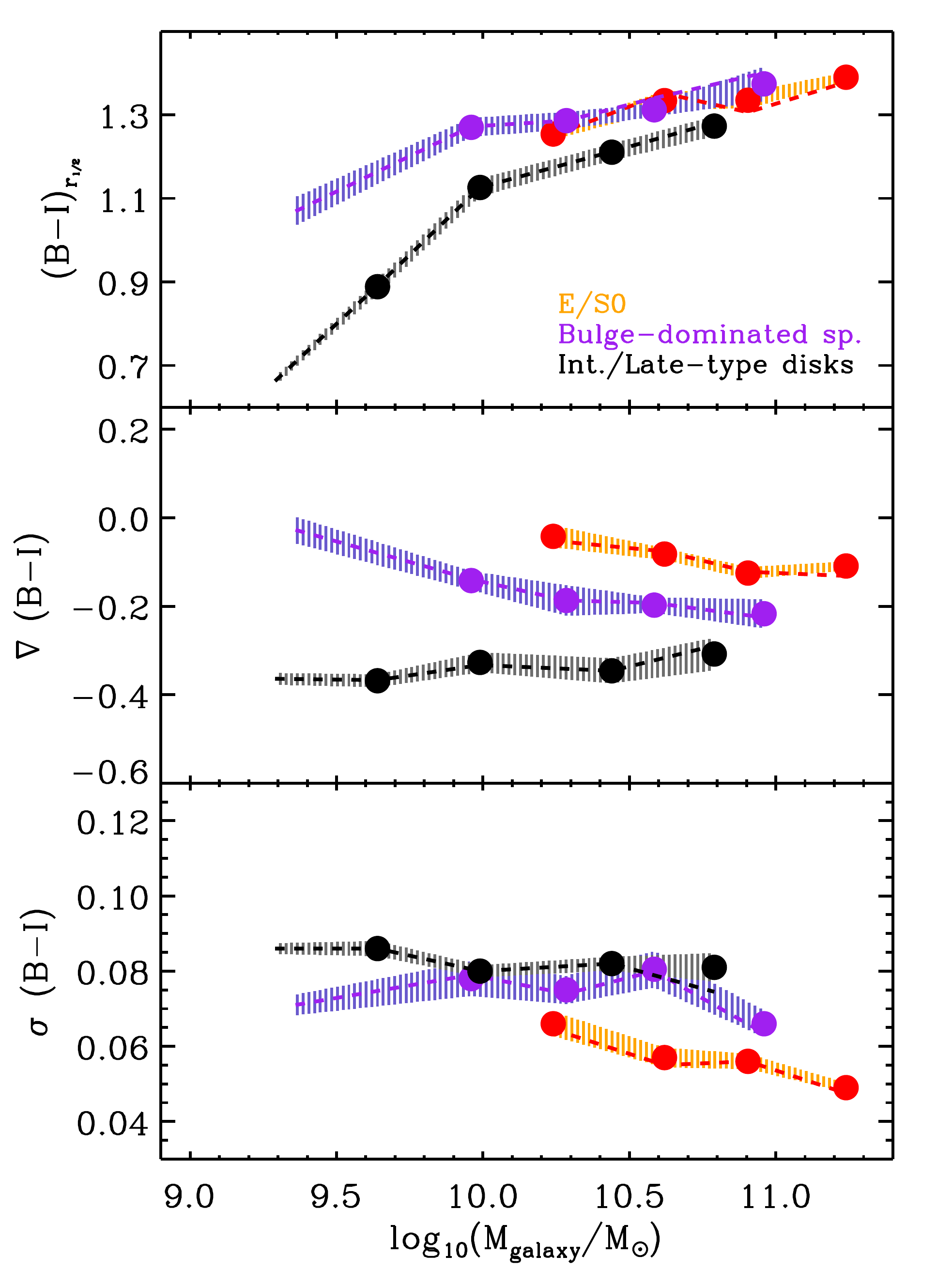}
\end{center}
\caption{\label{fig:ColorGrad} For the ZENS galaxies, from top to bottom, median  colors at the half-light radii, color gradients, and r.m.s. 
scatter around the best-fit  color profiles  as a function of  galaxy stellar mass. Colors 
correspond to the three morphological classes as identified in the legend (and defined in Paper II). 
The shaded areas with filled points show the median relations and the 1$\sigma$ confidence levels for the entire ZENS sample, with no distinction among centrals and satellites. 
The left-most  filled symbol in each curve identifies the stellar mass completeness threshold  for the given galaxy population. 
The dashed lines are  for satellite galaxies only.\\
(A color version of this figure is available in the online journal.) }
\end{figure*}

Figure \ref{fig:ColorGrad} shows, from top to bottom,  the median values of the $(B-I)_{r_{1/2}}$, $\nabla(B-I)$   and $\sigma(B-I)$  distributions as a function of galaxy stellar  mass and at fixed  Hubble type. 
We remind that we  use the GIM2D-based estimates for  $(B-I)_{r_{1/2}}$, $\nabla(B-I)$, and in the following drop the subscript ``GIM2D" for simplicity of notation. 
The colors at the half-light radii show the expected trends: they get redder with increasing galaxy mass and progressively earlier Hubble types. Also not surprising  is the  strong dependence of the color gradients on the morphological type:
E/S0 galaxies show the flattest profiles, an indication that these galaxies have rather
homogeneous stellar populations from their inner to their outer regions;  
late(r) morphological types show stronger color gradients, mostly as a consequence of redder bulges and segregation of star-forming regions in the outer disk regions. 

Within each Hubble type, color gradients are however rather insensitive to stellar mass variations,  and are typically 
$\nabla(B-I)\sim-0.3$ for intermediate- and late-type disks, and $\nabla(B-I)\sim-0.1/-0.2$ for earlier morphological types.
Also the  color dispersion around the average color profile, $\sigma(B-I)$, is  largely independent of galaxy mass for intermediate-type  and late-type disk galaxies. 
 E/S0 galaxies instead show  a net increase in  color scatter around the average color gradient with decreasing galaxy stellar mass.  
We note  that, in the mass bins $M\gtrsim10^{10.5}\Msol$, about half of the disk-dominated galaxies are also central galaxies in our sample. When the latter are  excluded from the analysis, there is a hint (at the $2\sigma$ level between $\sim 10^{9.5} \Msol$ and $M\gtrsim10^{10.5}\Msol$) for a modest  decline in  the color dispersion with increasing galaxy stellar mass also in this morphological type; this is shown by the dashed lines in Figure \ref{fig:ColorGrad}. 
The inclusion of central galaxies has instead no significant impact for the earlier morphological types.
  
\begin{figure}
\begin{center}
\includegraphics[width=70mm,angle=90]{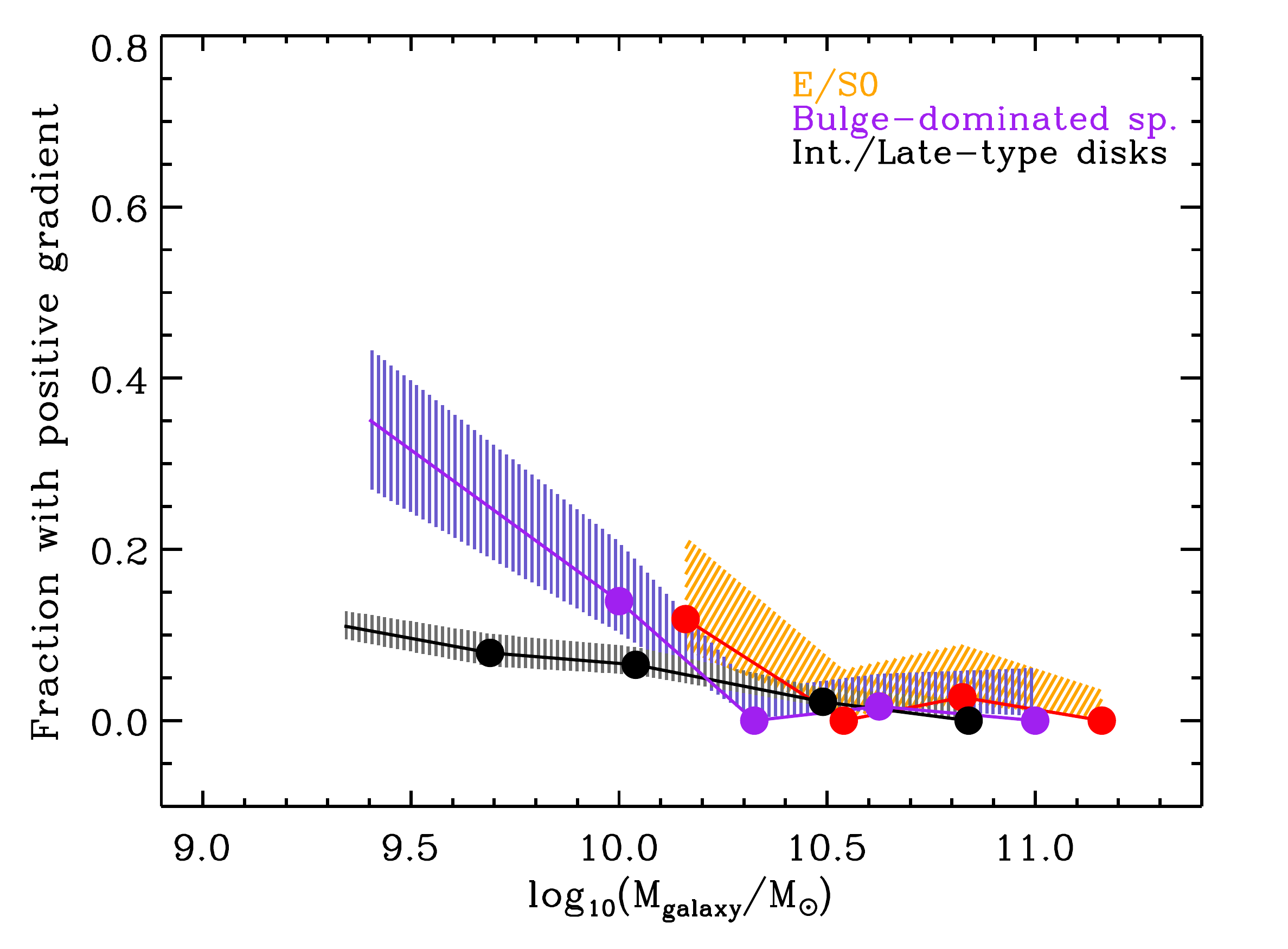}
\end{center}
\caption{\label{fig:FracPositive} Fraction of galaxies with positive (``inverted") color gradients, i.e., with  blue nuclei and red outskirts,  as a function of galaxy stellar mass at fixed  morphological type. Shaded areas show the $1\sigma$ confidence levels on the fractions, derived using the beta distribution quantile technique \citep[see][]{Cameron_2011}. The fractions are corrected for 2dFGRS spectroscopic incompleteness.  The left-most filled symbol in each curve identifies the stellar mass completeness threshold  for the given galaxy population. \\
(A color version of this figure is available in the online journal.) }
\end{figure}

Also noticeable is that a non negligible fraction of galaxies have   positive color gradients, i.e., bluer  centers relative to redder  outer regions.
Specifically, we divide the sample into three bins of color gradients, i.e., 
$\nabla (B-I)<-0.1$ (negative/normal color gradients),  $-0.1<\nabla (B-I)<0.1$ (flat color gradients) and  $\nabla (B-I)>0.1$ (positive/inverted color gradients), respectively. 
Figure \ref{fig:FracPositive} shows the frequency of galaxies with positive 
gradient as a function of galaxy stellar mass, again at fixed morphological type as before.
The fraction of galaxies with inverted color gradients  increases with decreasing galaxy stellar mass:  above $10^{10.5}\Msol$ virtually no galaxies have inverted color gradients, while at $10^{9.5-10}\Msol$,  between $10\%$ and $20\%$ of  galaxies have such inverted color profiles, with the higher fraction interestingly detected for earlier morphological types (in particular, bulge-dominated spirals, which are well sampled down to $\sim10^{9.8} M_\odot$ in our sample).

We explore the  spectral types  of  galaxies with $\nabla (B-I)>0.1$ which are found at all  group-halo masses in  the ZENS sample.
In absolute terms the number of elliptical and S0 galaxies with inverted color profiles is negligible, thus we focus in the following on the  bulge-dominated spirals and intermediate/late-type disks; furthermore, as blue core galaxies are rare at  high galaxy masses   (see Figure \ref{fig:FracPositive}), we consider only the $10^{9.5-10.3}\Msol$  mass interval. We find that 70\% of bulge-dominated spirals with inverted profiles are classified as either moderately or strongly star-forming (for a median specific star-formation rate $<$log(sSFR/yr$^{-1}$)$>$=-10.67). While we should keep  in mind the caveat  that quenched and intermediate spectral type galaxies are somewhat incomplete in ZENS at  masses $\lesssim10^{10}\Msol$, these values are meaningful in a relative sense, i.e., in  comparison with galaxies of similar mass and morphological class and ``normal" color gradients; these have a star-forming fraction of only 24\% and   $<$log(sSFR/yr$^{-1}$)$>$=-12.22).
 This indicates  an enhancement of  SFR in  bulge-dominated galaxies with blue cores. In contrast,  late-type disks  with $\nabla (B-I)>0.1$ or $\nabla (B-I)<0.1$ have  overall similar star-forming fractions and SFRs. Other authors have already commented  that tidal perturbations and/or   ram pressure removal of the outer gas reservoir in galaxies in groups may  induce an enhancement of  star-formation in the galaxy cores \citep[e.g.][]{Koopmann_et_al_1998,Moss_Whittle_2000,Bartholomew_et_al_2001,Rose_et_al_2001}. We defer to a subsequent paper a more detailed  investigation of these inverted-color-gradient galaxies, in particular in relation to internal color differences between satellite galaxies, and between merging/interacting galaxies and their non-interacting counterparts.

\section{Colors, color gradients and color dispersion of disk satellite galaxies in different environments} \label{sec:results}

We finally investigate, for the  population  of \emph{disk satellite galaxies}, how  the colors, color gradients and color dispersion around these gradients depend, at constant stellar mass, on  the total mass of the host group halo $M_\mathrm{GROUP}$, the distance of the satellites from the centers of their host groups ($R/R_{200}$) and the large scale (over)density field $\delta_\mathrm{LSS}$.  In ZENS, disk morphologies account for 80\%  of the satellite population at $M< 10^{10} \Msol$ and 53\% at  $M\geqslant10^{10}\Msol$, providing a statistical sample that is large enough for this straightforward environmental  analysis  over about two dex in stellar mass. We postpone to a follow-up ZENS study the comparative analysis of centrals and satellites as well as the inclusion of elliptical and irregular morphologies, as both require a more elaborated discussion concerning  biases  toward one or another of these galaxy populations in both the stellar mass and halo mass ranges of our survey.

We first remark (without showing in a figure) that  both  bulge-dominated and disk-dominated satellites  have {\it nuclear $(B-I)$ colors}, i.e,   colors measured from the PSF-corrected profiles at 0.1$r_{1/2}$, which are fairly insensitive to  any of the environmental parameters:  all disk satellite populations under investigation (i.e., different bulge-to-disk ratios and different galaxy stellar masses) have similar central colors, ranging from $\sim$1.2  to $\sim1.4$,  which become redder with increasing galaxy mass, but with environmental variations at constant mass that are smaller than $\sim 0.05$ magnitudes.    

In Figure \ref{fig:ColorGradGrp} we show, from top to bottom,  median values for the $(B-I)$ colors (integrated within the Petrosian aperture defined in Section \ref{sec:Phot}),  color gradients $\nabla(B-I)$,  and color dispersions $\sigma(B-I)$,  as a function of galaxy stellar mass for the satellite galaxies.  Red is used  for the  broad morphological bin of   {\it bulge-dominated} galaxies\footnote{The broad bin of bulge-dominated galaxies includes disk galaxies that are classified as S0  or bulge-dominated spirals in Paper II, to which we refer for details on the ZENS morphological classification.}; blue indicates 
 the broad morphological bin of  {\it disk-dominated} galaxies\footnote{The broad bin of disk-dominated galaxies  includes the  two narrower morphological bins of intermediate-type and late-type disks defined in Paper II.}. 

All  median values were calculated in three bins of stellar mass using a  box of width 0.4 dex  for the disk-dominated galaxies and, considering the smaller size of the sample, in two  mass bins below 
and above $10^{10.5}\Msol$ for the bulge-dominated galaxies.  In all bins in each panel,  the used galaxy samples are  above the mass completeness limits of the relevant morphological types, apart from a very marginal incompleteness in the lowest mass bin for the disk-dominated galaxies, as the intermediate-type disks (which are summed together to the late-type disks to build this broader morphological late-type bin) are formally complete only above a slightly higher stellar mass bin; see Section \ref{sec:MassCompleteness}. We checked however that this small effect does not impact our results, as also discussed below. 

 In Figure \ref{fig:ColorGradGrp}, filled and empty symbols show respectively ``dense" and ``sparse" environments, i.e., from left to right, group masses $M_\mathrm{GROUP}$ above and below $10^{13.5}\Msol$  (left panels), group-centric distances larger and smaller than $R=0.6R_{200}$ (central panels), and  LSS overdensities  above and below $\log_{10}(1+\delta_\mathrm{LSS})=0.65$, respectively. Note that the $\delta_\mathrm{LSS}$ analysis includes only groups with $M_\mathrm{GROUP}<10^{13.5}\Msol$, to minimize the degeneracy between group mass and LSS density: as discussed in Paper I, while  by definition  massive groups are located in high $\delta_\mathrm{LSS}$ regions,  low mass group cover a large range of LSS density regimes, thus enabling the disentanglement of effects due to one or the other of these two environmental parameters.
  All ZENS groups were otherwise used in this study, to increase the statistics, including the  groups defined as unrelaxed in Paper I (see also Section \ref{sec:surveyData});  we checked  that, with slightly larger error bars, consistent results hold when excluding the  unrelaxed groups from the analysis. 
However, following the same approach as in Paper II,  we limit the sample to galaxies with $R\leqslant1.2R_{200}$ to minimize  contamination from interlopers and/or small sub-haloes at large radii  (see Paper I for a discussion).
 
For clarity,  we split below the discussion on the environmental trends seen in   Figure  \ref{fig:ColorGradGrp}  separately for the bulge- and disk-dominated satellite populations.

\subsection{Environmental impact on bulge-dominated satellites}

\begin{figure*}[htbp]
\begin{center}
\includegraphics[width=160mm]{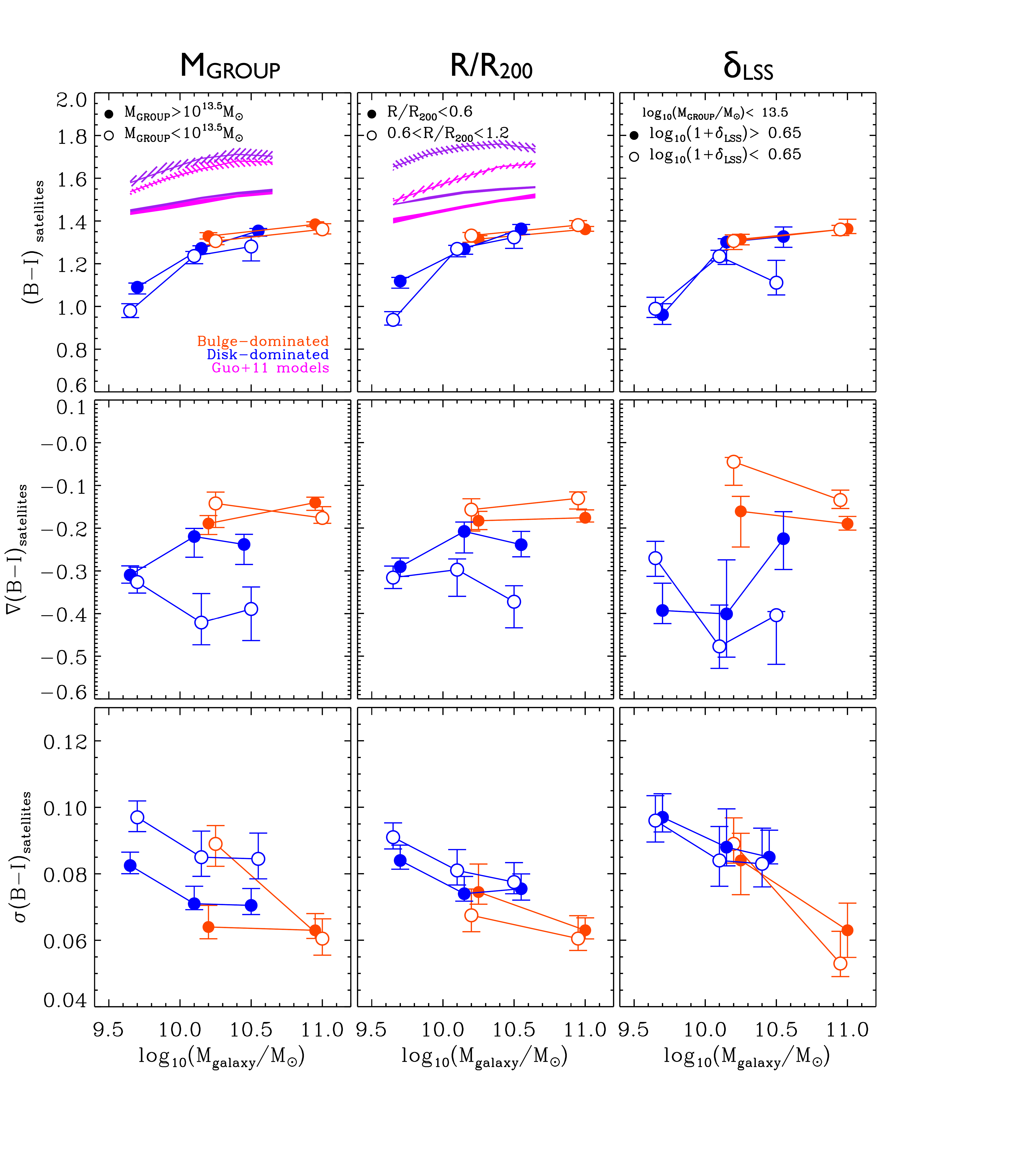}
\end{center}
\caption{\label{fig:ColorGradGrp}   $(B-I)$ color (top panels), color gradients (central panels) and
 rms scatter around the best-fit  color profiles (bottom panels) as a function of the galaxy mass for \emph{disk satellites}   in the ZENS groups. Red  and blue identify respectively bulge-dominated satellites and disk-dominated satellites.  From left to right, the galaxy sample is split in two environmental bins of, respectively group mass  $M_\mathrm{GROUP}$, group-centric distance   $R/R_{200}$,  and  LSS overdensity  $\delta_\mathrm{LSS}$.   Error bars correspond to the error on the median derived from the
 16th and 84th percentiles. In magenta and purple we show the predictions of the \citet{Guo_et_al_2011} semi-analytical model for the  sparser  (low group mass and high group-centric distances) and  denser  (high group mass and low group-centric distances)  environments, respectively;   dashed and solid curves are for models with and without dust reddening.
Strongly star-forming galaxies with red colors (i.e.,  dust-reddened galaxies, see Section \ref{sec:color_color_cut} and Figure \ref{fig:color-color}) are excluded from this analysis; the effects of their inclusion are in any case minimal. Trends with the $\delta_\mathrm{LSS}$ field  are investigated  restricting the sample to groups with  $M_\mathrm{GROUP}<10^{13.5}\Msol$,  to avoid spurious effects with increasing $\delta_\mathrm{LSS}$ due to the degeneracy between high $\delta_\mathrm{LSS}$ values  and high group mass.
(A color version of this figure is available in the online journal.) }
\end{figure*}

The most evident environmental effect  we observe in Figure \ref{fig:ColorGradGrp}  for  bulge-dominated satellites is the fact that they show systematically shallower color gradients at low (i.e., in the group outskirts and at low $\delta_\mathrm{LSS}$) relative to high  environmental densities (i.e., inner group regions and at high $\delta_\mathrm{LSS}$).
This effect is seen in our data at a  3-$\sigma$ level  in the bin at high galaxy stellar mass centered around $10^{11} M_\odot$ for $\delta_\mathrm{LSS}$. 
 As indeed evident from  Figure  \ref{fig:ColorGradGrp}, however, the significance of the effect depends, for each environment, on the specific galaxy mass scale and, in some of the mass bins, it goes down to the $\lesssim2$-$\sigma$ level; still, its systematic nature makes it worth highlighting and exploring it further. 
The global $(B-I)$ colors of bulge-dominated satellites  are in contrast largely independent of environment. 
Likewise no clear environmental influence is seen for the dispersion around the color profile.

Finally, and not expectedly, in any environment,  at the galaxy mass scales of our study, bulge-dominated  satellites  are redder,  have shallower color gradients  and lower color dispersions relative to  disk-dominated satellites.

\subsection{Environmental impact on disk-dominated satellites}

The colors  of {\it disk-dominated satellites} are more significantly affected by the ``local" environment of the host group.  
At low galaxy stellar masses, $\sim 10^{9.5} M_\odot$,  
the $(B-I)$ colors of disk-dominated satellites are $\sim 0.2$ magnitudes redder in the centers of galaxy groups ($R<0.6R_{200}$), relative to similar satellites in the outer regions of the group potentials. 
 This difference is significant at the  95\% level as quantified by Kolmogorov-Smirnov test.  At the same mass scale disk-dominated galaxies are about 0.1mag bluer in low mass groups. No variation with group-centric distance (or any other environment) is seen in the colors of disk-dominated satellites at higher stellar masses. 

 The median bulge-to-total ratios, integrated above $10^{10} M_\odot$ in the disk-dominated satellite samples at $R<0.6R_{200}$ and $R>0.6R_{200}$, respectively, are 0.33  and  0.29; the median integrated $(B-I)$ colors of, separately,  the disk and bulge components of these two samples, are respectively 1.28 and 1.41 at $R<0.6R_{200}$ and 1.22 and 1.46 at $R>0.6R_{200}$. All these values are thus very consistent with each other, within the uncertainties, indicating that the lack of an environmental effect at these high masses is not explained either by bulge-to-total variations or by variations in bulge and/or disk colors between the samples that are being compared.  We note however that the median bulge half-light radii and disk scale lengths, are respectively
 $1.30_{-0.06}^{+0.10}$ kpc and $2.38_{-0.08}^{+0.13}$ kpc within $0.6R_{200}$, and $1.60_{-0.10}^{+0.19}$ kpc and $3.14_{-0.13}^{+0.20}$ kpc in the outer group regions, indicating that, in particular,  disks of $>10^{10} M_\odot$ disk-dominated satellites in the inner cores of groups are $\approx 20\%$ smaller in size, and thus their bulges are more prominent, i.e.,  less ``embedded within the disks",  than it is the case  at large group-centric distances. This difference is observed at the $>3\sigma$ level; a Kolmogorov-Smirnov test excludes a common parent distribution at the 99\% level. The sum of  the bulge contribution to the integrated colors may thus dilute the environmental effect with $R_{200}$ on the $(B-I)$ color that is   seen at lower masses. We defer  a deeper investigation of  the colors and sizes of  bulge and disk components of ZENS galaxies to a forthcoming publication.
 
The strongest environmental effect on the disk-dominated satellite population is seen on the color gradients, which are, 
at fixed galaxy mass $\geq10^{10} M_\odot$,  shallower  by  $\approx-0.15$ in high-mass groups  and in the dense innermost group regions    relative to less massive groups and to the  outer  group regions. Precisely, $\nabla(B-I)\sim-0.25$  in the more massive groups and in the denser groups regions, and $\nabla(B-I)\sim-0.4$ in the less massive groups and in the outer regions of groups.   Also worth noticing for this population  is a hint for a systematic lower  color scatter in massive groups  relative to lower mass groups (at an integrated statistical significance for $M>10^{9.5}\Msol$ of 2.5$\sigma$); a similar trend, seen at a slightly lower significance in our data,  may also be there with group-centric distance, for a lower color scatter in the inner and denser  regions of groups relative to the less dense outer group regions.
 
In the following we investigate the implications of the strongest environmental effects that we have detected, namely 
the  color reddening of low-mass disk-dominated satellites inside the cores (relative to the outskirts) of groups,   as well as the dependence of their color gradients on both group mass and group-centric distance. In particular we discuss these environmental effects for the disk-dominated satellite population  in the context of stellar population synthesis models, semi-analytic  models of galaxy formation, and the ``continuity equation" approach of \citealt{Peng_et_al_2010,Peng_et_al_2012}. 

 \subsection{Interpreting with stellar population models the environmental differences in  colors and color gradients of disk-dominated satellites}
 
Despite  the well-known  degeneracy of  optical colors towards  age/metallicity/dust and SFHs, the color difference  that we have observed in low-mass disk-dominated satellites  as a function of group-centric distance contains information on the origin of such environmental effect.   In particular,  while  dust\footnote{We remark that we have not attempted any correction for  intrinsic dust absorption within galaxies. In  interpreting colors as stellar population (histories) effects,  we  exclude ZENS galaxies which are mostly affected by internal reddening effects, i.e., galaxies which are strongly star-forming but have red colors  (see Section \ref{sec:color_color_cut}).} and, to a lesser extent, metallicity, do no doubt play a role,  the $(B-I)$ color is quite sensitive to stellar populations age variations in spiral galaxies \citep[e.g.][]{MacArthur_et_al_2004,Carollo_et_al_2007}.   We thus 
 assumed that the $(B-I)$ colors and color gradients that we have measured are mostly determined by the age of the underlying stellar populations. Despite the substantial uncertainty introduced in the derived age differences by this assumption, the differences provide a useful benchmark to rank stellar population properties, in a relative manner, between and within galaxies. In particular, since we will be comparing galaxy populations of similar Hubble type and stellar mass, the assumption that they have similar dust content and/or SFHs is less uncomfortable than when comparing very different galaxy populations. 

In constructing stellar population models for the comparison with the data, we  constrained  the stellar metallicity  using  the relation between galaxy stellar mass and stellar metallicity of 
 \citet{Gallazzi_et_al_2005}, which, at the  galaxy masses of our ZENS study, gives     $Z_{\log M\in [9.5-10]}=0.004-0.007$,  $Z_{\log M\in[10.-10.5]}=0.01-0.02$ and  $Z_{\log M>10.5}=0.02-0.03$.
  To model the observed colors we  used  BC03 templates with the above metallicities, a  \citet{Chabrier_2003} IMF  and the same SFHs 
 as in Table \ref{tab:ZEBRA}; note that the reddest ($B-I$) colors observed in Figure \ref{fig:ColorGradGrp} are  obtained only for models with  $\tau<4$ Gyr. 
 We also assessed the impact of different galaxy template models by comparing the BC03-based results with the equivalent estimates based on the 2007 updated version of the original \citet{Bruzual_Charlot_2003} libraries (CB07 hereafter, see  \citet{Bruzual_2007}).   
 
The observed color difference between $\sim10^{9.5} \Msol$  disk-dominated satellites within and outside $0.6R_{200}$, respectively, is consistent with a median $\sim2$ Gyr age difference using either the CB07 and  the BC03 templates, in the sense of ``inner-group" disk-dominated satellites being older than ``outer-group" disk-dominated satellites. As a consequence of the inclusion of the thermally pulsating asymptotic giant branch stars, the CB07 models   reach redder colors at shorter times relative to the BC03 models, thus the spread around the median value is larger for the former   ($\sim 2$ Gyr) than for latter ($\sim0.7$ Gyr). 

An alternative way to interpret the observed color difference between $\sim10^{9.5} \Msol$ disk-dominated satellites in the inner and outer group regions is in terms of a variation of the time scale over which star-formation occurs, i.e., in terms of  $\tau$ variations within a the family of  exponentially-declining SFHs (after fixing the start of star formation activity to the same redshift, $z_f,$ for all galaxies).   Using the  mass-metallicity relation above, $z_f$ set to 10 Gyr ago and the BC03 models,  the redder colors of these low-mass disk-dominated satellites in the inner-group regions correspond to $\approx$2 Gyr shorter  $\tau$ values relative to similar satellites in the outer -group regions (i.e., $\tau_{R>0.6 R_{200}}\sim6$Gyr vs.\, $\tau_{R<0.6 R_{200}}\sim3-4$Gyr). We note that the CB07 models would give about twice as large $\tau$ differences between inner and outer groups regions. 

With a typical galaxy crossing time within a ZENS group 
$t_{cr}=R_{200}/v\sim 10^9\mathrm{yr}\left( \frac{R_{200}}{\mathrm{Mpc}} \right)\left(\frac{\sigma_{los}}{1000 \mathrm{km} \mathrm{s}^{-1}} \right)^{-1}\sim1-3$ Gyr  -- where $R_{200}=\sqrt{3}\sigma_{los}/(10H(z))$, $H(z)$ is the expansion rate  of the Universe at the median redshift $z=0.055$ of the ZENS groups,   and $\sigma_{los}$, i.e., the line-of-sight velocity dispersion of the group,  is typically $\sim 200-400$ km s$^{-1}$ -- the estimated $\tau$ differences between low-mass disk-dominated satellites in the outer and inner group regions are comparable with the timescales of  ``migration" of these galaxies across their host  groups.
This interpretation   thus  supports  a picture in which disk-dominated galaxies that plunge into galaxy groups, and become satellites  within these potentials, see their star formation quenched, on timescales $\tau_\mathrm{quench}\sim2$ Gyr, by physical processes that act on the satellites as they migrate towards the cores of the groups. This is in agreement  with several other studies (see e.g., \citealt{Rasmussen_et_al_2012} and references therein).

Following a similar approach, the  color  gradients of disk-dominated satellites imply differences between the inner and outer galactic regions of:  $(i)$  $\Delta\tau \sim$ 1.5 Gyr and $\sim$3 Gyr in the gas-consumption/star-formation  time scales  or  $(ii)$  $\Delta$Age$\sim$3 Gyr and $\sim$4 Gyr in the typical age, for galaxies in the cores and outskirts of groups, respectively. In light of their similar nuclear colors (and thus stellar populations), independent of any environment,  these comparatively different age variations between inner and outer galactic regions of inner- and outer-group satellites may well  indicate quenching  of the  outer-disk components of satellites in the cores of groups.

Together, the above considerations on the colors and  change in   strength of the color gradients  of disk-dominated satellites lead to a picture where  ``satellite-quenching" preferentially acts on the outer parts of the disks, and occurs on  extended timescales of order $\sim 2$ Gyr as satellites journey towards the centers of their group halos. 
 This is consistent with the picture discussed in Section \ref{sec:trendswithmass}, in which  gas is removed from the outer regions of satellite galaxies in groups causing an enhancement of the central star-formation \citep{Koopmann_et_al_1998,Moss_Whittle_2000}.
 
\subsection{Comparisons with a state-of-the-art Semi-Analytic model}\label{sec:modelsComp}

Semi-analytic models (SAMs) offer an alternative avenue  to understand the color difference that we have observed between disk-dominated satellites in the cores and outskirts of groups. Several models have been developed; our scope here is however not to perform a complete comparison to identify the model which best fits the data, but, rather, to learn some lessons  either from similarities or from differences between observations and a state-of-the-art SAM, which incorporates recipes to ameliorate known deficiencies in previous model generations. 
We thus focus our data-model comparison entirely on the Guo et al. 2011 model (hereafter G11) which, relative to previous generations of SAMs, substantially improves the description of environmental effects leading to the evolution of satellite galaxies within group halos. In particular, the G11 model implements a relatively less abrupt stripping of gas from satellites and a recipe for disruption of satellites that are stripped of their dark matter halo.

The G11 model was run on the Millennium simulation \citep{Springel_et_al_2005}, on a box size of 685 Mpc, and has a resolution in galaxy stellar mass of a few $10^{9}\Msol$, well-suited for a comparison with the ZENS galaxies. G11 used a Chabrier IMF and included  a treatment for dust  to build the model galaxies. We created 10 random halo samples from all the haloes in the G11 volume, matching the  mass function of ZENS. From these halo samples we selected satellite galaxies, which were classified into two broad morphological types according to a stellar-mass  bulge-to-total ratio threshold of B/T=0.5, i.e., $B/T\geq0.5$ for bulge-dominated galaxies, and $B/T<0.5$ for disk dominated galaxies. We note that a similar B/T numerical  threshold is used to classify our ZENS sample in similar types, however in $I$-band light rather than stellar mass; thus, there might be a  slight   difference in morphological separation when comparing  models and ZENS data, which we will keep in mind.
Since resolved color gradients are not available for the model, we limit the comparison  with the ZENS galaxies to the  integrated $(B-I)$ colors.  Also,  the default G11  colors are provided in the SDSS $ugriz$ system;  hence a conversion to $(B-I)$ was obtained using Table 2 of \citet{Blanton_Roweis_2007}.

In virtually all semi-analytic computations, galaxies are found to be too red compared to real galaxies, a red-color  excess that is interpreted  as arising from the combination of a too fast gas consumption through 
star formation in the early phases of  galaxy evolution relative to  star formation histories of real galaxies, coupled with an excessive  stripping  of the hot gas when the galaxies are accreted onto a group halo \citep[e.g.][]{Weinmann_et_al_2006,Font_et_al_2008,Kimm_et_al_2009,Fontanot_et_al_2009,Weinmann_et_al_2010}. 
As mentioned above, the G11 model includes an improved  treatment of  stripping of hot gas experienced by galaxies once they become satellites of larger haloes.  As in other SAMs, in G11 the hot-gas stripping happens after the galaxy enters the virial radius of the group. 
 However, in contrast with earlier generations of models, in which the whole hot gas reservoir is instantaneously removed, in the G11 model only the gas  outside a characteristic radius $R_{stripping}$ within the satellite's own subhalo is removed (with $R_{stripping}$ set equal to the smallest between the tidal radius, in the group potential, of the dark matter component of the satellite's subhalo, and the radius at which ram pressure stripping within the group potential equals the internal galaxy pressure).
Satellites can thus retain a  reservoir of hot gas in the G11 model, which keeps sustaining star formation activity after they enter the group virial radius. This helps reducing  the (progressively larger
at progressively  lower masses) excess of  red quenched galaxies  that is typically observed with respect to the observations in  semi-analytic computations. 

Galaxies in the  G11 model remain nevertheless too red relative to real galaxies, as shown in the upper panels of Figure \ref{fig:ColorGradGrp}, where the model colors are shown in comparison with our ZENS data. 
Specifically, magenta and purple lines are   the model median  colors for disk-dominated satellites outside and  within $0.6R_{200}$, respectively; dashed and solid curves describe models with and without dust reddening.
The hatched lines are  equivalent to the $1\sigma$ dispersion around the median, obtained from the 16th and 84th percentiles.

It is a success of the model, and a valuable insight for understanding our results, that the G11 model colors of disk-dominated satellites show  similar environmental trends  with group mass  and 
group-centric distance as similar  ZENS satellites. In the model, at constant galaxy mass, these satellites show colors that are almost independent of group mass;  they become however substantially redder towards the group centers relative to the groups outskirts. 
Although not explicitly discussed in G11, this is likely due to a feature that seems to be
common to several SAMs (including some earlier versions of the G11 model, on which the latter is based; e.g.,
\citealt{DeLucia_Blaizot_2007}), as well as to dark-matter-only simulations \citep{Gao_et_al_2004}: 
satellites that inhabit the group centers  have entered the group virial radius at earlier
times $t_{infall}$ relative to satellites which are found in the outskirts of the groups. This ``time delay" is typically of order $\Delta t_\mathrm{infall} \sim 2-4 Gyr$ (at least in the \citealt{DeLucia_Blaizot_2007}  model, as analyzed by \citealt{DeLucia_et_al_2012});  this implies a relatively larger gas consumption and/or loss in the inner-group relative to the outer-group satellites. Similar trends for satellites' $t_\mathrm{infall}$ as a function of group-centric distance are also found in hydrodynamical   simulations reproducing $z\sim 0$, $\sim 10^{13} M_\odot$ galaxy groups within a $\Lambda$CDM cosmology \citep{Feldmann_et_al_2011}.

The G11 model furthermore also reproduces the fact that the color-reddening of 
disk-dominated satellites in inner- (relative to outer-) group regions is  marginally larger at lower (relative to higher) galaxy stellar masses. 
In the models this can be  explained with a $\Delta t_{infall}$ that decreases with increasing mass
(from $\sim 4$ Gyr at $\sim 10^9 M_\odot$ to $\sim 2$ Gyr at $\sim 10^{11} M_\odot$ in the \citealt{DeLucia_et_al_2012} analysis), although 
a more efficient dynamical friction for high mass galaxies quickly moves  
these  toward the inner group regions. In other words, low(er) mass galaxies are subject to
the effects of ram pressure stripping for a factor  $> 2$  longer time periods than the more massive satellites.

There are however also  important discrepancies between the G11  model and the ZENS data, which are as illuminating as the successes highlighted above.

\begin{enumerate}

\item  First, as already stressed above,  the  G11 model  predicts colors that are still significantly  
redder than real satellite galaxies at any galaxy mass of our study.  In the models with no dust reddening and at $M<10^{10}\Msol$, the discrepancy is of $\approx 0.4-0.5$ mag, and grows to $\approx0.7$ mag for the models with dust.  A strictly quantitative comparison between data and  models  is difficult due to the slightly different morphological selection criteria, and the intervening conversion factions for the colors discussed above. The size of the discrepancy is however substantial; furthermore,  a similar discrepancy between model and SDSS data is already highlighted for the $u-i$ color in Figure 12 of  Guo e al.\, especially at stellar masses $\sim10^{9.5}\Msol$. We are thus confident of a genuine shift in color between the model and the data, in the direction of the model satellites being indeed redder than  the ZENS satellites. This residual reddening of the G11 model relative to the ZENS data indicates that further    revision of the model is required, most likely in the past  SFHs of present-day galaxies, so as to avoid gas over-consumption at early times while at the same time preventing excessive residual gas in galaxies at these mass scales at redshift $z=0$.

We note that \citet{DeLucia_et_al_2012} have recently stressed the importance of properly relating galaxies properties to their past environmental history, i.e. to the evolution of the host halo environment of galaxies with different masses and redshifts of infall into the final halos. 
These authors  infer the timescales for environmental quenching of satellite galaxies using mergers trees to link the observed passive  fraction with the fraction  of galaxies of a given mass that spend a given amount of time in the group/cluster environment. Predictions for  colors of galaxies in this model are not available. The  employed merger trees  are however similar to those adopted in the G11 model; thus, we suspect that, despite this model offers alternative ways to measure the timescale of environmental quenching and to interpret observed trends with environment, it will likely show a similar color discrepancy relative to our data.

\item Furthermore, the model  shows a stronger flattening of the color vs.\, galaxy mass relation than the real data: model colors remain essentially constant over the whole mass range of our analysis, with at most a $\lesssim0.1$ mag reddening towards higher masses, while ZENS satellites  show a reddening of $\sim 0.3$ mag over the same mass range.  This indicates that the modification to the SFHs in model galaxies needs to be a function of galaxy stellar mass.

\item Finally, on quantitative ground, while the amount by which  inner-group satellites are redder than  outer-group satellites
 for low mass $\sim 10^{9.5} M_\odot$ satellites  is similar ($\sim 0.2$ mag) between the observations and the model,
 at higher galaxy masses the latter shows a larger color difference between these two populations than the ZENS data.  Possible interpretations of this discrepancy may be that, in the model,  ram-pressure effects on massive galaxies,  and/or their dynamical friction timescales, are over-estimated  compared with ram pressure and time scales in the Universe. Alternatively, 
massive  disk-dominated  satellites in the real Universe may continue accreting gas also when close to the group centers, in contrast with the model-assumption that no gas accretion on satellites takes place after they enter the virtualized group regions.
\end{enumerate}

\subsection{Interpreting environmental trends in colors and color gradients of disk-dominated satellites within a continuity-equation approach}
 
The \citealt{Peng_et_al_2010} continuity approach to analyzing galaxy evolution has produced an impressively self-consistent picture of the overall evolution of the active and quenched galaxy populations since $z \sim 2$.   Within the framework of this approach, \citealt{Peng_et_al_2012} studied a large sample of groups in the SDSS.  While this methodology has been hitherto applied primarily to the integrated quantities of mass and SFR, rather than morphology, it is nevertheless worth comparing the results of this section, and especially those concerning the overall $(B-I)$ colors of the galaxies (the upper panels in Figure 17), with those in \citet{Peng_et_al_2012}.   

The independence of overall $(B-I)$ color for disk galaxies on group halo mass and the dependence, especially at low masses,  on group-centric distance  are  consistent with the finding by \citet{Peng_et_al_2012} that the red fraction of satellites depends on local density within a group, but not on halo mass (at fixed density).  The increasing impact of the group-centric radius environment with decreasing stellar mass of the galaxies is also consistent with the different characteristics of Peng et al.\,'s ``mass-quenching'' and ``environment-quenching" processes.  The rate of the former depends on the SFR, which is broadly proportional to the mass of star-forming galaxies, whereas the latter is, from the separability of the red fractions, independent of the stellar mass.  Environment quenching thus dominates over mass-quenching at lower stellar masses, and especially so below $10^{10}$M$_{\odot}$.  

The precise morphological mix in the quenched fractions of satellites in different environments will be further discussed in Carollo et al.\ (2013b, in preparation).

\section{Summary and concluding remarks}\label{sec:summary}

We have presented   photometric measurements  for the 1484 galaxies in the  sample of 141 ZENS groups. All measurements presented in this paper are included in the ZENS catalogue published in Paper I. Photometric galaxy parameters include  total $(B-I)$ colors, color gradients, color dispersions, and galaxy stellar  masses and SFRs,  as well as  bulge and disk $(B-I)$ colors and stellar masses, the latter based on the parametric bulge+disk decompositions in $B$ and $I$ presented in Paper II. Voronoi-tesselated two-dimensional $(B-I)$ color maps extending out to $\sim2$ galaxy half-light radii were also obtained.   ZENS galaxies were classified in the three spectral types of quenched, moderately star forming, and strongly  star forming systems, based on a spectral classification scheme using the combination of  spectral absorption/emission features,   broad-band (FUV) NUV-optical colors and SED fits to the available photometric data. The combination of multiple probes of star formation enables an efficient separation of  red, quenched systems from dust-reddened star forming galaxies; at  $\sim10^{10}\Msol$, the latter are found to contribute up to $50\%$ to the $(B-I)$ red-sequence.
 
In the ZENS sample as in others (see, e.g., \citealt{Peletier_et_al_1990,deJong_1996, Jansen_et_al_2000,Taylor_et_al_2005, Gonzalez_et_al_2011, Tortora_et_al_2010}):

\begin{enumerate}
\item the average $(B-I)$ colors at the half-light radius steadily increase with galaxy stellar mass and towards earlier types;  
 \item  the intensity of $(B-I)$ color gradients correlates with morphological type, as defined using the structural parameters and morphological classification of Paper II: E/S0 galaxies have $\nabla (B-I)\sim-0.1$ and bulge-dominated spiral galaxies instead $\nabla (B-I)\sim-0.2$; such color gradients are substantially shallower than those of later-type, disk-dominated  galaxies [$\nabla (B-I)\sim-0.3$]; 
 
 \item most galaxies have negative color gradients (centers redder than outskirts); a non-negligible fraction of galaxies shows however flat or positive color gradients. At $M\sim10^{10} \Msol$,  about $20\%$ of galaxies of all  morphological types have cores that are bluer than the outer galactic regions;
the  fraction of galaxies with inverted color gradients increases with decreasing galaxy stellar mass.
Interestingly, in bulge-dominated  galaxies the inversion of the color profiles is also accompanied by an enhancement of the median sSFR 
with respect to galaxies with the same morphology but normal (non-inverted) profiles; in contrast, positive and negative color gradient in disk-dominated galaxies have comparable sSFRs.  
Evidence for centrally concentrated star-formation in groups and clusters, especially relative to the field population, has been  suggested to be the outcome of tidal perturbations or   ram pressure removal of the outer gas reservoir, leading to an enhancement of the star-formation in the galaxy cores \citep[e.g.][]{Koopmann_et_al_1998,Moss_Whittle_2000,Bartholomew_et_al_2001,Rose_et_al_2001}.  \citet{Bahe_et_al_2013} have recently suggested that ram-pressure stripping may be already at work even outside of the group's virial radius.
\end{enumerate}

We find that, above the level of stellar mass completeness of our sample, the median mass of galaxies of a given morphological type, and especially of disk-dominated satellite galaxies,  does not depend substantially on any environment, including  the halo mass.  This is in agreement with the constancy of the characteristic $M^*$ in the Schechter \citep{Schechter_1976} mass function of satellite galaxies over a wide range of halo masses, which indicates that   mass-quenching  of satellites is independent of halo mass.  

Focusing on {\it disk satellites} only,  we have investigated, at fixed stellar mass and Hubble type, the dependence on  group  mass,   distance from the group center, and the LSS over-density  of  $(B-I)$ colors, color gradients and color dispersions around the mean gradients. Our findings are summarized as follows:

\begin{enumerate}
\item Bulge-dominated satellites have redder  colors, shallower color gradients and  smaller color scatter than disk-dominated satellites of similar masses in similar environments. Our data  suggest  that color gradients of bulge-dominated satellites may be shallower in regions of low environmental density  -- i.e., outer  group regions and low LSS over densities --  compared to the same Hubble type satellites  in the corresponding denser environments.

\item The group environment has  a noticeable effect on both colors and color gradients  of disk-dominated satellites. These satellites  have shallower $(B-I)$ color gradients, by about -0.2 dex,   in the inner   group regions relative to regions at larger group-centric distances, and  at high group masses  relative to lower group masses. 
Furthermore,  at galaxy masses $<10^{10} M_\odot$,  disk-dominated satellites in our sample have  $\sim0.2$ mag redder $(B-I)$ colors  in the inner group regions relative to similar satellites in the outer group regions.
\end{enumerate}

A stellar population analysis applied to the environmental differences in $(B-I)$ colors and color gradients  detected for disk-dominated ZENS satellites indicates  star formation timescales $\tau\sim2$ Gyr shorter, or  $\sim$2 Gyr  older stellar populations, for low-mass disk-dominated satellites in the cores  of groups relative to the group outskirts.  These timescales are comparable with the timescales of  satellite  infall within the inner group regions.

Similar  long $\tau_\mathrm{quench}$ to those inferred above are  found by other independent  studies,  which also estimate of order a couple of  billion years for physical processes within the groups to act   \citep[e.g.][]{Balogh_et_al_2000,Wang_et_al_2007,Weinmann_et_al_2009, von_der_Linden_et_al_2010,McGee_et_al_2011,Kovac_et_al_2010,Feldmann_et_al_2010,Feldmann_et_al_2011,Rasmussen_et_al_2012}. 
 In the semi-analytic study of \citet{DeLucia_et_al_2012}, the observations can be reproduced if  the passivization occurs on even longer timescales of $\sim5-7$ Gyrs. 
 Also numerical simulations of gas stripping in groups and cluster  support a scenario in which galaxies are depleted of their hot gas in a gradual manner while they orbit in the group potential:  as shown by \citealt{McCarthy_et_al_2008}  satellites can retain a substantial fraction of the hot halo and thus sustain some level of star formation up to 10 Gyr after the infall in the group.

We note that  \citet{Wetzel_et_al_2012,Wetzel_et_al_2012b}  report shorter time scales for suppression of star formation relative to this previous work and our own estimate. In particular,  \citet{Wetzel_et_al_2012b}  suggest a  two-phase  satellite-quenching process,  characterized by a $\sim2-4$ Gyr period after a galaxy enters the virialized group potential, in which its star-formation is unaffected by the group environment,  followed by a rapid truncation  of star formation in a short $<$0.8 Gyr event.  These authors  use an {\it indirectly} derived strong redshift evolution of the fraction of quenched satellite galaxies, grounded on the combination of heterogeneous datasets at low and high redshifts. This strong evolution is however at odds with several other  studies that self-consistently sample this evolution within  large, homogenous datasets \citep[e.g.][]{Knobel_et_al_2012,Ilbert_et_al_2013}. We suspect that inputting  in the Wetzel et al.\ analysis   a milder evolution of the fraction of quenched satellites at fixed stellar mass, as directly indicated by the data, could reconcile their estimates for the   quenching timescales with those inferred from the present analysis and the several other quoted works.

Semi-analytical model predictions  also point indeed at  inner-group satellites to have  an $\sim$2 Gyr earlier infall time   -- and consequently an  earlier star formation quenching through loss of the hot  gas reservoir, and gas consumption in star formation  -- than outer-group satellites.  A quantitative comparison shows however that  even the Guo et al.\, (2011) model, which incorporates a progressive effect of ram pressure, still produces too red colors of satellite galaxies at the explored mass scales and Hubble types relative to the observations. A modification of the model is required so as to avoid  excessive gas over-consumption at early times and, simultaneously,  excessive residual gas in galaxies at redshift $z=0$; most likely,  both the SFHs of  galaxies and  the feedback recipe need modifications. Such modifications need to be a function of galaxy stellar mass in order to reproduce the trends observed in the ZENS data.

The mass scales $\lesssim10^{10}\Msol$ at which  disk-dominated satellites     most evidently show a reddening of their $(B-I)$  colors with decreasing distance from the group center are those shown by \citet{Peng_et_al_2010} to be, at $z=0$, mostly affected by environment-quenching. This quenching process has been further shown by \citet{Peng_et_al_2012} to be practically entirely due to  environmental effects on  galaxies as they become satellites of virialized groups:   redder colors in high density environments  are explained through quenching of star formation in the satellite galaxy population, not in central galaxies (i.e., by``satellite-quenching").  The ZENS satellites' analysis that we have presented adds further evidence to this scenario. In particular, our analysis supports that the environment affects  galaxy evolution below a galaxy mass scale of $\approx10^{10.5} M_\odot$  through  physics that occurs within the group halos  \citep[][]{van_den_Bosch_2008,Weinmann_et_al_2009,Guo_et_al_2009,Peng_et_al_2010}. The fact that the strongest variation in the colors of low-mass disk-dominated satellites is observed as a function  of the  location within  groups  points to a physical mechanism  for  the ``satellite-quenching" that either depends    on  the local density rather than the global halo potential, or affects  the gas inflow, outflow and/or content of galaxies  as they sink as satellites in   the potentials of their host  groups.   

\section*{acknowledgments}
A.C., E.C.  and C.R. acknowledge support from the Swiss National Science Fundation.
This publication makes use of data from ESO Large Program 177.A-0680.
This publication makes use of data products from the Two Micron All Sky Survey, which is a joint project of the University of Massachusetts and the Infrared Processing and Analysis Center/California Institute of Technology, funded by the National Aeronautics and Space Administration and the National Science Foundation.
$GALEX$ (\emph{Galaxy Evolution Explorer}) is a NASA Small Explorer, launched in April 2003. We gratefully acknowledge NASA's support for construction, operation, and science analysis for the $GALEX$ mission.


\clearpage


\appendix


 \section{Reliability of the SED fits}\label{sec:MassAccuracy}

In this Appendix we present the tests that we performed  to asses the reliability of the 
photometric datasets  which is used as input for the SED fits, and the robustness of  the resulting mass estimates provided by ZEBRA+.

\subsection{Impact of photometric errors on derived quantities}

\begin{figure*}
\begin{center}
\includegraphics[width=170mm]{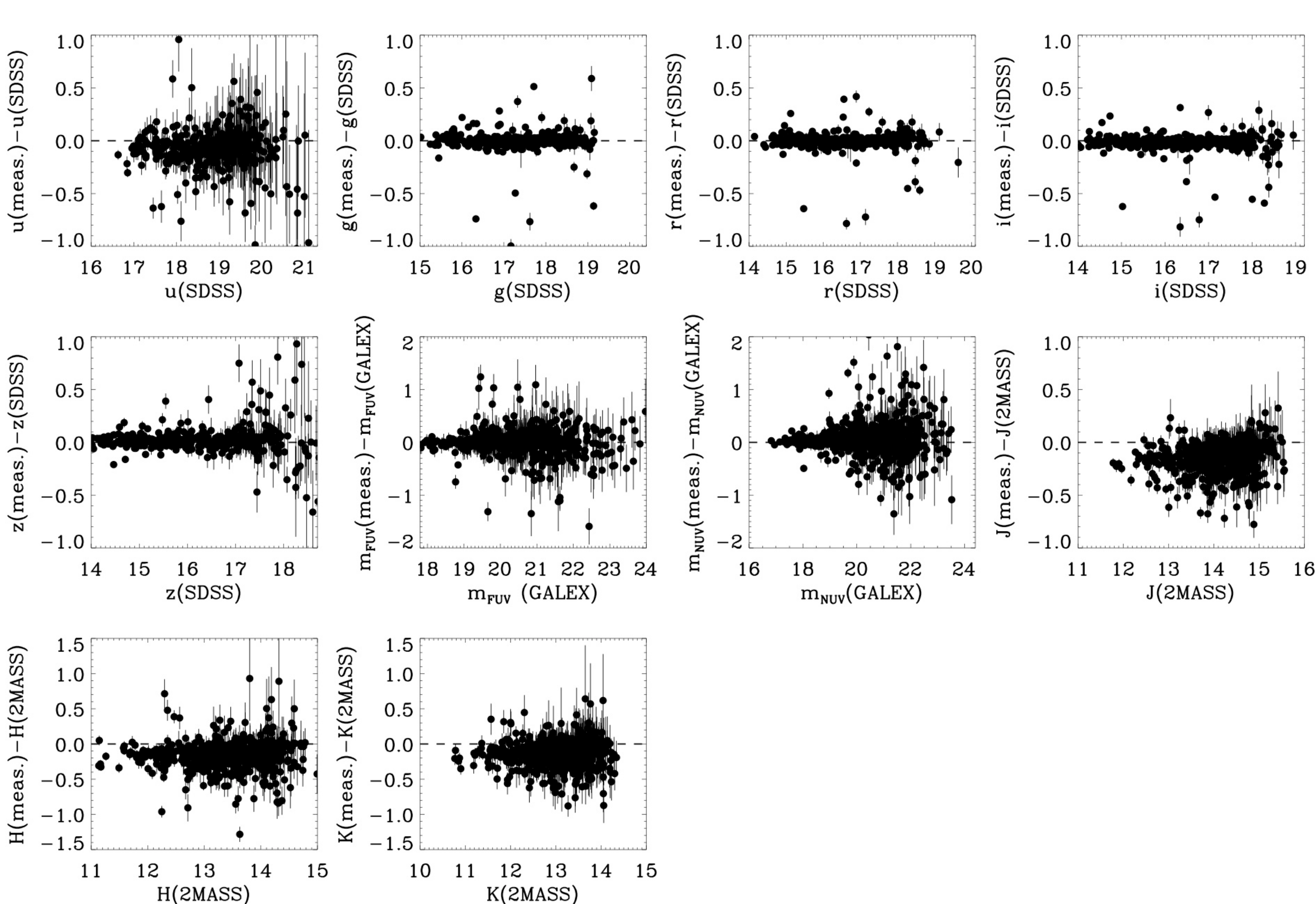}
\end{center}
\caption{\label{fig:PhotometryComp} 
Comparison between the magnitudes that we measured on the public $GALEX$, SDSS and 2MASS images (indicated with ``meas.")   and the magnitudes published by the survey teams (indicated with the names of the various surveys) for the  ZENS galaxies for which such data are available.  The mag($GALEX$) and mag(2MASS) data points refer to Kron \citep{Kron_1980} magnitudes, and the mag(SDSS) data points to Petrosian fluxes; the latter are based on circular apertures. Our own measurements are performed on elliptical apertures with a semi-major axis equal to twice the  largest between the WFI $B-$ and $I-$band Petrosian radii. } 
\end{figure*}

In Figure \ref{fig:PhotometryComp} we compare the magnitudes that we obtained by applying the aperture photometry described in Section \ref{sec:Phot} with the public measurements from the  various surveys databases. 
The agreement is generally good, and deviations are in all cases smaller than 0.1 magnitudes. There are however some systematic effects: 
our SDSS and 2MASS magnitudes are slightly  brighter  
than the published values.  We have tested  that this is partly due to our use of a larger aperture, indicating some flux losses in the published measurements. We also note that SDSS magnitudes are calculated using a circular aperture and hence part of the observed scatter stems from different aperture geometry.  In fact, some of the most highly deviant
points occur for highly inclined galaxies, for which the SDSS circular aperture was observed to miss a large fraction of the flux.  
Note also that a larger scatter in the magnitudes is observed for  the $GALEX$-UV and SDSS $u$/$z$ filters, which are   noise-dominated.

Given the heterogeneous nature and origin of our photometry, the size of the PSF FWHM varies rather strongly across the probed wavelength range: from $\sim 5^{\prime\prime}$ for the $GALEX$ ultra-violet data to sub-arcsecond resolution under the best observing condition for our own $I$-band imaging.
In principle a matching of the images from the different surveys to a common PSF scale would be required to derive consistent color measurements. The use of a relatively large aperture  ($2\times R_p$) in the calculation of the magnitudes, however, effectively reduces PSF related biases.   Indeed we tested on simulated data that degrading the quality of the images with the best PSF to the worst PSF increased rather than decreased the size of the errors on the derived quantities. Furthermore, we also checked on  simulated images that using variable apertures in the different passbands does not improve the determination of stellar masses and other derived quantities, due to two effects, namely the increased noise in the degraded images, and a residual contamination from nearby companions.   As a final  test for the reliability of our derived magnitudes, we smoothed the WFI  $I-$band images -- which have, on average, the best resolution -- with a Gaussian kernel to reach a homogeneous 5 arcseconds PSF. 
We then re-measured on these smoothed images the flux inside the same Petrosian aperture used to derive the stellar masses in Section \ref{sec:Phot} and compared the resulting luminosities with the ones obtained from the unconvolved images. For about three quarters of the sample, the effect of PSF blurring causes flux variations that are within the average photometric error in the $I$ filter;  furthermore,  the difference remains within 0.1 magnitudes for about 90\% of the sample. In the remaining few cases, differences of up to about 0.3 magnitudes were observed, both in the positive and in the negative direction, indicating that both flux losses and increased noise have an impact in this measurement. In these cases, the approach  of deriving  the galaxy fluxes on the original images guarantees higher quality measurements. 
 We are therefore confident that the adopted scheme for measuring galaxy fluxes  delivers stellar masses and related quantities which are robust to PSF variations across the spectral region of the used imaging dataset.

\subsection{Errors on the ZEBRA+ stellar masses and (s)SFRs, and comparisons with the literature}

\begin{figure*}
\begin{center}
\includegraphics[width=70mm,height=160mm,angle=90]{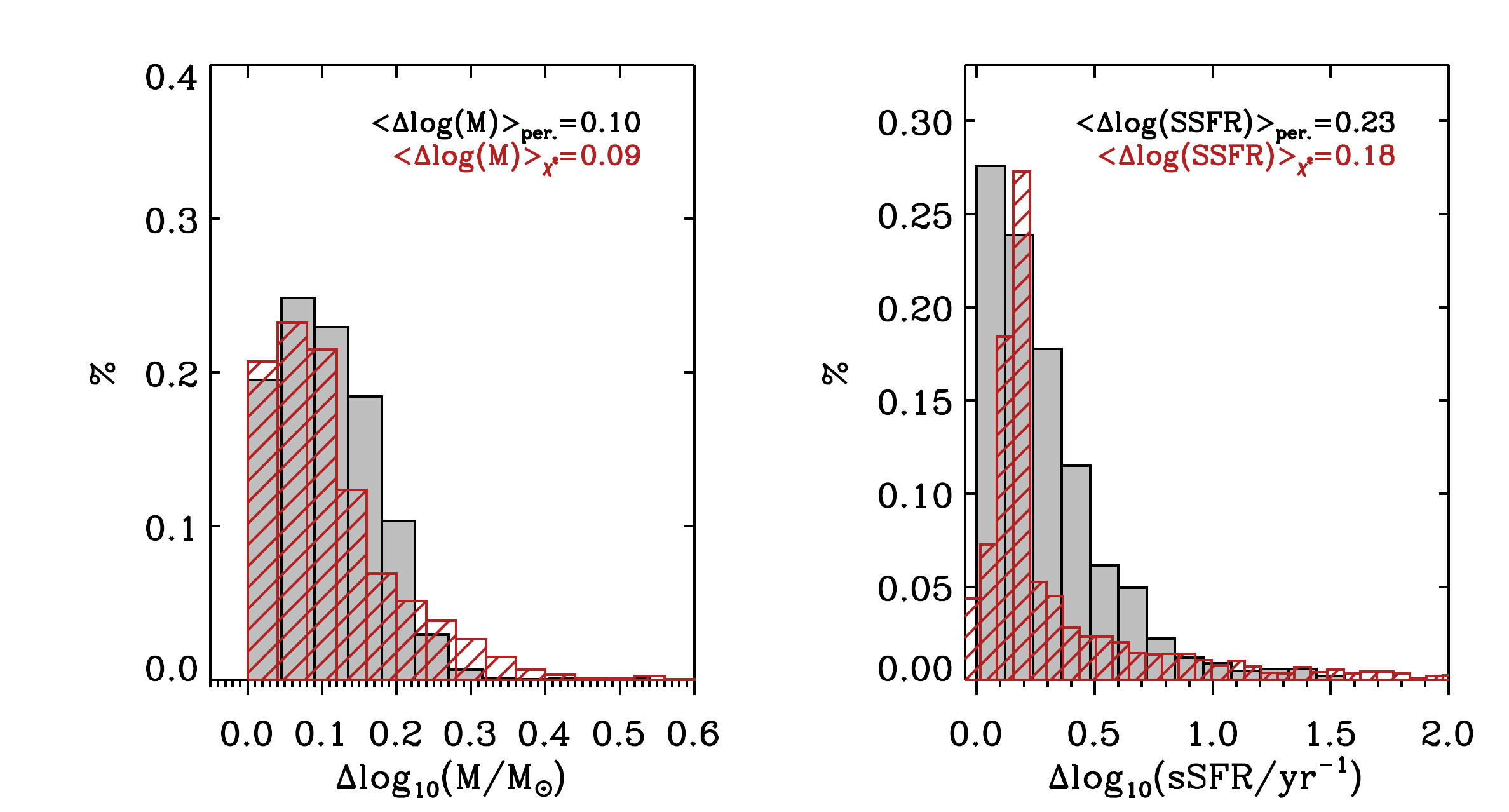}
\end{center}
\caption{\label{fig:ZEBRAerrors}Distributions of the errors for the stellar masses and sSFRs  obtained from the ZEBRA+ SED fits.
The gray histograms are for errors on the parameters calculated from the 16th and 84th percentiles of 
the likelihood distributions of the complete set of templates used in the fit. The black hatched histograms  (red in the online version) show the typical
uncertainties that are obtained by determining the values of the considered parameters that correspond to an increase of $50\%$ in the value of $\chi^2$
with respect to the minimum  $\chi^2$ of the best-fit templates.
 The corresponding median errors are indicated  at the top of each panel.\\
 (A color version of this figure is available in the online journal.)} 
\end{figure*}

To derive an uncertainty on the parameters obtained from the ZEBRA+ fits we utilized the likelihood and $\chi^2$ distributions of the relevant quantities
 (masses, SFRs, sSFRs, etc.)  from the full set of the employed templates.
Specifically,  we derived the errors in two different ways, i.e., $(1)$ using the 16th and 84th percentiles of the global likelihood distributions as our uncertainty measure, and $(2)$ deriving the values of the considered parameters that bracket an increase of $50\%$ in $\chi^2$ with respect to the $\chi^2$ 
 of the best fit template.
We show the calculated errors in Figure \ref{fig:ZEBRAerrors} for galaxy masses and SFRs. 
The median error on the galaxy mass is $\sim 0.1$ dex and $\sim 0.2$ dex for the sSFRs.
As expected, (specific) SFRs are subject to larger uncertainties than the masses, 
hence our careful inspection of the spectral lines and color-color diagrams outlined in Section \ref{sec:ActiveQuiescent} to define the quenched
and star-forming populations substantially improves the purity of these samples relative to a simple cut in (SED-fit-based) sSFR. 

\begin{figure}
\begin{center}
\includegraphics[width=85mm,height=140mm,angle=90]{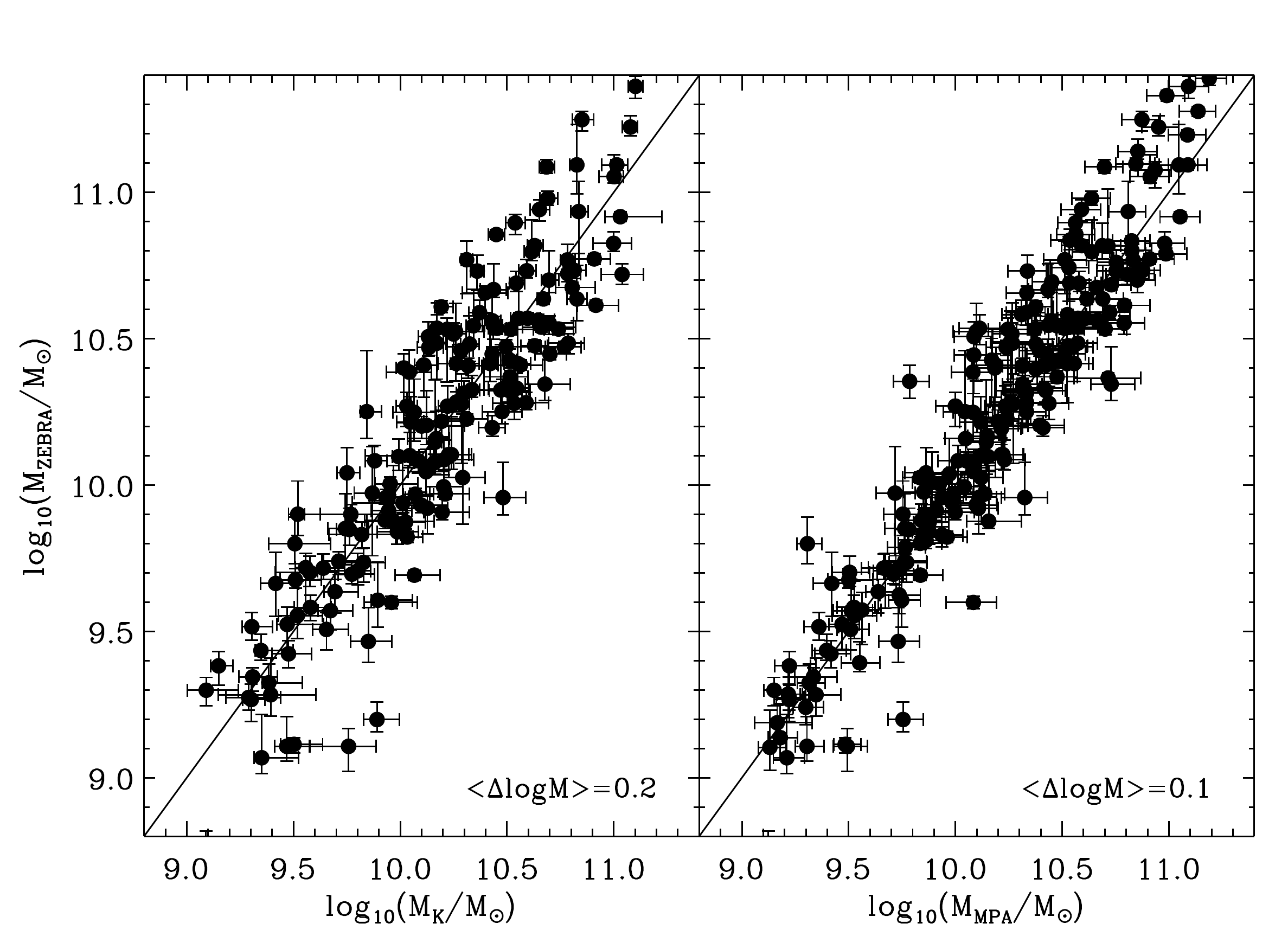}
\end{center}
\caption{\label{fig:my_k_masses} Comparison between stellar masses available in the literature and  best fit masses measured by us with ZEBRA+ for a subsample of ZENS galaxies. The left panel shows the comparison with the
SDSS DR4  \citep{Kauffmann_et_al_2003b} mass estimates based on the analysis of spectra absorption lines. 
The right panel shows the comparison with the DR7 masses based on photometric fits \citep[see][]{Salim_et_al_2007}. 
The median mass difference between our ZEBRA+ masses and the published ones is  in both cases
reported  at the bottom-right corner of the panel.}
\end{figure}

As a further indication of the robustness of our stellar masses, we can use published measurements for a direct comparison.
For about 200  galaxies in our sample, stellar masses are available from SDSS DR4 \citep[derived from stellar absorption lines]{Kauffmann_et_al_2003b} and DR7
\citep[calculated from fits to the photometry]{Salim_et_al_2007} analyses. 
 In Figure \ref{fig:my_k_masses} we show  the comparison between these published masses  and ours, estimated with ZEBRA+ SED fits.
Figure \ref{fig:my_k_masses} shows a good agreement between our masses and the values in the literature, with typical scatter of about $0.1-0.2$ dex, consistent with the
estimate calculated from the ZEBRA likelihoods. At a conservative $3\sigma$ level, the scatter in the stellar mass estimates is $\sim 0.6$, i.e., remains within a factor of four in the worst cases.

 \subsection{Inclination effects}
 
The derived photometric quantities need to be checked for a possible dependence on inclination (i.e., ellipticity).
The light absorption by the interstellar dust  is more severe at shorter wavelengths than in the red part of the spectrum and hence a poor treatment of the internal dust extinction  could lead to fitting highly inclined galaxies with preferentially older, less active -- hence redder -- templates than those used for face-on galaxies.
\citet{Maller_et_al_2009}, for example, show a systematic offset in the estimates of stellar masses  for edge-on and face on galaxies in  the SDSS analysis of  Kauffmann et al. (2003b; we do not detect this effect in our comparison above due to the rather limited number of galaxies in the comparison). 

\begin{figure*}
\begin{center}
\includegraphics[width=70mm,height=160mm,angle=90]{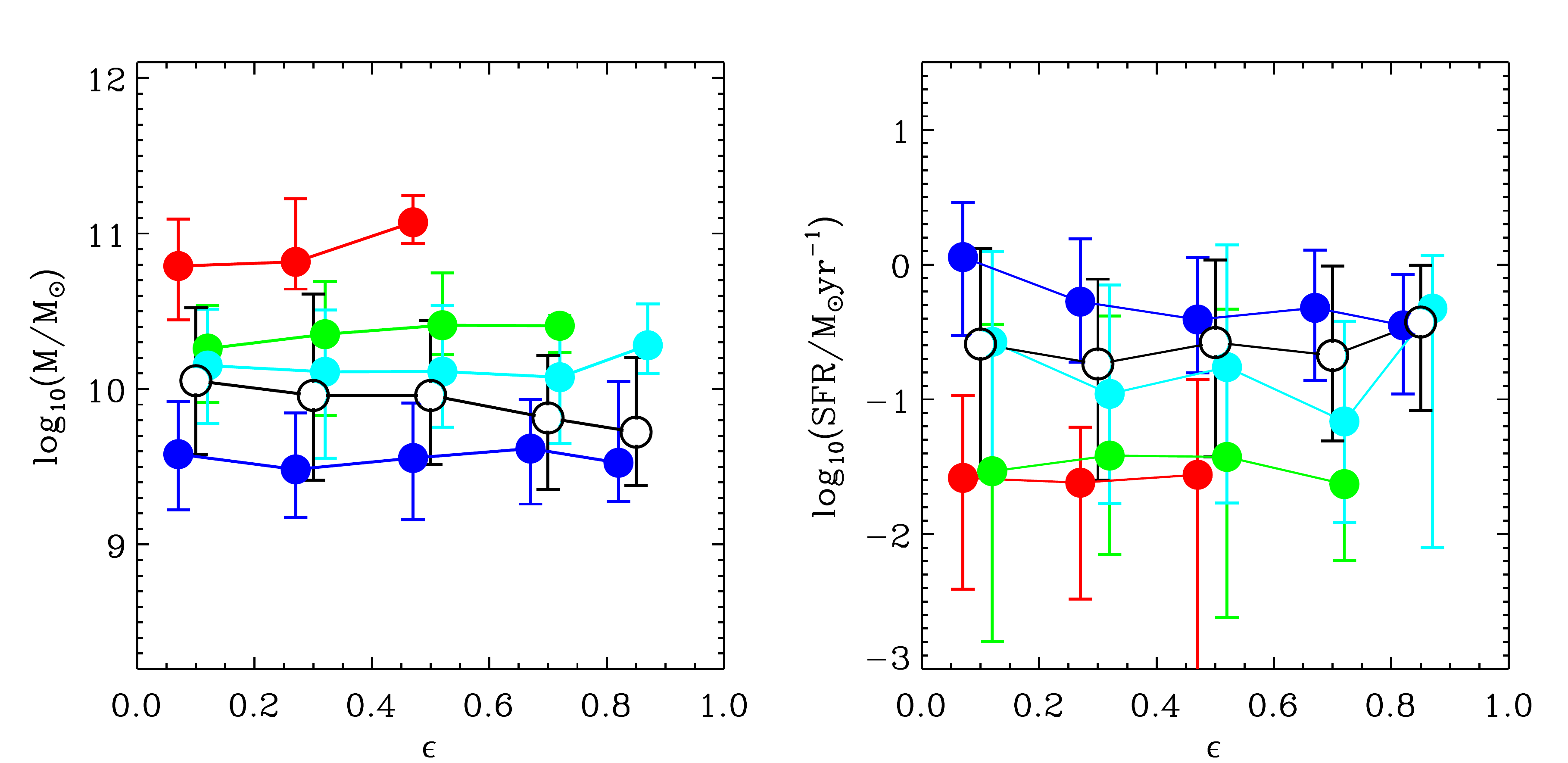} 
\end{center}
\caption{\label{fig:MassEllip}Median galaxy stellar mass and SFR as a function of ellipticity. Symbols with error bars are the median, 
in five ellipticity bins, for the global galaxy sample (stars, black in the online version), and for morphologically-split samples of E/S0 galaxies (black, red in the online version),
 bulge-dominated spirals (dark gray, green in the online version), intermediate-type disks (gray, cyan in the online version) and late-type disks (empty circles, blue in the online version). 
 Error-bars show the 25th and 75th percentiles of the distributions. Ellipticities and morphological classes are taken from Paper II.\\
(A color version of this figure is available in the online journal.) } 
\end{figure*}

Using the structural parameters and morphological classifications presented in Paper II, we show in Figures \ref{fig:MassEllip} the ZEBRA+ galaxy masses and SFRs as a function of the measured galaxy ellipticity, both for the full sample of galaxies and also splitting the sample by morphological type.
Considering the global sample we find a slight decrease of stellar masses with ellipticity: looking at the morphological type separately, however, we see that the median mass is remarkably flat with increasing galaxy elongation and the aforementioned trend is simply a consequence of a changing
morphological mix at a given axis ratio plus a different mass function of galaxy types.
Regarding the SFR, there is a very modest decrease of the median SFR with ellipticity
for late disk galaxies only, although the distributions for face-on and edge-on galaxies are statistically consistent.
The other morphological types  do not show any trends of masses and SFRs with ellipticity.
We  conclude that our estimates of (s)SFRs and stellar masses are not significantly biased with ellipticity.

\begin{thebibliography}{}
\bibitem[Abazajian et al.(2009)]{Abazajian_et_al_2009} Abazajian, K.~N., Adelman-McCarthy, J.~K., Ag{\"u}eros, M.~A., et al.\ 2009, \apjs, 182, 543 
\bibitem[Annibali et al.(2010)]{Annibali_et_al_2010} Annibali, F., Bressan, A., Rampazzo, R., Zeilinger, W.~W., Vega, O., \& Panuzzo, P.\ 2010, \aap, 519, A40 
\bibitem[Bah{\'e} et al.(2013)]{Bahe_et_al_2013} Bah{\'e}, Y.~M., McCarthy, I.~G., Balogh, M.~L., \& Font, A.~S.\ 2013, \mnras, 430, 3017 
\bibitem[Baldry et al.(2006)]{Baldry_et_al_2006} Baldry, I.~K., Balogh, M.~L., Bower, R.~G., et al.\ 2006, \mnras, 373, 469 
\bibitem[Baldwin et al.(1981)]{Baldwin_et_al_1981} Baldwin, J.~A., Phillips, M.~M., \& Terlevich, R.\ 1981, \pasp, 93, 5 
\bibitem[Balogh et al.(2000)]{Balogh_et_al_2000} Balogh, M.~L., Navarro, J.~F., \& Morris, S.~L.\ 2000, \apj, 540, 113 
\bibitem[Balogh et al.(2004)]{Balogh_et_al_2004} Balogh, M.~L., Baldry, I.~K., Nichol, R., Miller, C., Bower, R., \& Glazebrook, K.\ 2004, \apjl, 615, L101 
\bibitem[Bamford et al.(2009)]{Bamford_et_al_2009} Bamford, S.~P., et al.\ 2009, \mnras, 393, 1324 
\bibitem[Bartholomew et al.(2001)]{Bartholomew_et_al_2001} Bartholomew, L.~J., Rose, J.~A., Gaba, A.~E., \& Caldwell, N.\ 2001, \aj, 122, 2913 
\bibitem[Bell et al.(2004)]{Bell_et_al_2004} Bell, E.~F., McIntosh, D.~H., Barden, M., et al.\ 2004, \apjl, 600, L11 
\bibitem[Bertin \& Arnouts(1996)]{Bertin_et_Arnouts_1996} Bertin, E., \& Arnouts, S.\ 1996, \aaps, 117, 393 
\bibitem[Biviano et al.(2002)]{Biviano_et_al_2002} Biviano, A., Katgert, P., Thomas, T., \& Adami, C.\ 2002, \aap, 387, 8 
\bibitem[Blanton \& Roweis(2007)]{Blanton_Roweis_2007} Blanton, M.~R., \& Roweis, S.\ 2007, \aj, 133, 734 
\bibitem[Bolzonella et al.(2010)]{Bolzonella_et_al_2010} Bolzonella, M., Kova{\v c}, K., Pozzetti, L., et al.\ 2010, \aap, 524, A76 
\bibitem[Brinchmann et al.(2004)]{Brinchmann_et_al_2004} Brinchmann, J., Charlot, S., White, S.~D.~M., Tremonti, C., Kauffmann, G., Heckman, T., \& Brinkmann, J.\ 2004, \mnras, 351, 1151 
\bibitem[Bruzual(2007)]{Bruzual_2007} Bruzual, A.~G.\ 2007, IAU Symposium, 241, 125 
\bibitem[Bruzual \& Charlot(2003)]{Bruzual_Charlot_2003} Bruzual, G., \& Charlot, S.\ 2003, \mnras, 344, 1000 
\bibitem[Bundy et al.(2006)]{Bundy_et_al_2006} Bundy, K., Ellis, R.~S., Conselice, C.~J., et al.\ 2006, \apj, 651, 120 
\bibitem[Bundy et al.(2010)]{Bundy_et_al_2010} Bundy, K., Scarlata, C., Carollo, C.~M., et al.\ 2010, \apj, 719, 1969 
\bibitem[Caldwell(1984)]{Caldwell_1984} Caldwell, N.\ 1984, \pasp, 96, 287 
\bibitem[Calvi et al.(2012)]{Calvi_et_al_2012} Calvi, R., Poggianti, B.~M., Fasano, G., \& Vulcani, B.\ 2012, \mnras, 419, L14 
\bibitem[Calzetti et al.(2000)]{Calzetti_et_al_2000} Calzetti, D., Armus, L., Bohlin, R.~C., Kinney, A.~L., Koornneef, J., \& Storchi-Bergmann, T.\ 2000, \apj, 533, 682 
\bibitem[Cameron(2011)]{Cameron_2011} Cameron, E.\ 2011, PASA, 28, 128 
\bibitem[Cappellari \& Copin(2003)]{Cappellari_Copin_2003} Cappellari, M., \& Copin, Y.\ 2003, \mnras, 342, 345 
\bibitem[Cardelli et al.(1989)]{Cardelli_et_al_1989} Cardelli, J.~A., Clayton, G.~C., \& Mathis, J.~S.\ 1989, \apj, 345, 245
\bibitem[Carollo et al.(2007)]{Carollo_et_al_2007} Carollo, C.~M., Scarlata, C., Stiavelli, M., Wyse, R.~F.~G., \& Mayer, L.\ 2007, \apj, 658, 960  
\bibitem[Carollo et al.(2013)]{Carollo_et_al_2013a} Carollo, C.~M., Cibinel, A., Lilly, S.~J., et al.\ 2013, \apj, 776, 71 (Paper I)
\bibitem[Cattaneo et al.(2006)]{Cattaneo_et_al_2006} Cattaneo, A., Dekel, A., Devriendt, J., Guiderdoni, B., \& Blaizot, J.\ 2006, \mnras, 370, 1651 
\bibitem[Cen(2011)]{Cen_2011} Cen, R.\ 2011, \apj, 741, 99 
\bibitem[Chabrier(2003)]{Chabrier_2003} Chabrier, G.\ 2003, \pasp, 115, 763
\bibitem[Cibinel et al.(2013)]{Cibinel_et_al_2013} Cibinel, A., Carollo, C.~M., Lilly, S.~J., et al.\ 2013, \apj, 776, 72 (Paper II) 
\bibitem[Colless et al.(2001)]{Colless_et_al_2001} Colless, M., et al.\ 2001, \mnras, 328, 1039 
\bibitem[Couch \& Sharples(1987)]{Couch_Sharples_1987} Couch, W.~J., \& Sharples, R.~M.\ 1987, \mnras, 229, 423 
\bibitem[Crain et al.(2009)]{Crain_et_al_2009} Crain, R.~A., Theuns, T., Dalla Vecchia, C., et al.\ 2009, \mnras, 399, 1773 
\bibitem[Croton et al.(2005)]{Croton_et_al_2005} Croton, D.~J., Farrar, G.~R., Norberg, P., et al.\ 2005, \mnras, 356, 1155 
\bibitem[Daigle et al.(2006)]{Daigle_et_al_2006} Daigle, O., Carignan, C., Amram, P., et al.\ 2006, \mnras, 367, 469 
\bibitem[Davis et al.(2005)]{Davis_et_al_2005} Davis, M., Gerke, B.~F., Newman, J.~A., \& Deep2 Team 2005, Observing Dark Energy, 339, 128 
\bibitem[Davoodi et al.(2006)]{Davoodi_et_al_2006} Davoodi, P., Pozzi, F., Oliver, S., et al.\ 2006, \mnras, 371, 1113 
\bibitem[de Jong(1996)]{deJong_1996} de Jong, R.~S.\ 1996, \aap, 313, 377 
\bibitem[Dekel \& Birnboim(2006)]{Dekel_Birnboim_2006} Dekel, A., \& Birnboim, Y.\ 2006, \mnras, 368, 2 
\bibitem[De Lucia \& Blaizot(2007)]{DeLucia_Blaizot_2007} De Lucia, G., \& Blaizot, J.\ 2007, \mnras, 375, 2 
\bibitem[De Lucia et al.(2012)]{DeLucia_et_al_2012} De Lucia, G., Weinmann, S., Poggianti, B.~M., Arag{\'o}n-Salamanca, A., \& Zaritsky, D.\ 2012, \mnras, 423, 1277 
\bibitem[Diehl \& Statler(2006)]{Diehl_Statler_2006} Diehl, S., \& Statler, T.~S.\ 2006, \mnras, 368, 497 
\bibitem[Dressler(1980)]{Dressler_1980} Dressler, A.\ 1980, \apj, 236, 351 
\bibitem[Dressler \& Gunn(1983)]{Dressler_Gunn_1983} Dressler, A., \& Gunn, J.~E.\ 1983, \apj, 270, 7 
\bibitem[Ebeling et al.(2006)]{Ebeling_et_al_2006} Ebeling, H., White, D.~A., \& Rangarajan, F.~V.~N.\ 2006, \mnras, 368, 65 
\bibitem[Eke et al.(2004)]{Eke_et_al_2004} Eke, V.~R., Baugh, C.~M., Cole, S., et al.\ 2004, \mnras, 348, 866 
\bibitem[Feldmann et al.(2006)]{Feldmann_et_al_2006} Feldmann, R., Carollo, C.~M., Porciani, C., et al.\ 2006, \mnras, 372, 565 
\bibitem[Feldmann et al.(2010)]{Feldmann_et_al_2010} Feldmann, R., Carollo, C.~M., Mayer, L., Renzini, A., Lake, G., Quinn, T., Stinson, G.~S., \& Yepes, G.\ 2010, \apj, 709, 218 
\bibitem[Feldmann et al.(2011)]{Feldmann_et_al_2011} Feldmann, R., Carollo, C.~M., \& Mayer, L.\ 2011, \apj, 736, 88 
\bibitem[Ferreras et al.(2005)]{Ferreras_et_al_2005} Ferreras, I., Lisker, T., Carollo, C.~M., Lilly, S.~J., \& Mobasher, B.\ 2005, \apj, 635, 243 
\bibitem[Font et al.(2008)]{Font_et_al_2008} Font, A.~S., Bower, R.~G., McCarthy, I.~G., et al.\ 2008, \mnras, 389, 1619 
\bibitem[Fontanot et al.(2009)]{Fontanot_et_al_2009} Fontanot, F., De Lucia, G., Monaco, P., Somerville, R.~S., \& Santini, P.\ 2009, \mnras, 397, 1776 
\bibitem[Gallazzi et al.(2005)]{Gallazzi_et_al_2005} Gallazzi, A., Charlot, S., Brinchmann, J., White, S.~D.~M., \& Tremonti, C.~A.\ 2005, \mnras, 362, 41 
\bibitem[Gao et al.(2005)]{Gao_et_al_2005} Gao, L., Springel, V., \& White, S.~D.~M.\ 2005, \mnras, 363, L66 
\bibitem[Gao et al.(2004)]{Gao_et_al_2004} Gao, L., White, S.~D.~M., Jenkins, A., Stoehr, F., \& Springel, V.\ 2004, \mnras, 355, 819 
\bibitem[Gil de Paz et al.(2007)]{GildePaz_et_al_2007} Gil de Paz, A., Boissier, S., Madore, B.~F., et al.\ 2007, \apjs, 173, 185 
\bibitem[Giodini et al.(2012)]{Giodini_et_al_2012} Giodini, S., Finoguenov, A., Pierini, D., et al.\ 2012, \aap, 538, A104 
\bibitem[Gonzalez-Perez et al.(2011)]{Gonzalez_et_al_2011} Gonzalez-Perez, V., Castander, F.~J., \& Kauffmann, G.\ 2011, \mnras, 411, 1151 
\bibitem[Gunn \& Gott(1972)]{Gunn_Gott_1972} Gunn, J.~E., \& Gott, J.~R., III 1972, \apj, 176, 1 
\bibitem[Guo et al.(2009)]{Guo_et_al_2009} Guo, Y., McIntosh, D.~H., Mo, H.~J., et al.\ 2009, \mnras, 398, 1129 
\bibitem[Guo et al.(2011)]{Guo_et_al_2011} Guo, Q., White, S., Boylan-Kolchin, M., et al.\ 2011, \mnras, 413, 101
\bibitem[Hahn et al.(2007)]{Hahn_et_al_2007} Hahn, O., Porciani, C., Carollo, C.~M., \& Dekel, A.\ 2007, \mnras, 375, 489 
\bibitem[Hahn et al.(2009)]{Hahn_et_al_2009} Hahn, O., Porciani, C., Dekel, A., \& Carollo, C.~M.\ 2009, \mnras, 398, 1742  
\bibitem[Haines et al.(2008)]{Haines_et_al_2008} Haines, C.~P., Gargiulo, A., \& Merluzzi, P.\ 2008, \mnras, 385, 1201 
\bibitem[Heckman(1980)]{Heckman_1980} Heckman, T.~M.\ 1980, \aap, 87, 152 
\bibitem[Ilbert et al.(2010)]{Ilbert_et_al_2010} Ilbert, O., Salvato, M., Le Floc'h, E., et al.\ 2010, \apj, 709, 644 
\bibitem[Ilbert et al.(2013)]{Ilbert_et_al_2013} Ilbert, O., McCracken, H.~J., Le Fevre, O., et al.\ 2013, arXiv:1301.3157 
\bibitem[Jansen et al.(2000)]{Jansen_et_al_2000} Jansen, R.~A., Franx, M., Fabricant, D., \& Caldwell, N.\ 2000, \apjs, 126, 271 
\bibitem[Kauffmann et al.(2003)]{Kauffmann_et_al_2003} Kauffmann, G., et al.\ 2003, \mnras, 346, 1055 
\bibitem[Kauffmann et al.(2003b)]{Kauffmann_et_al_2003b} Kauffmann, G., Heckman, T.~M., White, S.~D.~M., et al.\ 2003, \mnras, 341, 33 
\bibitem[Kawata \& Mulchaey(2008)]{Kawata_Mulchaey_2008} Kawata, D., \& Mulchaey, J.~S.\ 2008, \apjl, 672, L103 
\bibitem[Khochfar \& Ostriker(2008)]{Khochfar_Ostriker_2008} Khochfar, S., \& Ostriker, J.~P.\ 2008, \apj, 680, 54 
\bibitem[Kimm et al.(2009)]{Kimm_et_al_2009} Kimm, T., Somerville, R.~S., Yi, S.~K., et al.\ 2009, \mnras, 394, 1131
\bibitem[Knobel et al.(2012)]{Knobel_et_al_2012} Knobel, C., Lilly, S.~J., Kovac, K., et al.\ 2012, arXiv:1211.5607 
\bibitem[Koopmann \& Kenney(1998)]{Koopmann_et_al_1998} Koopmann, R.~A., \& Kenney, J.~D.~P.\ 1998, \apjl, 497, L75  
\bibitem[Kova{\v c} et al.(2010)]{Kovac_et_al_2010}Kova{\v c}, K., Lilly, S.~J., Knobel, C., et al.\ 2010, \apj, 718, 86 
\bibitem[Kron(1980)]{Kron_1980} Kron, R.~G.\ 1980, \apjs, 43, 305 
\bibitem[Larson et al.(1980)]{Larson_et_al_1980} Larson, R.~B., Tinsley, B.~M., \& Caldwell, C.~N.\ 1980, \apj, 237, 692 
\bibitem[Lemaux et al.(2010)]{Lemaux_et_al_2010} Lemaux, B.~C., Lubin, L.~M., Shapley, A., Kocevski, D., Gal, R.~R., \& Squires, G.~K.\ 2010, \apj, 716, 970
\bibitem[Lilly et al.(2007)]{Lilly_et_al_2007} Lilly, S.~J., Le F{\`e}vre, O., Renzini, A., et al.\ 2007, \apjs, 172, 70 
\bibitem[Lilly et al.(2009)]{Lilly_et_al_2009} Lilly, S.~J., Le Brun, V., Maier, C., et al.\ 2009, \apjs, 184, 218 
\bibitem[MacArthur et al.(2004)]{MacArthur_et_al_2004} MacArthur, L.~A., Courteau, S., Bell, E., \& Holtzman, J.~A.\ 2004, \apjs, 152, 175 
\bibitem[Macchetto et al.(1996)]{Macchetto_et_al_1996} Macchetto, F., Pastoriza, M., Caon, N., Sparks, W.~B., Giavalisco, M., Bender, R., \& Capaccioli, M.\ 1996, \aaps, 120, 463
\bibitem[Madgwick et al.(2002)]{Madgwick_et_al_2002} Madgwick, D.~S., Lahav, O., Baldry, I.~K., et al.\ 2002, \mnras, 333, 133 
\bibitem[Maller et al.(2009)]{Maller_et_al_2009} Maller, A.~H., Berlind, A.~A., Blanton, M.~R., \& Hogg, D.~W.\ 2009, \apj, 691, 394 
\bibitem[Martin et al.(2005)]{Martin_et_al_2005} Martin, D.~C., Fanson, J., Schiminovich, D., et al.\ 2005, \apjl, 619, L1 
\bibitem[Maulbetsch et al.(2007)]{Maulbetsch_et_al_2007} Maulbetsch, C., Avila-Reese, V., Col{\'{\i}}n, P., et al.\ 2007, \apj, 654, 53 
\bibitem[McCarthy et al.(2008)]{McCarthy_et_al_2008} McCarthy, I.~G., Frenk, C.~S., Font, A.~S., et al.\ 2008, \mnras, 383, 593 
\bibitem[McGee et al.(2011)]{McGee_et_al_2011} McGee, S.~L., Balogh,M.~L., Wilman, D.~J., et al.\ 2011, \mnras, 413, 996 
\bibitem[Moss \& Whittle(2000)]{Moss_Whittle_2000} Moss, C., \& Whittle, M.\ 2000, \mnras, 317, 667 
\bibitem[Newman et al.(2013)]{Newman_et_al_2013} Newman, J.~A., Cooper, M.~C., Davis, M., et al.\ 2013, \apjs, 208, 5 
\bibitem[Oemler(1974)]{Oemler_1974} Oemler, A., Jr.\ 1974, \apj, 194, 1 
\bibitem[Oesch et al.(2010)]{Oesch_et_al_2010} Oesch, P.~A., Carollo, C.~M., Feldmann, R., et al.\ 2010, \apjl, 714, L47 
\bibitem[Oke(1974)]{Oke_1974} Oke, J.~B.\ 1974, \apjs, 27, 21 
\bibitem[Pannella et al.(2009)]{Pannella_et_al_2009} Pannella, M., Gabasch, A., Goranova, Y., et al.\ 2009, \apj, 701, 787 
\bibitem[Peletier et al.(1990)]{Peletier_et_al_1990} Peletier, R.~F., Davies, R.~L., Illingworth, G.~D., Davis, L.~E., \& Cawson, M.\ 1990, \aj, 100, 1091
\bibitem[Peng et al.(2010)]{Peng_et_al_2010} Peng, Y.-j., Lilly, S.~J., Kova{\v c}, K., et al.\ 2010, \apj, 721, 193 
\bibitem[Peng et al.(2012)]{Peng_et_al_2012} Peng, Y.-j., Lilly, S.~J., Renzini, A., \& Carollo, M.\ 2012, \apj, 757, 4 
\bibitem[Phillips et al.(1986)]{Phillips_et_al_1986} Phillips, M.~M., Jenkins, C.~R., Dopita, M.~A., Sadler, E.~M., \& Binette, L.\ 1986, \aj, 91, 1062 
\bibitem[Poggianti et al.(2009)]{Poggianti_et_al_2009} Poggianti, B.~M., Arag{\'o}n-Salamanca, A., Zaritsky, D., et al.\ 2009, \apj, 693, 112 
\bibitem[Poggianti et al.(1999)]{Poggianti_et_al_1999} Poggianti, B.~M., Smail, I., Dressler, A., et al.\ 1999, \apj, 518, 576 
\bibitem[Pozzetti et al.(2007)]{Pozzetti_et_al_2007} Pozzetti, L., Bolzonella, M., Lamareille, F., et al.\ 2007, \aap, 474, 443 
\bibitem[Pozzetti et al.(2010)]{Pozzetti_et_al_2010} Pozzetti, L., Bolzonella, M., Zucca, E., et al.\ 2010, \aap, 523, A13 
\bibitem[Pracy et al.(2005)]{Pracy_et_al_2005} Pracy, M.~B., Driver, S.~P., De Propris, R., Couch, W.~J., \& Nulsen, P.~E.~J.\ 2005, \mnras, 364, 1147 
\bibitem[Rasmussen et al.(2012)]{Rasmussen_et_al_2012} Rasmussen, J., Mulchaey, J.~S., Bai, L., et al.\ 2012, \apj, 757, 122 
\bibitem[Rose et al.(2001)]{Rose_et_al_2001} Rose, J.~A., Gaba, A.~E., Caldwell, N., \& Chaboyer, B.\ 2001, \aj, 121, 793 
\bibitem[Salim et al.(2007)]{Salim_et_al_2007} Salim, S., Rich, R.~M., Charlot, S., et al.\ 2007, \apjs, 173, 267 
\bibitem[Sanders \& Fabian(2001)]{Sanders_Fabian_2001} Sanders, J.~S., \& Fabian, A.~C.\ 2001, \mnras, 325, 178 
\bibitem[Schechter(1976)]{Schechter_1976} Schechter, P.\ 1976, \apj, 203, 297 
\bibitem[Schlegel et al.(1998)]{Schlegel_et_al_1998} Schlegel, D.~J., Finkbeiner, D.~P., \& Davis, M.\ 1998, \apj, 500, 525 
\bibitem[Sersic(1968)]{Sersic_1968} Sersic, J.~L.\ 1968, Cordoba, Argentina: Observatorio Astronomico, 1968
\bibitem[Sheth \& Tormen(2004)]{Sheth_Tormen_2004} Sheth, R.~K., \& Tormen, G.\ 2004, \mnras, 350, 1385 
\bibitem[Simard et al.(2002)]{Simard_et_al_2002} Simard, L., Willmer, C.~N.~A., Vogt, N.~P., et al.\ 2002, \apjs, 142, 1 
\bibitem[Skibba(2009)]{Skibba_2009} Skibba, R.~A.\ 2009, \mnras, 392, 1467 
\bibitem[Skrutskie et al.(2006)]{Skrutskie_et_al_2006} Skrutskie, M.~F., Cutri, R.~M., Stiening, R., et al.\ 2006, \aj, 131, 1163 
\bibitem[Springel et al.(2005)]{Springel_et_al_2005} Springel, V., White, S.~D.~M., Jenkins, A., et al.\ 2005, \nat, 435, 629 
\bibitem[Taylor et al.(2005)]{Taylor_et_al_2005} Taylor, V.~A., Jansen, R.~A., Windhorst, R.~A., Odewahn, S.~C., \& Hibbard, J.~E.\ 2005, \apj, 630, 784 
\bibitem[Tortora et al.(2010)]{Tortora_et_al_2010} Tortora, C., Napolitano, N.~R., Cardone, V.~F., Capaccioli, M., Jetzer, P., \& Molinaro, R.\ 2010, \mnras, 407, 144
\bibitem[Tremonti et al.(2004)]{Tremonti_et_al_2004} Tremonti, C.~A., et al.\ 2004, \apj, 613, 898 
\bibitem[Tresse et al.(1996)]{Tresse_et_al_1997} Tresse, L., Rola, C., Hammer, F., Stasi{\'n}ska, G., Le Fevre, O., Lilly, S.~J., \& Crampton, D.\ 1996, \mnras, 281, 847 
\bibitem[van den Bosch et al.(2008)]{van_den_Bosch_2008} van den Bosch, F.~C., Aquino, D., Yang, X., Mo, H.~J., Pasquali, A., McIntosh, D.~H., Weinmann, S.~M., \& Kang, X.\ 2008, \mnras, 387, 79 
\bibitem[van der Wel(2008)]{van_der_Wel_2008} van der Wel, A.\ 2008, \apjl, 675, L13 
\bibitem[Veilleux \& Osterbrock(1987)]{Veilleux_Osterbrock_1987} Veilleux, S., \& Osterbrock, D.~E.\ 1987, \apjs, 63, 295 
\bibitem[Veron et al.(1997)]{Veron_Cetty_1997} Veron, P., Goncalves, A.~C., \& Veron-Cetty, M.-P.\ 1997, \aap, 319, 52 
\bibitem[von der Linden et al.(2010)]{von_der_Linden_et_al_2010} von der Linden, A., Wild, V., Kauffmann, G., White, S.~D.~M., \& Weinmann, S.\ 2010, \mnras, 404, 1231
\bibitem[Wang et al.(2007)]{Wang_et_al_2007} Wang, L., Li, C., Kauffmann, G., \& De Lucia, G.\ 2007, \mnras, 377, 1419  
\bibitem[Weinmann et al.(2006)]{Weinmann_et_al_2006} Weinmann, S.~M., van den Bosch, F.~C., Yang, X., \& Mo, H.~J.\ 2006, \mnras, 366, 2 
\bibitem[Weinmann et al.(2009)]{Weinmann_et_al_2009} Weinmann, S.~M., Kauffmann, G., van den Bosch, F.~C., Pasquali, A., McIntosh, D.~H., Mo, H., Yang, X., \& Guo, Y.\ 2009, \mnras, 394, 1213 
\bibitem[Weinmann et al.(2010)]{Weinmann_et_al_2010} Weinmann, S.~M., Kauffmann, G., von der Linden, A., \& De Lucia, G.\ 2010, \mnras, 406, 2249 
\bibitem[Wetzel et al.(2012a)]{Wetzel_et_al_2012} Wetzel, A.~R., Tinker, J.~L., \& Conroy, C.\ 2012a, \mnras, 424, 232 
\bibitem[Wetzel et al.(2012b)]{Wetzel_et_al_2012b} Wetzel, A.~R., Tinker, J.~L., Conroy, C., \& van den Bosch, F.~C.\ 2012b, arXiv:1206.3571 
\bibitem[Williams et al.(2009)]{Williams_et_al_2009} Williams, R.~J., Quadri, R.~F., Franx, M., van Dokkum, P., \& Labb{\'e}, I.\ 2009, \apj, 691, 1879
\bibitem[Woo et al.(2013)]{Woo_et_al_2013} Woo, J., Dekel, A., Faber, S.~M., et al.\ 2013, \mnras, 428, 3306  
\bibitem[Wyder et al.(2005)]{Wyder_et_al_2005} Wyder, T.~K., Treyer, M.~A., Milliard, B., et al.\ 2005, \apjl, 619, L15 
\bibitem[Yan et al.(2006)]{Yan_et_al_2006} Yan, R., Newman, J.~A., Faber, S.~M., Konidaris, N., Koo, D., \& Davis, M.\ 2006, \apj, 648, 281
\bibitem[York et al.(2000)]{York_et_al_2000} York, D.~G., Adelman, J., Anderson, J.~E., Jr., et al.\ 2000, \aj, 120, 1579 
\bibitem[Zabludoff et al.(1996)]{Zabludoff_et_al_1996} Zabludoff, A.~I., Zaritsky, D., Lin, H., et al.\ 1996, \apj, 466, 104 
\bibitem[Zibetti et al.(2009)]{Zibetti_et_al_2009} Zibetti, S., Charlot, S., \& Rix, H.-W.\ 2009, \mnras, 400, 1181 
\end{thebibliography}
 \end{document}